\documentclass[9pt]{article}
\usepackage[margin=0.9in]{geometry}
\usepackage[english]{babel}
\usepackage[utf8]{inputenc}

\usepackage{subfiles}
\usepackage{graphicx}
\usepackage{caption}
\usepackage{subcaption}
\usepackage{soul}
\usepackage{color}
\usepackage{float}
\usepackage{amsmath}
\usepackage{tikz}
\usetikzlibrary{shapes}
\usepackage{authblk}
\usepackage[nottoc]{tocbibind}
\usepackage{mathrsfs}
\usepackage{multirow}
\usepackage{diagbox}
\usepackage{xcolor}

\newcommand\Tstrut{\rule{0pt}{2.9ex}}         
\newcommand\Bstrut{\rule[-1.2ex]{0pt}{0pt}}   
\newcommand\TBstrut{\Tstrut\Bstrut}           

\title{Predicting human cooperation: \\ sensitizing drift-diffusion model to interaction and external stimuli}

\date{November 2024}

\author[1]{Lucila G. Alvarez-Zuzek$^{\star}$
\thanks{lalvarezzuzek@fbk.eu}}
\author[2]{Laura Ferrarotti$^{\star}$
\thanks{lferrarotti@fbk.eu}}
\author[2]{Bruno Lepri}
\author[1]{Riccardo Gallotti}
\affil[1]{Complex Human Behavior Laboratory, Fondazione Bruno Kessler, Trento, Italy}
\affil[2]{Mobile and Social Computing Laboratory, Fondazione Bruno Kessler, Trento, Italy}

\begin{document}

    \maketitle
    \def\thefootnote{$\star$}\footnotetext{The authors contributed equally to the work, and they are alphabetically sorted.}
    
    \begin{abstract}
        As humans perceive and actively engage with the world, we adjust our decisions in response to shifting group dynamics and are influenced by social interactions. This study aims to identify which aspects of interaction affect cooperation-defection choices. Specifically, we investigate human cooperation within the Prisoner's Dilemma game, using the Drift-Diffusion Model to describe the decision-making process. We introduce a novel Bayesian model for the evolution of the model's parameters based on the nature of interactions experienced with other players. This approach enables us to predict the evolution of the population's expected cooperation rate. We successfully validate our model using an unseen test dataset and apply it to explore three strategic scenarios: co-player manipulation, use of rewards and punishments, and time pressure. These results support the potential of our model as a foundational tool for developing and testing strategies aimed at enhancing cooperation, ultimately contributing to societal welfare.

    \end{abstract}
    
    \newpage
    \section*{Introduction}
    
    Decision-making, intended as the fundamental capability of selecting the most favorable course of action among various options is an essential part of our daily life \cite{baron2023thinking}. Although fundamental, the human decision-making process is indeed difficult to model, due (among other factors) to the complexity of brain processes \cite{bassett2011understanding, krohn2023spatiotemporal}, the influence of cognitive biases \cite{hilbert2012toward, korteling2020cognitive}, emotions \cite{george2016affect, yukalov2022quantification}, social and environmental factors \cite{bruch2017decision, danziger2011extraneous}, unconscious mechanisms \cite{harmon1999cognitive, kahneman2011thinking}, and individual differences \cite{franken2005individual, appelt2011decision}. The study of human decision-making \cite{edwards1954theory,evans2011dual} is still an open and vast multidisciplinary field that draws on insights from various scientific disciplines, including psychology \cite{kahneman2002representativeness, mishra2014decision}, neuroscience \cite{fellows2004cognitive,yoon2012decision}, economics \cite{dale2015heuristics, american1966theories} and public health \cite{dube2013vaccine, benin2006qualitative}.
    
    Our work focuses on {\it social decisions}, intended as choices or judgments made by individuals within the context of social interactions or group dynamics \cite{bruch2017decision, ajzen1996social}. Social decisions can encompass a wide range of behaviors \cite{stangor2015social}, including cooperation \cite{gong2017social, rand2013human}, competition \cite{keddy2012competition, toma2013strategic, garcia2013psychology}, altruism \cite{batson2010altruism, tusche2021neurocomputational}, reciprocity \cite{rossetti2024direct, solanas2009measuring}, conformity \cite{romano2017reciprocity, constant2019regimes, hertz2016influence}, and conflict resolution \cite{aureli2000natural, danesh2002has, kelman1990interactive}. The study of social decisions often involves examining how individuals navigate social situations, make choices that affect themselves and others, and respond to the influence of social factors on their decision-making processes \cite{cikara2014neuroscience}. Social decisions are integral to understanding the dynamics of human societies and the complex interplay between individual preferences and social structures \cite{lee2013how, rilling2011neuroscience, sanfey2007social}. Particularly, given the importance of cooperation as an asset in human societies \cite{rand2013human,boyd2006solving}, being able to deploy policies that enhance it as a preferred choice is crucial. Cooperation is considered fundamental in the development of successful human societies, through fostering social harmony and stability \cite{kwon2019peace, maas2013conflict}, mutual aid and sharing of resources \cite{delay2019mutual, bouma2008trust}, economic prosperity \cite{fomina2018industrial, strachan2018relationship}, cultural and technological advancement \cite{boyd2009culture, johnson2013cooperation}, social evolution and adaptation \cite{van2000cooperation}, among others. Conversely, the lack of cooperation might cause compromising scenarios with respect to societal welfare \cite{suarez2021prevalence, swire2020public}.
    
    In this manuscript, we explore the social decision-making process of cooperation-defection in the Multiplayer Iterated Prisoner’s Dilemma game (MIPD) \cite{carroll1988iterated}. MIPD is an extended version of the classic two-player non-zero-sum Iterated Prisoner’s Dilemma game (IPD), a benchmark among iterative social dilemmas, providing a model for situations in which individual rationality leads often to sub-optimal collective equilibria \cite{basu1977information}. Iterative social dilemmas \cite{lee2008game} is a controlled and versatile framework for studying social decision-making and cooperation, as shown for instance by \cite{rand2016cooperation}. These scenarios offer insights into the strategic considerations, incentives, and dynamics that shape individual and collective behaviors, making them invaluable for researchers seeking to understand the foundations of cooperation \cite{wood2016cooperation, zeng2016risk, li2016changing}. Moreover, the iterative nature of these games allows researchers to observe the evolution of cooperation over time. By studying repeated interactions, it is possible to explore how cooperation behavior develops and adapts within a population \cite{raihani2011resolving, ishibuchi2005evolution}.
    
    We employ a sequential sampling model, {\it Drift Diffusion Model} (DDM) \cite{ratcliff2016diffusion,ratcliff1978theory}, to represent the human binary decision process, when choosing either cooperation or defection. Starting with the model proposed and developed by Ratcliff in \cite{ratcliff1978theory}, a DDM assumes a one-dimensional random walk representing the accumulation in our brain of noisy evidence in favor of one of the two alternative options. The process models the temporal evolution of a random variable $x(t)$ from a point $x(0) = z \cdot a$, where $z \in [\, 0, \, 1 \,]$ represents the initial bias of the individual towards one of the two possible alternatives, embodied by the extremes of the interval $[\, 0, \, a \, ]$. At each time step, as information is randomly collected in favor of one of the two choices, this accumulation is represented as increases or decreases in $x(t)$ (with information accumulation rate $\nu$), up until a decision is made when one of the boundaries $x=a$ or $x=0$ is reached (see Materials and Methods for a mathematical description of the model). From a neuroscience perspective, the DDM free parameters (i.e., $z$, $a$, $\nu$, $t_0$) have the following interpretation: the bias $z$ is related to the prior inclination (the pre-existing opinion) the individual has about the two options, the height of the barrier $a$ quantifies cautiousness in response, the drift rate $\nu$ is related to the quality and velocity of the information extracted from the stimulus, hence to the task complexity, and the non-decision time $t_0$ models the preparation time that individuals need before starting the decision process. The accuracy of this model is measured with the response time probability density functions (PDFs). By fitting the response time PDFs from the data, we can obtain an estimation of the free parameters characterizing the DDM.
    
    The universality of DDMs allowed scientists to use it in a wide spectrum of different topics, in order to understand the cognitive mechanisms behind human decision-making in different contexts. Previous research used the model to describe and explore human cooperation-defection in the IPD-MIPD scenarios \cite{gallotti2019quantitative}. DDMs have also been used to describe human decisions in the context of altruism \cite{hutcherson2015neurocomputational}, food choices \cite{krajbich2015common}, moral judgments \cite{andrejevic2022response}, and more. Notwithstanding this, in the mentioned literature, such models were employed as descriptive models. By employing the classic DDM, we can describe a behavior present in the collected dataset, but we cannot simulate any changing behavior due to mutated environmental conditions and stimuli. The primary objective of our work is to design a predictive model capable of simulating how human behavior towards cooperation/defection choices evolves, in response to various social interactions. Interest in this model is driven by its potential to identify the conditions under which cooperative behavior emerges, as well as its utility as a tool for studying (via learning and/or testing) the impact of policies designed to enhance cooperation through targeted stimuli.
    
    As a first step, we developed a set of regressors that capture individuals' interactions with other players and their propensity to cooperate or defect during a given iteration of the MIPD. These regressors are linked to the parameters of the classic DDM, allowing the DDM parameters at each iteration to be described by a simple parametric function. The parameters of this function become the new free parameters of our model. Using Bayesian regression and a training dataset, we fit these parameters, enabling then, along with iteration-specific regressors, the prediction of the original DDM parameters over time, and thus the evolution of human cooperative behavior. As detailed in section Results, the parameters predicted by our model on a separate test dataset demonstrate to be extremely effective in replicating the response times PDFs at every iteration, as well as in predicting the distribution of the final payoff across the whole population (for a description of training and test datasets, see Materials and Methods). Finally, we explored the results in terms of evolving cooperation rate, simulating our model under three different types of tailored stimuli, obtained acting (i) on the co-players scenarios (either enforcing a specific level of cooperation or opportunely shuffling the playing group); (ii) on the MIPD gain matrices (rewarding cooperation or punishing defection); (iii) on the maximum response time (shortening it to evaluate our model in simulating more spontaneous and less strategic responses). This analysis, included in the section Results, connects well the behavior of our model with interesting findings in literature, related to human behavior.
     
    By modeling interactive human decision-making, we hope to lay down the first brick on the path towards the development of automated systems capable of tuning and enhancing the conditions under which a beneficial collective behavior is maximized, contributing as well to shed light on the understanding of human interaction and cooperation.

    
    \section*{Results}
        First, we explore the performance obtained by enriching the classic DDM with dynamics describing the evolution of time-varying parameters. The construction of such an augmented model is described in detail in Materials and Methods (see section ``Predictive Drift-Diffusion Model"), while here we recall only the fundamental equations. In particular, we propose a predictive model for the evolution of the cooperation-defection decision-making process of players involved in consecutive rounds of MIPD. The model is designed by enriching a DDM with dynamics of the following form
        \begin{equation}
            a_t = F_{a}(x_{t-1}^a, \, P_a), \qquad \nu_t = F_{\nu}(x_{t-1}^{\nu}, \, P_{\nu}), \qquad z_t = F_{z}(x_{t-1}^z, \, P_z), \qquad {t_0}_t = F_{t_0}(x_{t-1}^{{t_0}}, \, P_{t_0}),
        \end{equation}
        describing the variation in time of its parameters (initial bias $z$, barrier length $a$, drift $\nu$ and non-decision time $t_0$) depending on interactions among players involved in the same game (represented by regressors $x_{t-1}^a$, $x_{t-1}^z$, $x_{t-1}^{\nu}$, and $x_{t-1}^{{t_0}}$). After designing such dynamics, we fit the associated parameters $P_a$, $P_{\nu}$, $P_z$ and $P_{t_0}$ over a training dataset (see section ``Data" in Materials and Methods), using Bayesian Regression. The next section discusses the results of the augmented DDM at round $t$,
        \begin{equation}
            \begin{split}
                & d x(i)  =  \nu_t  \; di + \xi(i), \\
                & x({t_0}_t) = z_t \; a_t  
            \end{split}
             \label{DDM-eq-aug}
        \end{equation}
        in terms of accuracy. In the augmented DDM represented in \eqref{DDM-eq-aug}, the letter $i$ represents the model internal time of the decision, whose final value, once a decision is made, corresponds to the response time attained at round $t$. The accuracy of the model is firstly measured by evaluating results in terms of the response time PDFs \cite{bogacz2006physics,ratcliff2008diffusion}, separated for the two possible choices. We consider a test dataset (see section ``Data" in Materials and Methods) containing samples unseen by the model during the training phase: we compare the PDF of the response times empirically estimated from the test data and the PDF of the response times obtained by combining the DDM parameters predicted at each round $t$ of the test set by \eqref{DDM-eq-aug}, with the following theoretical formulas
        \begin{equation}
            \begin{split}
               &P_{\rm D}(i;\nu_t, a_t, z_t) = \frac{\pi}{a_t^2}\; exp \left(- \nu_t\;z_t\;a_t - \frac{\nu_t^2i}{2}\right) \times \sum^{\infty}_{k=1} \; k \; exp \left(-\frac{k^2\pi^2i} {2a_t^2} \right)\; sin(k \; z_t \; \pi), \\
               & P_{\rm C}(i;\nu_t, a_t, z_t) = \frac{\pi}{a_t^2}\; exp \left(\nu_t\;(1-z_t)\;a_t - \frac{\nu_t^2i}{2}\right) \times \sum^{\infty}_{k=1} \; k \; exp \left(-\frac{k^2\pi^2i} {2a_t^2} \right)\; sin(k \; (1-z_t) \; \pi),       
            \end{split}
            \label{RT-PDF-inResults}
        \end{equation}
        representing the probability of experiencing response time $i$ while defecting ($P_{\rm D}$) and cooperating ($P_{\rm C}$) (more details on the formulas in Materials and Methods, section ``Drift-Diffusion Model"). 
        
        Additionally, we test the accuracy of our model with respect to the capability of predicting the final earnings distribution achieved by the players of the MIPD. The considered final earning of an individual is the sum of the MIPD payoffs accumulated by an individual playing the game round after round (for a more detailed definition of the earnings see section ``Multiplayer Iterated Prisoner's Dilemma" in Materials and Methods). We compare the final earnings captured in the training dataset and test dataset, with the averaged final earnings obtained by simulating the behavior of our model on the same data many times and averaging the results for statistical significance.
        \begin{figure}[h]
            \centering
            \begin{subfigure}[b]{0.45\textwidth}
                \centering
                \includegraphics[width=\textwidth]{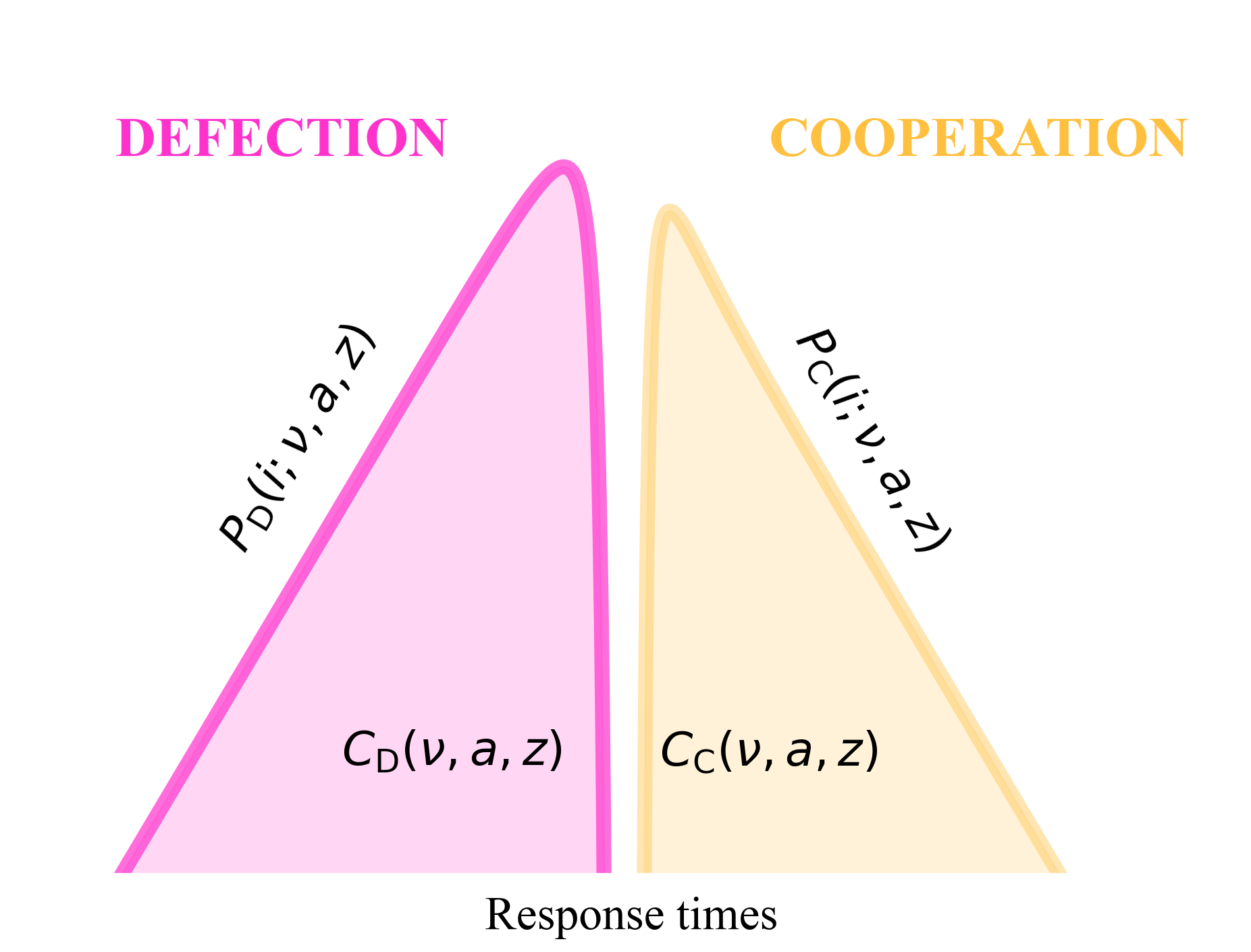}
                \caption{}
                \label{fig:schemePDF}
            \end{subfigure}
            \begin{subfigure}[b]{0.45\textwidth}
                \centering
                \includegraphics[width=\textwidth]{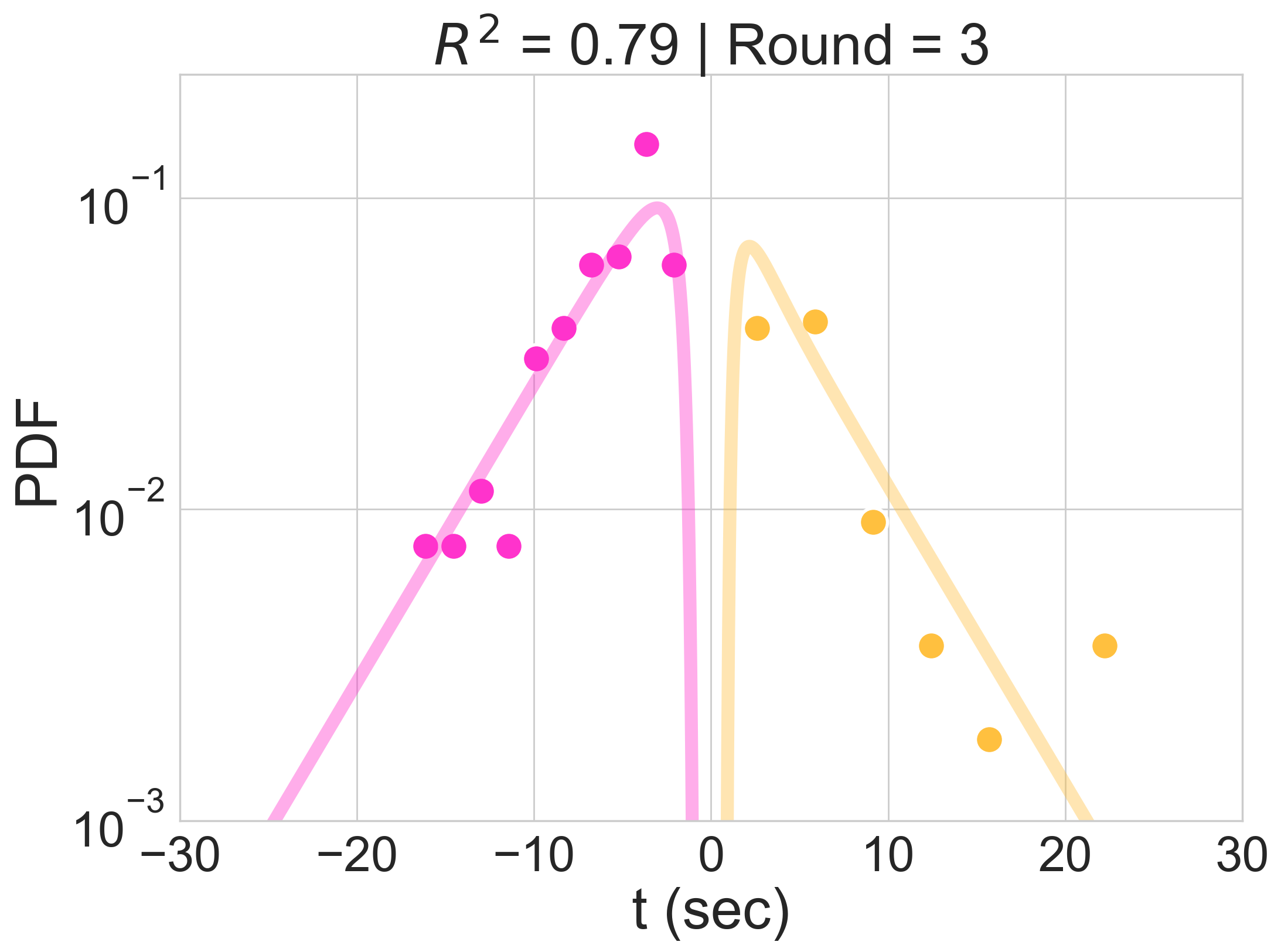}
                \caption{}
                \label{fig:PDF_3}
            \end{subfigure}
            \begin{subfigure}[b]{0.45\textwidth}
                \centering
                \includegraphics[width=\textwidth]{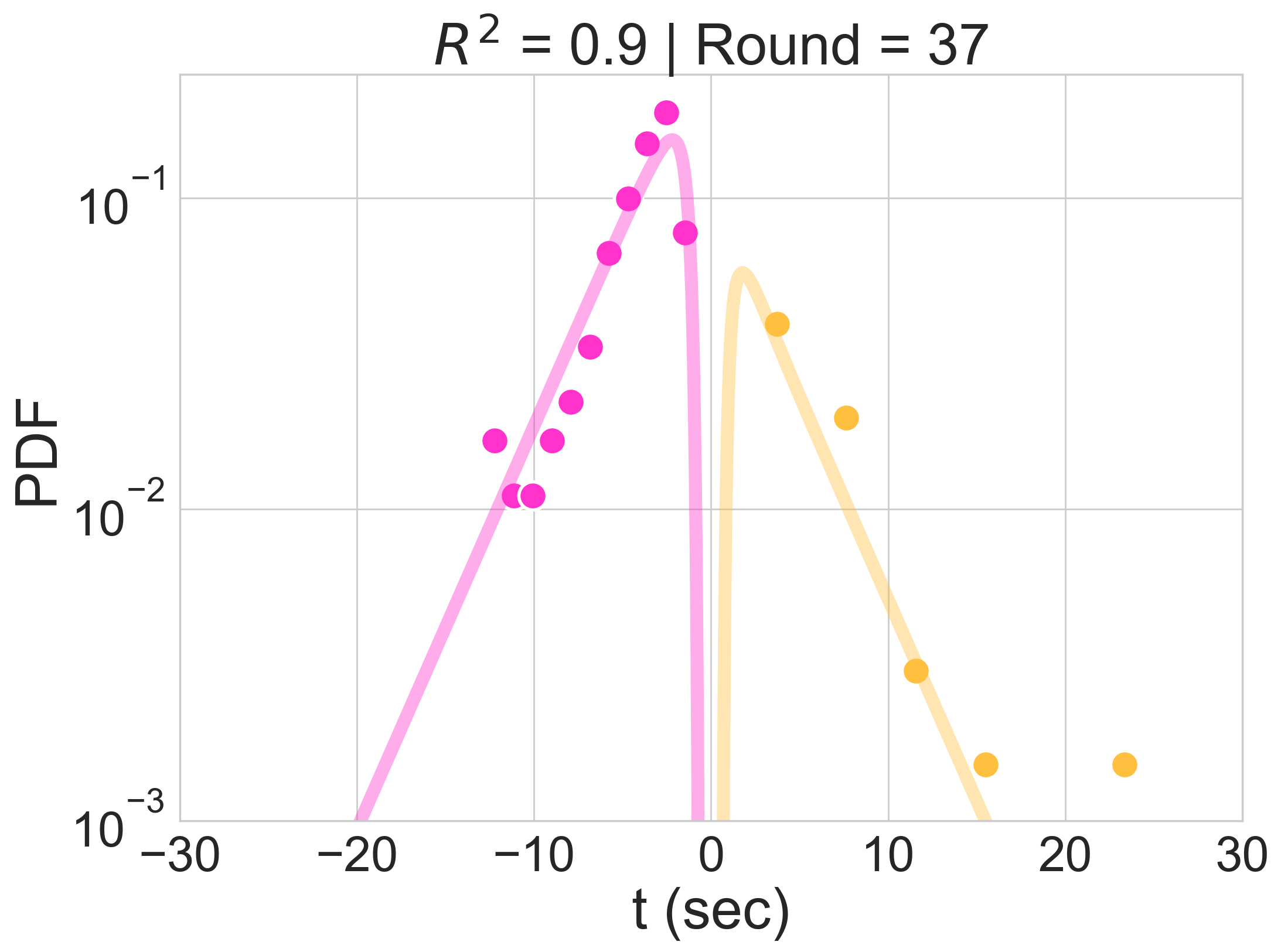}
                \caption{}
                \label{fig:PDF_37}
            \end{subfigure}
            \begin{subfigure}[b]{0.45\textwidth}
                \centering
                \includegraphics[width=\textwidth]{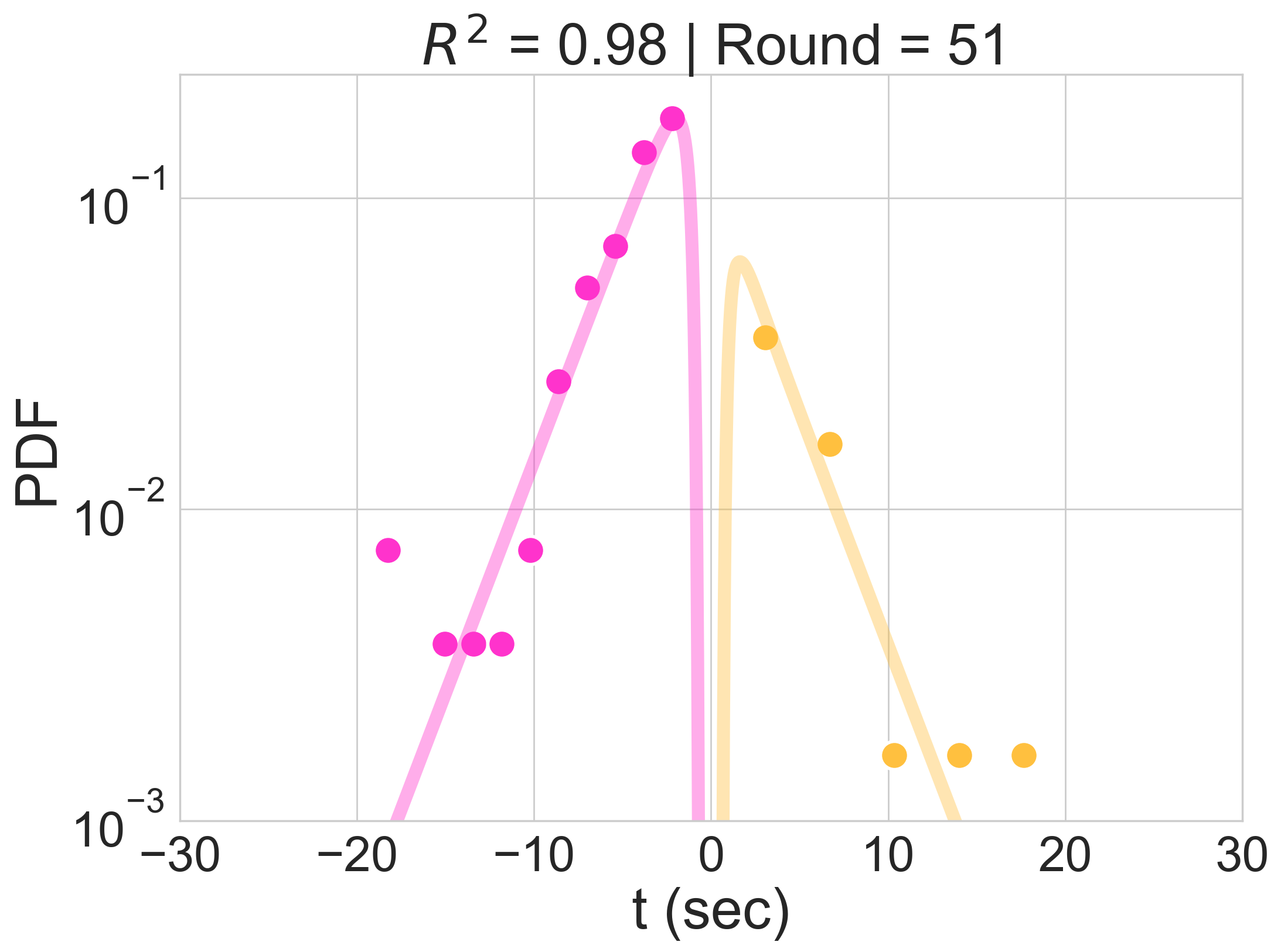}
                \caption{}
                \label{fig:PDF_51}
            \end{subfigure}
            \caption{\textbf{Accuracy over the testing set: PDFs.} \textbf{Panel \ref{fig:schemePDF}} illustrates how to read the plots: the response time PDFs in case of defection and cooperation are shown on the left side (pink) and right side (orange) respectively. Both curves are given by the PDFs equations (Eq.~\ref{RT-PDF-inResults}), and their integral (the area below the curve) represent the expected rate (see Eq.~\ref{expected-coop-eq} in Materials and Methods for more details). \textbf{Panels \ref{fig:PDF_3}-\ref{fig:PDF_37}-\ref{fig:PDF_51}} show the PDFs at rounds $3$ (up-right), $37$ (down-left), and $51$ (down-right). Results are obtained by: i) Eq.~\ref{RT-PDF-inResults} using i) data (dots), and ii) our predictive model (lines).} 
            \label{PDF-test}
        \end{figure}

        \subsection*{DDM with time-varying parameters links human interaction and cooperation}
        
        Fig.~\ref{PDF-test} shows the two PDFs $P_{\rm D}$ and $P_{\rm C}$ obtained at rounds $3$, $37$, and $51$ of the test set by fitting the empirical distributions from data and by employing the DDM formulas combined with the parameters predicted with our model. The three rounds represented in Fig.~\ref{fig:PDF_3} - \ref{fig:PDF_37} - \ref{fig:PDF_51} are chosen among all the rounds composing the test set, as examples of the best, average, and worst predictive performance of our model, but results related to all the rounds are included as additional material in Supplementary Information. It is noticeable how the PDFs predicted with our method are close to the actual ones obtained directly from the empirical data, hence predicting well the unseen batch of the decision-making behaviors of the population. 
        
        Although this insight is already offered by the direct observation of the included figures, in order to gain a quantitative estimation of the goodness of our prediction, we computed the R-squared index achieved by our model at each round of the test dataset. The R-squared index of a predictive model $f$ over a dataset $\lbrace (x_i, y_i) \rbrace_i$ (where $x_i$ is a vector of regressors, and $y_i$ is the variable to be predicted by $f$) is the statistic
        \begin{equation*}
            R^2 = 1 - \frac{\sum_{i} (y_i -f(x_i))^2}{\sum_{i} (y_i - \overline{y})^2}
        \end{equation*}
        where  $f(x_i) = \hat{y_i}$ and $\overline{y}$ indicates the average of the measured $y_i$ in the dataset. $R^2$ is used in the context of statistical models to provide a measure of how well-observed outcomes are replicated by the model based on the proportion of explained total variation of outcomes. The closer the $R^2$ is to $1$, the better the model is explaining the data \cite{lewis2015applied}. Fig.~\ref{fig:PDF_3} - \ref{fig:PDF_37} - \ref{fig:PDF_51} include the value of $R^2$ achieved by our model at the selected rounds $3$, $37$, and $51$, associated with low $R^2=0.79$, good $R^2=0.9$ and excellent $R^2=0.98$, respectively. \\
        
\begin{figure}[H]
            \includegraphics[width=\textwidth]{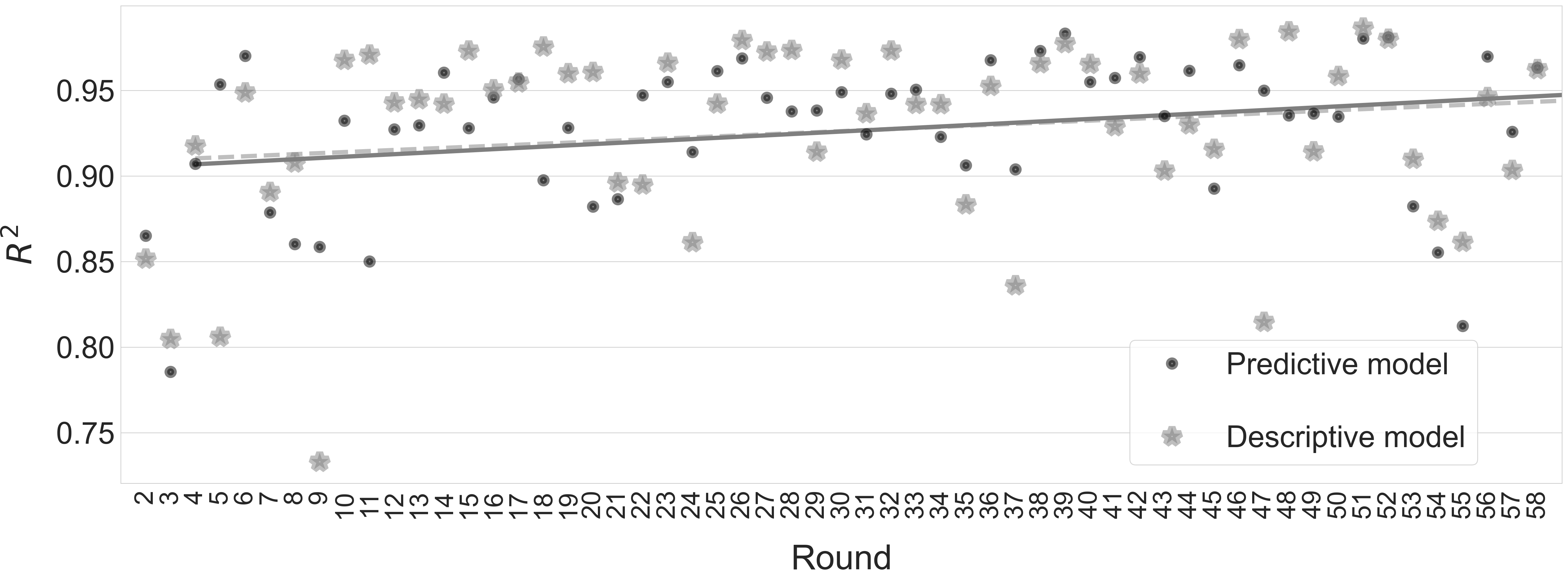}
            \caption{\textbf{Accuracy over the testing set: $\mathbf{R^2}$.} Values of $R^2$ index attained on the test dataset by PDFs employing parameters directly fitted at each round via Bayesian Regression (gray), and by PDFs employing parameters predicted by our model (black).} 
        \label{rsquared_testing}
        \end{figure}

        To summarize the quantitative evaluation of our method on the whole test set, we analyze the $R^2$ obtained at each round, represented in Fig.~\ref{rsquared_testing}. We can see that our model attains at every round $R^2$ values that are greater than $0.75$. Moreover, we consider as reference the $R^2$ values attained on the same dataset by PDFs employing parameters directly fitted at each round via Bayesian Regression, as done in literature by \cite{gallotti2019quantitative}, on the same data we employed. While Fig.~\ref{rsquared_testing} includes both the $R^2$ values achieved by our method (black dots and line) and by the direct fitting of parameters per round (gray stars and dashed line), it is important to highlight that the second method, included as reference, is a descriptive one, capable of extracting information about the parameters at round $t$ \textit{a posteriori}, based on data collected when round $t$ is terminated. Our method, instead, predicts the parameters at round $t$ using information collected up to round $t-1$, as detailed better in Materials and Methods, section ``Predictive Drift-Diffusion Model". Comparing the two results, we can see how the $R^2$ values achieved at each round by the two methods are comparable, as shown by the two lines underlying the $R^2$ trend. 
        \begin{figure}[H]
            \centering
            \includegraphics[width=1\linewidth]{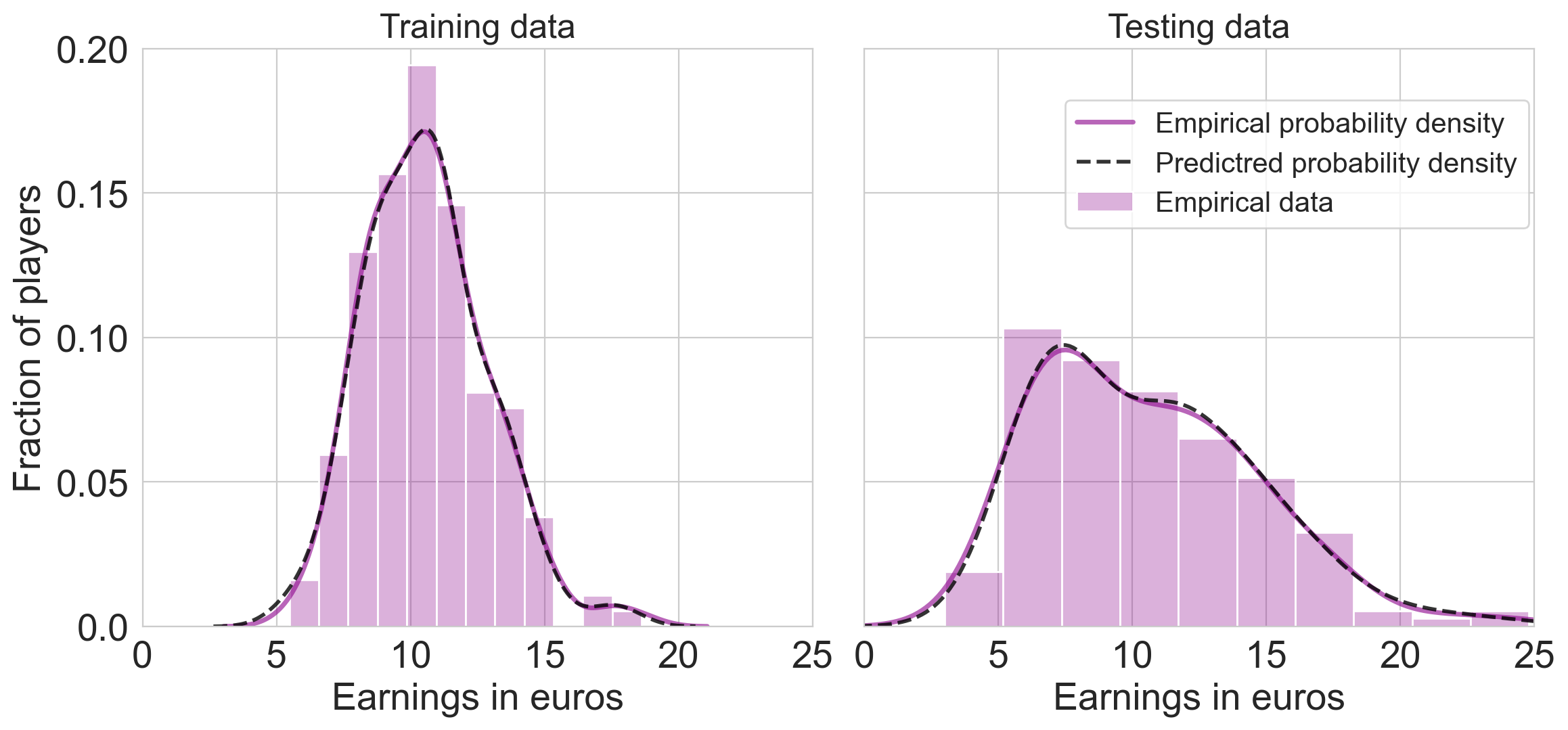}
            \caption{{\bf Final earnings probability distribution}. The considered final earning of an individual is the sum of the MIPD payoffs accumulated by an individual playing the game round after round, as a result of the payoffs for their decision. This figure shows the final earning distribution among the considered population. The black dashed line corresponds to the probability density predicted by our model, averaged over $50$ independent stochastic simulations, the violet line corresponds to the probability density estimated from the empirical final earnings of the population (violet histogram), from the training (left) and test dataset (right).}
            \label{PDF-earnings}
        \end{figure}
        
        The results presented up to this point demonstrate that we are capable of effectively predicting the response time PDFs at each round, in the case of both cooperation and defection, by using data related to the interaction among players collected at previous game iterations. Thanks to theoretical results on DDM, this allows us to derive conclusions on the performance of our methods in predicting the fraction of expected cooperation (or defection) at each round. Eq.~\ref{expected-coop-eq} (in Materials and Methods, section ``Drift-Diffusion Model") allow us to link the fraction of expected cooperation $C_{\rm C}(\nu, \, a , \, z)$ associated with a DDM characterized by parameters $(\nu, \, a , \, z)$ with the area under the curve described by the response time PDF $P_{\rm C}(i;\nu, a, z)$ in the cooperation case. Hence, by correctly predicting such a PDF, we ensure that the area under the curve approximates the actual expected cooperation well. Analogous consideration can be derived for the expected defection rate $C_{\rm D}(\nu, \, a , \, z)$ and $P_{\rm D}(i;\nu, a, z)$, given that our model predicts well both cooperation and defection response time PDFs.
        
        Another quantity that can help us measure the accuracy of our model is the final earnings. At each round, individuals receive a payoff depending on their decisions and those of their co-players. By the end of the game, players will have a final earn on their possession resulting from the accumulation of payoffs. In Fig.~\ref{PDF-earnings} we compare our predicted probability density (black dashed line) averaged over $50$ realizations, with the empirical one (violet line), obtained from the real earning values (violet histogram) from both training and testing datasets. The Kolmogorov-Smirnov distance between the predictive probability density and the empirical one is equal to $0.04$ ($p$-value $0.99$) over the test set and $0.02$ ($p$-value $0.99$) over the training set. 
        
        In a previous work \cite{grujic2010social}, where the data were collected, the authors developed a mathematical model in order to reproduce the distribution of the final earnings. They found that by dividing and partially analyzing individuals according to their cooperative/defective pre-orientation, results were approximately validated. However, we do not use such differentiation, and still, as we can see in Fig.~\ref{PDF-earnings}, our model is capable of capturing very well the patterns of the final earnings. We propose a different angle where we analyze the whole dataset and consider the propensity of individuals to cooperate or defect, as well as the context in which individuals are playing, which results in an essential asset to understanding the dynamic.
        \begin{figure}[t]
            \centering
            \includegraphics[width=1\textwidth]{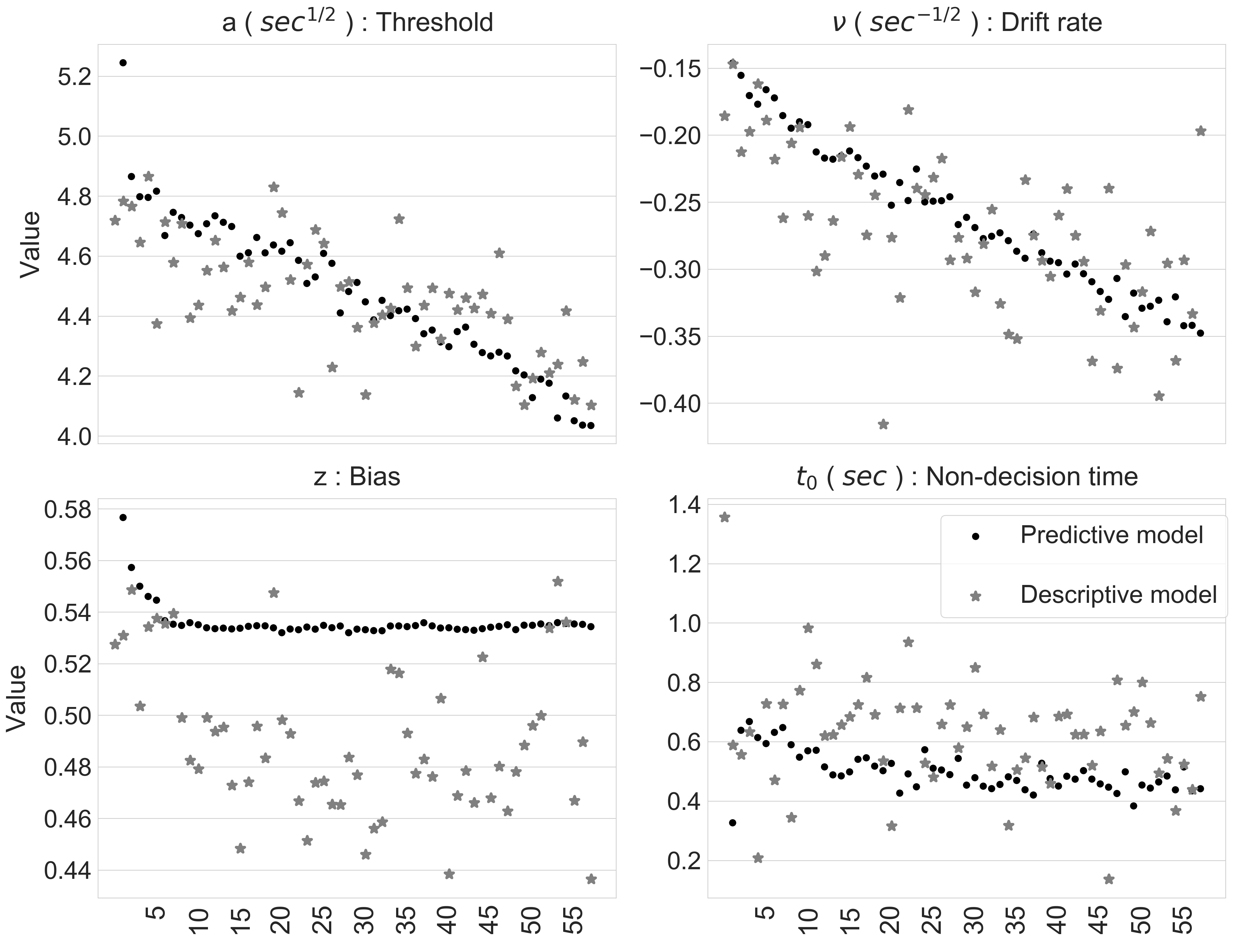}
            \caption{{\bf DDM parameters evolution over the testing set}. The figure includes parameters directly fitted at each round via Bayesian Regression (grey), and parameters predicted by our model (black).}
            \label{testing-param-comparison}
        \end{figure}   
        We include the evolution in time of the DDM parameters along the test dataset in Fig.~\ref{testing-param-comparison}. The figures show both the parameters fitted \textit{a posteriori} at each round as done in \cite{gallotti2019quantitative} (gray) and the parameters predicted (black). As we can see, although both methods individuate similar trends in terms of decreasing parameters, or comparable dimensions of the estimated parameters, they do not coincide round by round. This is not necessarily surprising or problematic: the four parameters together define the probability distributions associated with a DDM. It is potentially possible to change the value of one of such parameters and then find values for the remaining three so that the new model still represents the underlying data well. Particularly, observing the evolution of the bias parameter, although predictive and descriptive models show a similar trend, the values obtained differ substantially. This discrepancy can be explained by considering the level of rationality in responses, defined in Eq. \ref{eq-rationality}. In a regime of high rationally ($\mathcal{R} \to 1$), responses tend to be less intuitive: hence the role of the bias vanishes, while the drift and length of the barrier are the significant parameters for driving responses towards a preferred option. Looking at Fig.~\ref{fig:rationality} in the Supplementary Information, we can see that individuals start from a more neutral regime but rapidly enter into a rational regime. This effect impacts directly the accuracy of the estimation of the bias parameter.
        
        In summary, it is not necessarily true that it exists only one quadruple of parameter values $( \, a_t, \, \nu_t, \, z_t , \, {t_0}_t \, )$ capable of representing well the behavior of the DDM driving the decision-making process at round $t$. For completeness, we include in Supplementary Information the analysis of the response times PDFs predicted on the training set. 
        
        \section*{Evolving cooperation rate under stimuli: simulation and analysis}
        \label{simulations}
        
        Taking advantage of the predictive nature of our model, in this section, we conduct an analysis of its behavior by simulating it in different scenarios and studying the variations of the associated expected cooperation rate. We simulate in silico experimental conditions similar to the ones described in \cite{grujic2010social}, a large-scale social experiment testing the tendency of users to cooperate in a MIPD on a fixed network. We assume an average individual that was already involved in the first phase of the experiment (Experiment 1) where players are distributed on a fixed lattice and  are asked to cooperate (or not) with their neighbors (see section ``Data" in Materials and Methods). We simulate successive phases of such experiment, in particular alternatives to the observations of Experiment 2 (where originally the subjects played the same game with their positions reshuffled on the lattice), by changing the playing conditions and observing the evolution of the model. In tune with what was just described, the parameters of the DDMs involved in the simulations are initialized using values estimated via Bayesian Regression from data collected at the first round of Experiment 2. In particular, we simulate our model under three different groups of stimuli: in the first one, we control some characteristics of the co-players scenarios, either by enforcing a specific level of cooperation in the co-players neighborhood, or by opportunely shuffling the playing group; in the second one, we act on the gain matrices in order to reward cooperation and punish defection; lastly, in the third one, we impose shortened maximum response times to the players, to evaluate our model in simulating more spontaneous and less strategic responses. 
        
        In both Experiment 1 and 2 in the original study, the cooperation level progressively dropped as the game progressed. In all three scenarios we selected, the goal is to identify a treatment that allow sustain a higher cooperation level with respect of what has been observed experimentally in Experiment 2. 
        
        \subsubsection*{Scenario 1: Co-players manipulation}
        
\begin{figure}[t]
            \centering
            \includegraphics[width=0.8\textwidth]{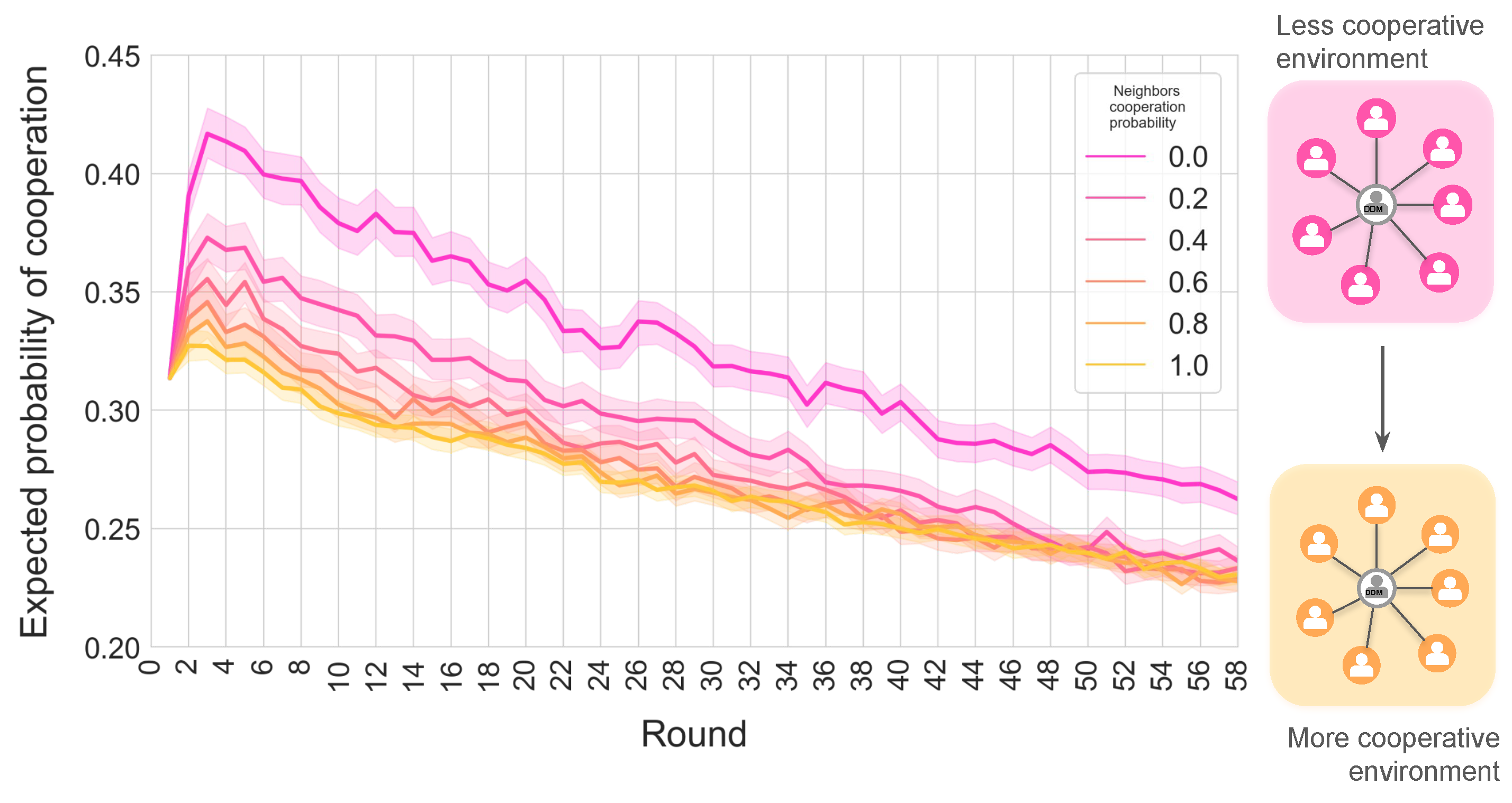}
            \caption{{\bf Impact of the level of cooperative behavior of neighbors}. Expected probability of cooperation of the focal individual exposed to different levels of neighboring cooperation, spanning from less to more cooperative environments. At every round, all seven neighbors have exactly the same probability of cooperating. Each curve corresponds to a different probability ranging from $0$ (pink) less cooperative to $1$ (yellow) more cooperative. Conversely, the focal individual decides with a probability given by our augmented DDM. All cases start with the same initial conditions and free parameter values estimated from the first round of Experiment 2. Notice that the expected probability of cooperation is only related to the focal individual.}
            \label{fig:sim_uniform_neighbors_p}
        \end{figure}

        In the first set of simulations, we consider a focal player whose decisions are driven by our predictive augmented DDM. We simulate the evolution of the individual decision process while playing with seven other co-players. We control the co-players decision-making behavior, modeling it by a probability $p$ of cooperating at every round. We study the expected cooperation rate of the focal individual across a MIPD composed by $58$ iterations, under different values of cooperation probability $p$ of the co-players, i.e., $p = 0, \, 0.2, \, 0.4, \, 0.6, \, 0.8, \, 1$. Each simulation is repeated for a number $N_{\text{realiz}} = 1000$ of realizations. The results in this first scenario are included in Fig.~\ref{fig:sim_uniform_neighbors_p}.
        As visible from the figure, the expected cooperation rate decreases in all the scenarios considered. Exactly like humans, our model evolves by learning that, in the considered MIPD, it is mostly convenient to defect \cite{rand2011dynamic}, at an individual level, as described in section  ``Multiplayer Iterated Prisoner's Dilemma", in Materials and Methods. The simulated result is coherent with the known literature and theory: on one side, our model tends to reach asymptotic values of cooperation that are similar to the ones highlighted from human data in \cite{grujic2010social} and \cite{traulsen2010human}. On the other hand, we can observe that, although mildly, the figure suggests that, in less cooperative environments, the expected cooperation rate is higher. This seems explainable by observing the Schelling diagram in section   ``Multiplayer Iterated Prisoner's Dilemma", in Materials and Methods: we can see that the gains matrix employed in the considered MIPD slightly favors cooperation in scenarios with very few cooperating neighbors.
        
        In this context, another interesting scenario to be considered is the impact of shuffling teams. Previous works \cite{mao2017resilient,gallotti2019quantitative} have shown the existence of a ``restart effect" in IPDs-MIPDs. Such effect consists in an increase of the cooperation rate at the beginning of a new set of game iterations conducted with different co-players, followed by a decrease or stabilization of said cooperation rate. In particular, in \cite{mao2017resilient}, players were asked to conduct $400$ games of IPD, playing $20$ games per day, on $20$ consecutive weekdays, each game composed of $10$ iterations. After a $10$ iterations game, the players were randomly reshuffled. The authors of  \cite{mao2017resilient} observed indeed a sharp jump in cooperation from the last round of a game to the first round of the next. In \cite{gallotti2019quantitative}, instead, the restarting effect is observed between three subsequent games MIPD-IPD-MIPD, with co-players reshuffling. 
        
\begin{figure}[t]
            \centering
            \includegraphics[width=1\textwidth]{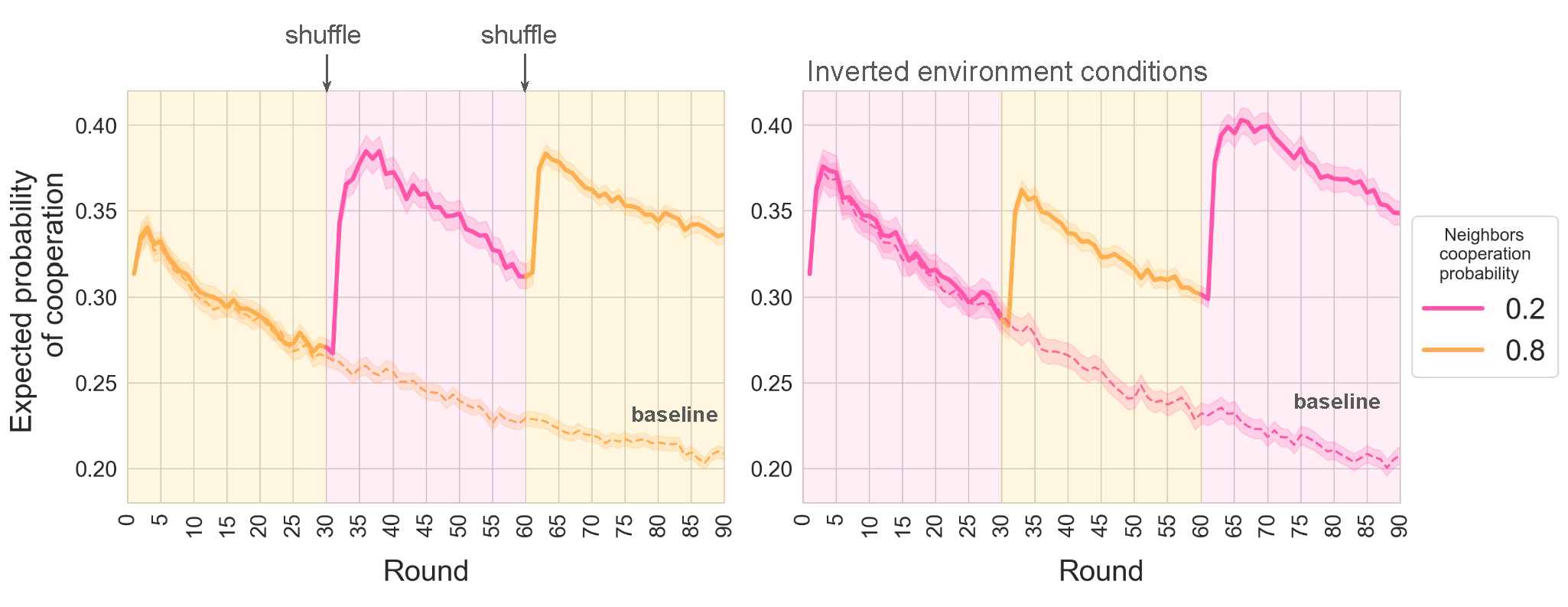}
            \caption{{\bf Impact of shuffling neighbors during the game}. Every $30$-round time window, a change of the co-players of the focal individual is made. At every round, all seven neighbors have exactly the same probability of cooperating, while the focal individual decides with a probability given by our augmented DDM. Initially, in the left panel, at every round, co-players cooperate with the same low probability, which is equal to $0.2$. At round $30$, a shuffling of neighbors is made for a more cooperative environment with a probability of cooperation of $0.8$. Lastly, at round $60$, another shuffle for a less cooperative co-player environment occurs. Conversely, the right panel displays the opposite scenario of the left panel. Notice that the expected probability of cooperation is only related to the focal individual.}
            \label{fig:sim_shuffling_neighbors}
        \end{figure}

        To explore if our model replicates this dynamic, we consider the case in which a focal player engages in a sequence of three MIPDs, each one composed by $30$ iterations, playing each match with a different set of co-players, characterized by different behaviors. In particular, in Fig.~\ref{fig:sim_shuffling_neighbors}, we can observe the expected cooperation rate resulting from consecutive supergames with a group $G_1$ of co-players, characterized by probability of cooperating $p_1=0.8$  at each round, then with a group $G_2$ having cooperating probability $p_2=0.2$, and finally with a group $G_3$ with cooperating probability again of $p_3=0.8$. This is compared with the evolution of the focal player's cooperation probability when engaging in one single MIPD of $90$ iterations, all played with the same group $G$ of co-players, having cooperating probability $p=0.8$. In the same figure on the right, we can observe the same simulation scenario with inverted environment conditions, that is, a focal player governed by a DDM initialized analogously, facing three sequential supergames with co-playing teams described by cooperation probabilities $p_1 = 0.2$, $p_2 = 0.8$, and $p_3 = 0.2$, compared with a single match of $90$ iterations, all played with the same group $G$ of co-players, having cooperating probability $p=0.2$. 
        
        We can see that shuffling teams have a strong impact on the final level of expected cooperation. In both cases, it sharply increases when compared with the case without shuffling (baseline), from $\sim 0.2$ (baseline) to $\sim 0.35$ (with shuffling), without a significant difference between the two considered shuffling scenarios. Besides, in accordance with \cite{mao2017resilient,gallotti2019quantitative}, we see at every shuffling a reset effect in which individuals tend to be more cooperative. Interestingly, this behavior does not hold in time, given that individuals become more defectors as they gain understanding and experience on the game and the context. 
        
        \subsubsection*{Scenario 2: Gains matrix design}
        Reward and punishment are essentially two different types of incentives and both have been proven to be highly effective in promoting cooperation \cite{balliet2011reward}. In our incentives scenario, we explore the impact of modifying the original gain matrix characterizing the MIPD (see section  ``Multiplayer Iterated Prisoner's Dilemma", in Materials and Methods), by adding and subtracting quantities to its elements, in order to reward cooperation or punish defection. We consider $N = 8$ players, each making decisions based on a copy of our predictive augmented DDM while they engage together in a MIPD of $58$ iterations. We simulate the evolution of each of their individual decision processes, considering $N_{\text{realiz}} = 125$ realizations. After fully simulating the scenario, we consider each of the eight players to be a realization of an average subject, and we obtain the associated expected cooperation rate by aggregating it across participants. The final result is, hence, based on $N_{\text{realiz}} \times N = 1000$ total realizations. Here, we test the effect of different modifications of the original gains $R=7$, $S=0$, $T=10$, and $P=0$ employed in the dataset used to train the model.
        \begin{figure}[H]
            \centering
             \begin{subfigure}[b]{0.49\textwidth}
                 \centering
                 \includegraphics[width=\textwidth]{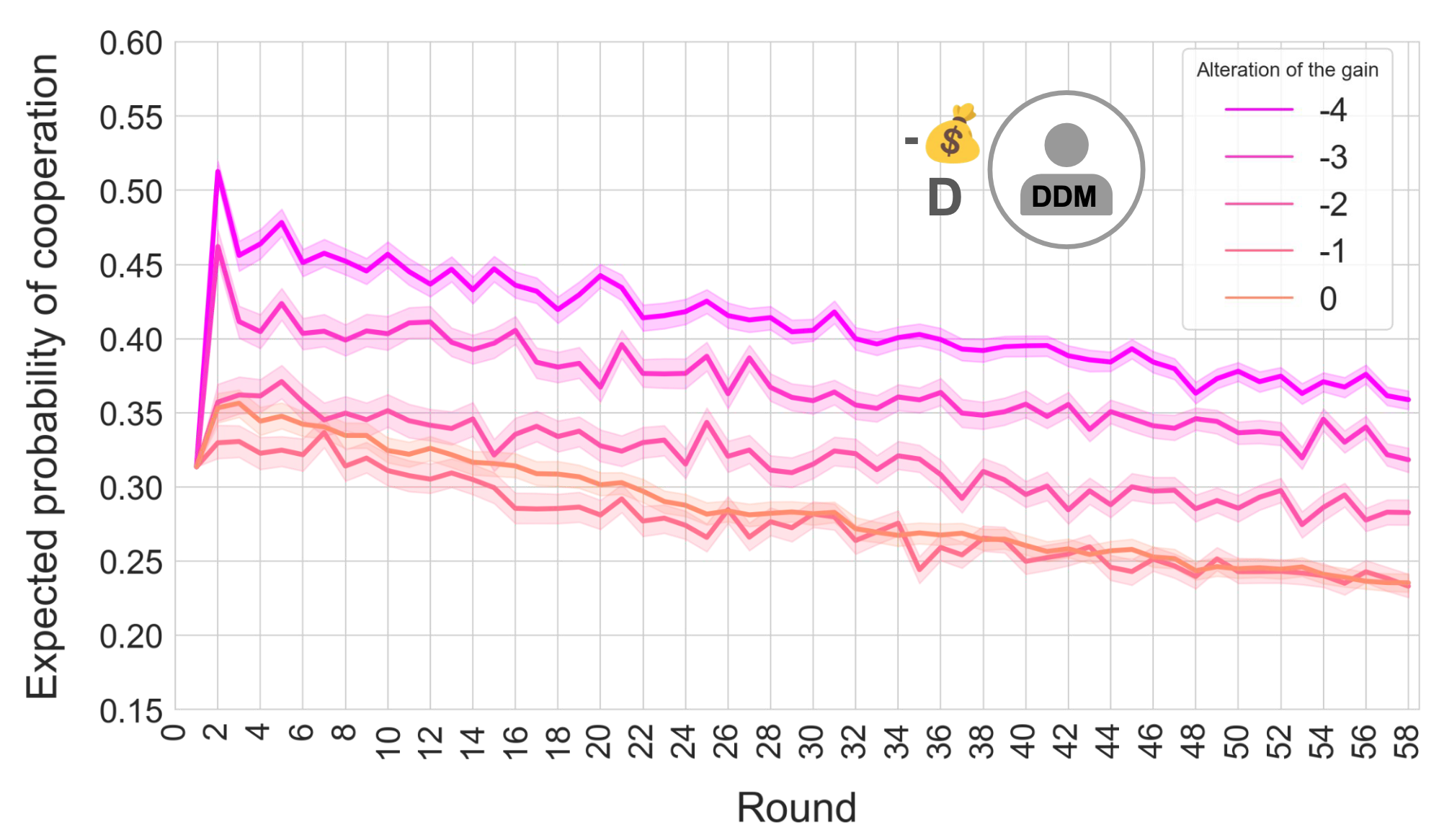}
                 \caption{}
             \end{subfigure}
             \begin{subfigure}[b]{0.49\textwidth}
                 \centering
                 \includegraphics[width=\textwidth]{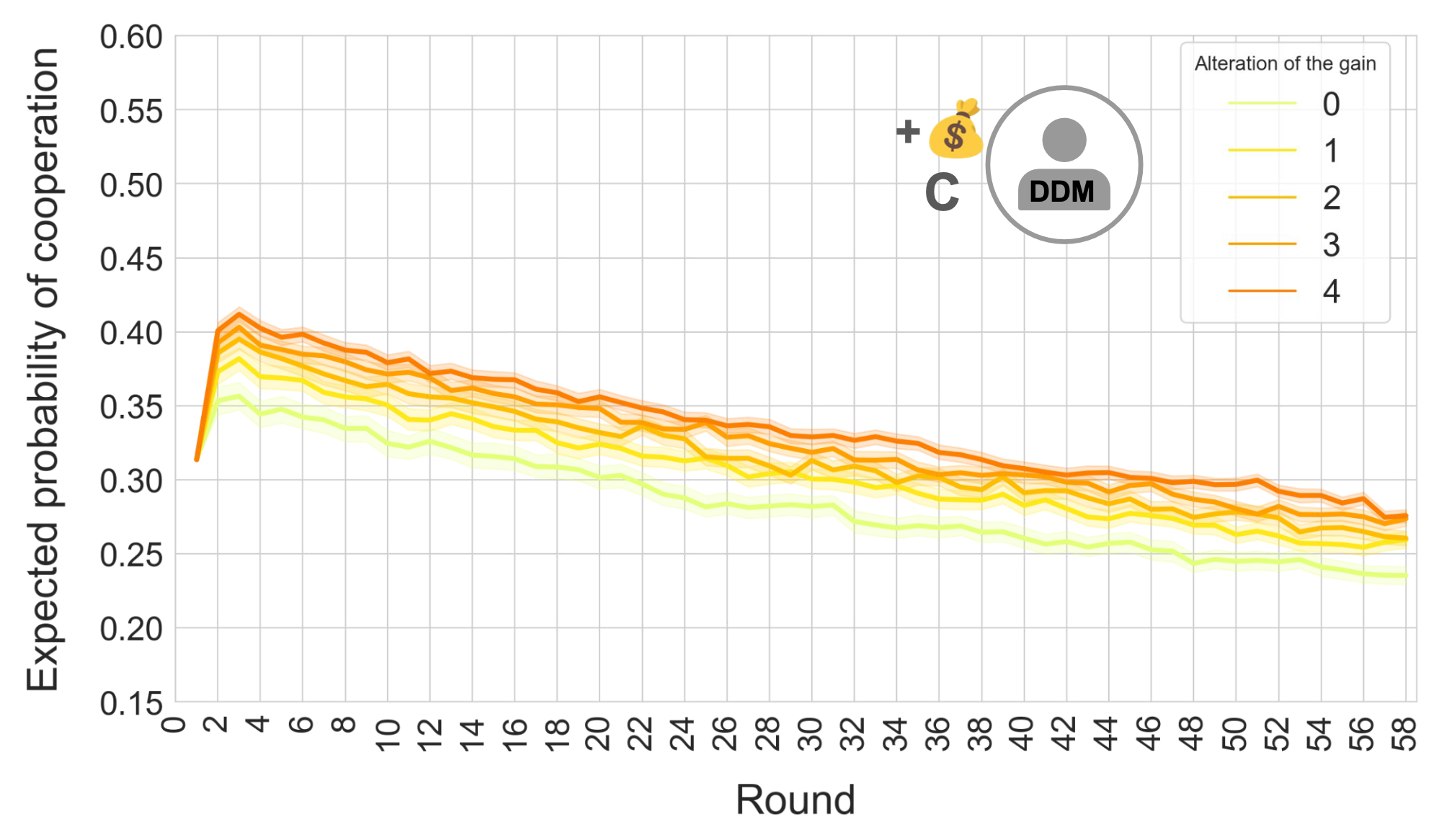}
                 \caption{}
             \end{subfigure}
            \caption{{\bf Rewarding cooperation vs punishing defection: impact of the payoff}. Each curve corresponds to a different value of gains alteration $\lambda$ for (a) punishing defection (pink), subtracting negative values to gains $T$ and $P$ and (b) rewarding cooperation (yellow), by adding positive values $\lambda$ to gains $R$ and $S$.}
            \label{fig:sim_perturb_gains}
        \end{figure}
        In particular, we consider two types of alteration of the gains: (i) \textit{punishing defection}, by subtracting a penalty $\lambda$ to the gains $T=10$ and $P=0$ assigned to defective decisions;
        (ii) \textit{rewarding cooperation}, realized by adding a reward $\lambda$ to the gains $R=7$ and $S=0$ assigned to cooperative decisions.
        
        In Fig.~\ref{fig:sim_perturb_gains}, a study of the two approaches is explored, including the expected cooperation rate for 
        \begin{itemize}
            \item $R = 7$, $\,\,S = 0$, $\,\,T = 10 - \lambda$, $\,\, P = 0 - \lambda$ (first case),
            \item $R = 7 + \lambda$, $\,\,S = 0 + \lambda$, $\,\,T = 10$, $\,\, P = 0$ (second case),
        \end{itemize}
        where $\lambda = 0, 1, 2, 3, 4$ (the case with $\lambda = 0$ corresponds in both cases to the original gains matrix).
        As we can see, both strategies positively impact the levels of expected cooperation by increasing it, in agreement with \cite{balliet2011reward}. However, our findings yield that punishing defection seems to have a slightly greater impact than rewarding cooperation.
        
        \subsubsection*{Scenario 3: Response times pressure}
        
        Pioneer's research has shown that decision time is heavily linked to the mechanisms underlying cooperation \cite{evans2019cooperation}. Intuitive responses are typically characterized by a short response time, less cognitive effort, and being more emotional, while deliberate responses tend to be slower, more controlled, and cognitively demanding. In \cite{rand2014social}, authors propose the theoretical framework  ``Social Heuristic Hypothesis" (SHH), which argues that deliberation favors behaviors that maximize the payoff in the current situation, thus defection, while intuition tends to favor behaviors that maximize payoffs in the long run, thus cooperation. These results have been also previously supported by several studies \cite{bear2016intuition,rand2012spontaneous,everett2017deliberation}. Moreover, \cite{rand2014social} presents strong evidence that, on average, time pressure results in more intuitive responses that tend to support cooperative behavior in naive individuals, while experienced individuals tend to favor selfishness, undermining cooperative intuitions. Therefore, manipulating the decision time through an external application of time pressure, in principle, could allow us to tune responses to be more intuitive or deliberate.
        \begin{figure}[H]
            \centering
            \includegraphics[width=0.8\textwidth]{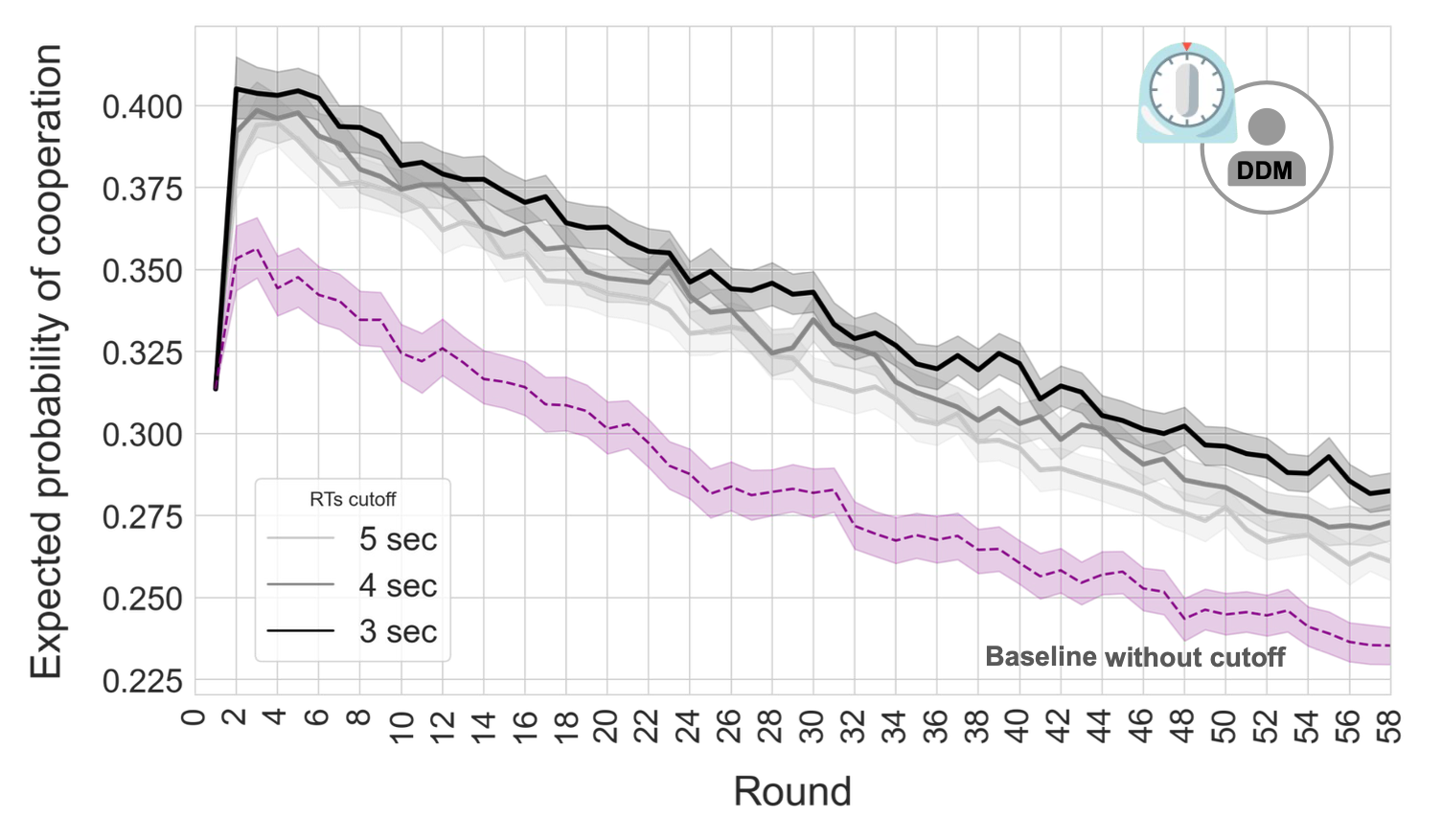}
            \caption{{\bf Impact of time pressure}. Each gray line corresponds to a different threshold value imposed on the response, spanning from $3 sec$ to $5 sec$. The lower the value, the faster individuals have to think and give an answer. The violet lines correspond to unlimited time given to players to respond, which corresponds to the null case.}
            \label{fig:sim_shortened_rt}
        \end{figure}
        In this section, we explore the impact of reducing the response time on the expected probability of cooperation among individuals predicted by our model. As in the previous simulation scenario, we consider $N = 8$ players based on augmented DDMs, interacting on a match of $58$ iterations, considering $N_{\text{realiz}} = 125$ realizations, and aggregate across the individuals so that the final result is again based on $N_{\text{realiz}} \times N = 1000$ realizations.
        
        We are interested in studying the effect of imposing a shorter response time on the decision-makers. In order to do so, we set a maximum decision time $T^d_{\text max}$: when the internal time of decision $i$ of the drift-diffusion walk described in Eq.\ref{DDM-eq-aug} reaches $T^d_{\text max}$, the decision process is interrupted, and we consider as given answer the one associated to the barrier that is closer to the current accumulation state $x$, i.e., if $x(T^d_{\text max})$ is closer to $0$ than to $a$ the individual will defect. Otherwise, it will cooperate. Fig.~\ref{fig:sim_shortened_rt} includes the results in terms of the expected cooperation rate obtained by imposing as maximum decision time $T^d_{\text max} = 3, 4, 5$ seconds, compared with the baseline case with unlimited decision time, purely driven by the dynamics of the augmented DDMs, without any response time constraint. Implementing time pressure indeed forces responses to be more intuitive, and as we can see in all cases, this impact increases the probability of cooperation when compared with unlimited response times (baseline curve) to almost $10\%$ in some cases. In all scenarios, the probability of cooperation decreases with rounds, in agreement with SHH \cite{rand2014social}: individuals at the beginning are naive in the game and gaining experience with rounds. Thus, even though responses remain intuitive, 
    
    \section*{Discussion}
    
        Modeling human cooperation is essential for understanding the dynamics of social interactions and developing strategies to promote collective welfare in societies. In this work, we tackle this matter by developing a mathematical approach to explain and replicate the evolution of human inclinations regarding cooperative and defective choices in the setup of the Multiplayer Prisoner's Dilemma games. It is well-known in the literature that the standard Drift Diffusion Model \cite{ratcliff2016diffusion} serves as a bridge between mathematics and neuroscience by describing the cognitive mechanisms behind individuals' choices, particularly in the context of game theory \cite{gallotti2019quantitative}. Yet, the model is only descriptive and is bound to controlled experiments, generally, with limited and specific data. Our model is a promising tool that deploys Bayesian Regression estimation to predict individuals' choices by incorporating their intrinsic nature toward cooperative behavior and the impact of group interaction in shaping decisions over time. The parameters predicted by our model have shown to be highly effective on unseen data. Specifically, the model successfully predicts both the PDFs of response times at each iteration (closely associated with the population's expected cooperation rate), and the distribution of the final payoff across the entire population. 

        Cooperation prediction on Prisoner's Dilemma games has been tackled from different angles in previous literature. In \cite{nay2016predicting}, authors present a computational model for the prediction of individual human behavior in two-player IPD via logistic regression, considering a multi-experiment dataset of individual decisions. In \cite{embrey2018cooperation}, instead, human decisions are predicted assuming that each individual has a prior over the strategy that its partner will use, chosen among a user-defined set of possible strategies, and iteratively updated after every IPD set of iterations. This scheme, although allowing to fit one model per individual, does not consider variations in the behavior of players across iterations inside one IPD, differently from our proposed methodology. Additionally, it requires to pre-design a set of possible strategies, a task requiring a not negligible amount of expertise on complex human behavior in the specific scenario. Case-Based Decision Theory is employed in \cite{guilfoos2016predicting}, where agents choose at every round between cooperation and defection, by maximizing the weighted average of the difference between the reward associated with each choice at previous rounds, and a target level of satisfactory reward (fitted from data). Both \cite{embrey2018cooperation} and \cite{guilfoos2016predicting} attempt to model strategic decision-making, neglecting more instinctive instances. A current branch of studies involves simulating and replicating behavioral trajectories through Reinforcement Learning, integrating learning-related parameters inspired by cognitive principles of the human brain \cite{lin2019split, lin2019story, lin2020unified, lin2021models}. While these models offer valuable, interpretable insights into human decision-making, their capability of predicting human action sequences in complex decision-making contexts, such as IPD, is still constrained \cite{lin2022online}. 
        
        All the described models focus on two-player IPDs, differently from our approach, which is based on regressors describing MIPD interactions with any number $N$ of players, and hence can be employed as well in the two-player IPD scenario (i.e., MIPD with $N = 2$). Moreover, while directly modeling either the individual probability of cooperation or directly the next decision, they do not provide interpretation for the underlying human neurocognitive characteristics. Instead, our work exploits DDMs as a decision mechanism and study the associated probability of cooperation through a theoretical framework. This modeling approach offers several advantages. First, it enables the representation of human decision-making as the accumulation of noisy evidence until a decision is reached, a concept supported by behavioral experiments, neurocognitive evidence, and clinical observations \cite{forstmann2016sequential,pardo2019mechanistic}). Furthermore, while simulating human decisions, it can provide human-like features, for instance, response times, that are fitted to the ones experienced in training. Additionally, this parametrization, while maintaining linear relationships concerning the regressors, is capable of generating a more complex stochastic behavior, by joining the drift-diffusion random walk with the predicted dynamics of the parameters. Moreover, the predicted parameters offer precious insights from a neurocognitive perspective of human decision-making. 
        
        The main interest in our proposed predictive model lies in its utility as a tool for studying and testing the effects of policies for enhancing cooperation through targeted interventions, without the need to carry out real-human experiments. To employ a model in this way, we argue that such a model should mimic as much as possible the sensitivity of human beings to the most diverse range of stimuli. First, we explore scenarios where we vary the levels of cooperative co-players, finding a good alignment with established literature. Like humans, our model learns that, within the considered MIPD scenario, defection is often more advantageous, leading to asymptotic cooperation rates similar to those observed in human studies \cite{grujic2010social, rand2011dynamic, traulsen2010human}. Consistently with studies on human behavior, our model exhibits a reset effect when playing teams are reshuffled, where individuals initially tend to be more cooperative. However, this behavior diminishes over time as individuals become more inclined to defect as they gain understanding and experience in the game. Secondly, we explore how our model is sensitive to modifications in the payoff matrix, with both rewards and punishments effectively increasing the levels of expected cooperation, particularly in the case of punishment. Lastly, we consider time pressure in responses, the model exhibits that individuals tend to have more intuitive responses, leading to a higher probability of cooperation compared to conditions with unlimited response times. However, as rounds progress, the probability of cooperation decreases, which is consistent with SHH.
        
        We augmented the classic DDM to directly sensitize it to interactions among individuals, as well as reward and punishment measures, or time pressure, coherently with what is known about human decisions. This aspect makes it different from the previously mentioned literature contributions. In particular, none of the mentioned works consider the effect of time pressure stimuli. In \cite{nay2016predicting, embrey2018cooperation} and \cite{lin2022predicting}, the only direct effect on these models is interaction with different players. This implies that stimuli related to modifying the payoff matrix of the game could have exclusively an indirect effect, requiring the introduction among the simulated playing agents of special players, sensitive to such aspects. This is not true for human decisions: we are affected by the payoffs, even without observing other players deciding based on it. Considering \cite{guilfoos2016predicting}, while this method might react to gains adaptation, team reshuffling can influence its behavior only if by design we assume that each agent is allowed to access the average cooperation rate of its opponent at each instant, while this information is not always available in real life. 

        All mentioned advantages of our model come with the limitations and requirements of datasets that not only document human cooperation and defection but also include detailed recordings of the response times. This additional layer of data is crucial for fitting the response PDFs at the base of our approach. Moreover, either to obtain more generalized models or to calibrate models specifically to individual behavior, larger datasets might become necessary considering the data requirements of Bayesian Regression estimation. Besides, in the case of individual models, to avoid the stress/cost of conducting repeated experimental sessions with the same participants, the need for data could be mitigated by Hierarchical Bayesian Regression approaches. By doing so, we can leverage between volume and richness of data needed to accurately capture the unique decision-making patterns of each individual and the associated costs.

        In conclusion, the findings described in this work underline the potential of our model as a tool for designing and testing strategies that foster cooperation, enhancing our ability to study, understand, and intervene in scenarios where individual goals conflict with collective welfare. In the future, we envision to design an artificial agent, trained on this model, generating stimuli affecting human decision-making, such as adaptive payoffs and response times. One appealing possibility is to build a social planner, capable of organizing players in more cooperative sub-groups. This scheme is chosen as well in \cite{mckee2023scaffolding}, demonstrating that AI is capable of generating reasonable stimuli, given an underlying unknown dynamic. In \cite{mckee2023scaffolding}, though, the AI social planner is trained based on a simple and not quite realistic human model: we argue that an artificial agent trained on more accurate simulacrum of human decision-making process, such as our model, could have a bigger impact as a building block for human-machine systems, in which artificial agents support and enhance reciprocal cooperation.
    
    \section*{Materials and Methods}
        \subsection*{Multiplayer Iterated Prisoner’s Dilemma }
        \label{method-multi-iPD}
        
        The Multiplayer Iterated Prisoner’s Dilemma (MIPD) game \cite{carroll1988iterated} is an extended version of the classic two-player Iterated Prisoner’s Dilemma (IPD) \cite{basu1977information}, considering repeated rounds of the game among a number $N>2$ of players. We call each iteration of a MIPD \textit{round} or \textit{game}, while we indicate the whole MIPD as \textit{supergame}. At each round $t$, each player is asked to make a single choice $d_t$ among two alternatives, expressing in favor either of Cooperation (C) or Defection (D). Such a choice, together with the choices of the other $N-1$ players in the same round, would determine the payoff received by the player. In particular, at round $t$ a player would earn a payoff,
        \begin{equation}
            R_t(\ell) = 
            \begin{cases}
            \quad R \, (\ell + 1) \,\, + \,\, S \, (N - (\ell + 1)) & \text{if } d_t = C,\\
            \quad T \, \ell  \,\, + \,\,  P \, (N - \ell) & \text{if } d_t = D,\\
            \end{cases}
            \label{eq:mipd-payoff}
        \end{equation}
        where $\ell \in \lbrace 0, 1, \dots , N-1 \rbrace$ denotes the number of the remaining $N-1$ players whose decision was to cooperate at round $t$. The payoff formula, in the  classic Prisoner's Dilemma game, can be summarized by the gains matrix in Table \ref{tab:gains_matrix}.
        \begin{table}
        \begin{center}
            \begin{tabular}{ c|c c}\\
                & {\bf Cooperation}  (others)& {\bf Defection} (others) \\\\
                \hline \\
                \multirow{2}{10em}{{\bf Cooperation} (you) \\ {\bf Defection} (you)} 
                & R & S \\
                & T & P \\
            \end{tabular}
        \end{center}
        \caption{Matrix summarizing the gains of players engaging in an MIPD round.}
        \label{tab:gains_matrix}
        \end{table}
        The upper table, although giving insights into the dynamics of a two-player game, is not fully descriptive of the multiplayer scenario. Hence, in order to study better this case, we employ the definitions related to $N$-player sequential social dilemmas \cite{walsh2002analyzing, wellman2006methods, leibo2017multi, hughes2018inequity}. 
        
        An $N$-player sequential social dilemma is defined as a tuple \((M, \Pi = \{\Pi_c, \Pi_d\})\), where \(M\) represents a Markov game and \(\Pi\) comprises two disjoint sets of policies: the cooperative policies \(\Pi_c\), and the defective ones \(\Pi_d\). Considering a strategy profile \((\pi_1^c, \ldots, \pi_n^c, \pi_1^d, \ldots, \pi_m^d) \in \Pi_n^c \times \Pi_m^d\) with \(n + m = N\), the associated payoffs for the cooperating and defecting policies are denoted as \(p_c(n)\) and \(p_d(n)\), respectively. In this context, the game can be analyzed by studying a \textit{Schelling diagram} \cite{schelling1973hockey, perolat2017multi}. Such diagram visually represents the curves \(R_c(\ell) = p_c( \ell + 1)\) and \(R_d(\ell) = p_d(\ell)\), that is, the possible payoffs for a player in case it cooperates or defects, assuming that \(\ell\) of the remaining \(N-1\) players choose to cooperate, while the rest defects.
        Using this notation, \textit{fear} and \textit{greed} are defined in literature \cite{hughes2018inequity}. The fear property is realized in scenarios in which mutual defection is preferable to be exploited by defectors, i.e., \(R_d(i) > R_c(i)\) for sufficiently small \(i\). Instead, the greed property states that exploiting a cooperator is preferable to mutual cooperation, that is, \(R_d(i) > R_c(i)\) for sufficiently large \(i\). The tuple \((M, \Pi)\) is considered a sequential social dilemma if the following conditions hold:
        \begin{itemize}
            \item Mutual cooperation yields a higher payoff than mutual defection: \(R_c(N-1) > R_d(0)\).
            \item Mutual cooperation is preferable to being exploited by defectors: \(R_c(N-1) > R_c(0)\).
            \item Either the fear property, the greed property, or both must be satisfied.
        \end{itemize}
        In the case of MIPD, \(R_c(\ell)\) and \(R_d(\ell)\) are defined as the payoff included in the Eq. \ref{eq:mipd-payoff}, in case of cooperation and defection, respectively, and the sequential social dilemma definition can be verified or not, depending on the values of the gains $R$, $S$, $T$, and $P$.
        
        \subsection*{Data}
        \label{method-dataset}
        
\begin{figure}[t]
            \centering
        \includegraphics[width=0.7\textwidth]{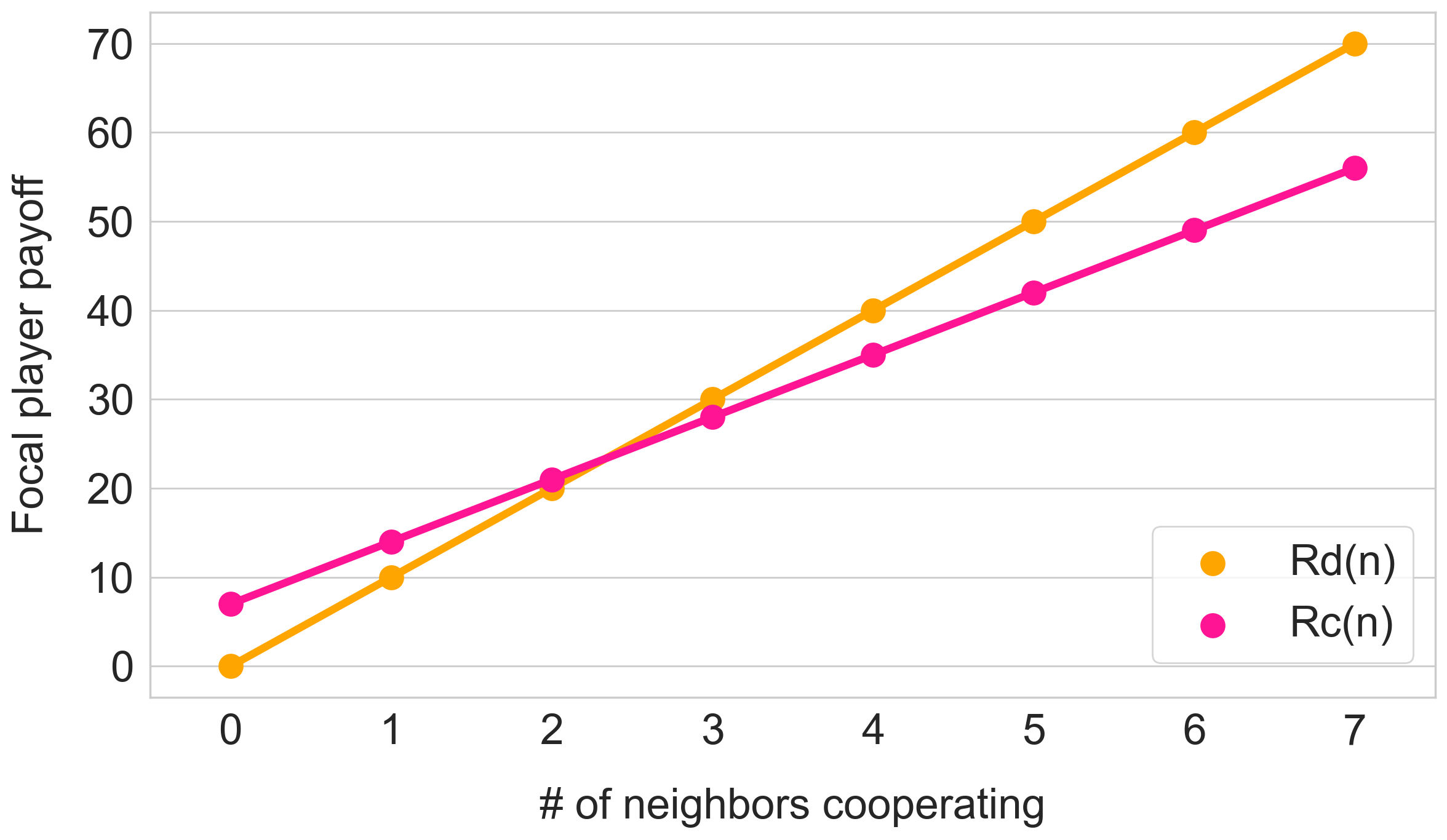}
            \caption{{\bf Multiplayer Prisoner's Dilemma game: Schelling diagram}. Game associated with the employed dataset, with $N = 8$ players and gains $R=7$, $S=0$, $T=10$, $P=0$.}
            \label{fig:payoff-IPD}
        \end{figure}

        The data was collected by \cite{grujic2010social} in a game theory experiment conducted among $N_{\text{subj}} = 169$ participants, with age ranges between $18$ and $26$ years old. They played a MIPD (see previous section) on a square lattice, simultaneously with $N=8$ neighbors (Moore’s neighborhood: up, down, left, right, and diagonally), with periodic boundary conditions. At the end of the experiment, the participants were rewarded proportionally to the payoff, computed as described in Eq. \ref{eq:mipd-payoff}, with values $R=7$, $S=0$, $T=10$, $P=0$. The game can be analyzed from the game-theoretical point of view by observing the Schelling diagram, introduced in the previous section, included in Figure \ref{fig:payoff-IPD}. 
        From this diagram we can observe that the three conditions defining sequential social dilemmas are indeed verified in this case: $R_c(7)>R_d(0)$, $R_c(7)>R_c(0)$, and \(R_d(i) > R_c(i)\) for \(i > 2\) (greedy condition). Hence, the game proposed to the participants of this experiment presents the characteristics of a dilemma, in which from the individual point of view, the rewarding strategy is to defect, while from a societal point of view, cooperation is a better option.
        
        The experiment was structured into three phases, however, we only focus on two: the initial phase, composed of $47$ rounds, denoted as Experiment 1 (fixed network during all the rounds), and the final one, including $58$ rounds, denoted as Experiment 2 (fixed network, however different from the network in the Experiment 1). At all stages, subjects knew the result of the previous round they played: the actions and payoffs of themselves and all of their neighbors. Participants did not have previous experience in the game before starting, yet they received an explanation of the rules. Notice that the different phases are not independent, given that the subjects participated in all three phases. For additional details on the experimental settings and the data collection see \cite{grujic2010social}. 
        
        \subsection*{Drift-Diffusion Model}
        \label{methods-ddm}
        
\begin{figure}
            \centering
            \includegraphics[width=0.85\textwidth]{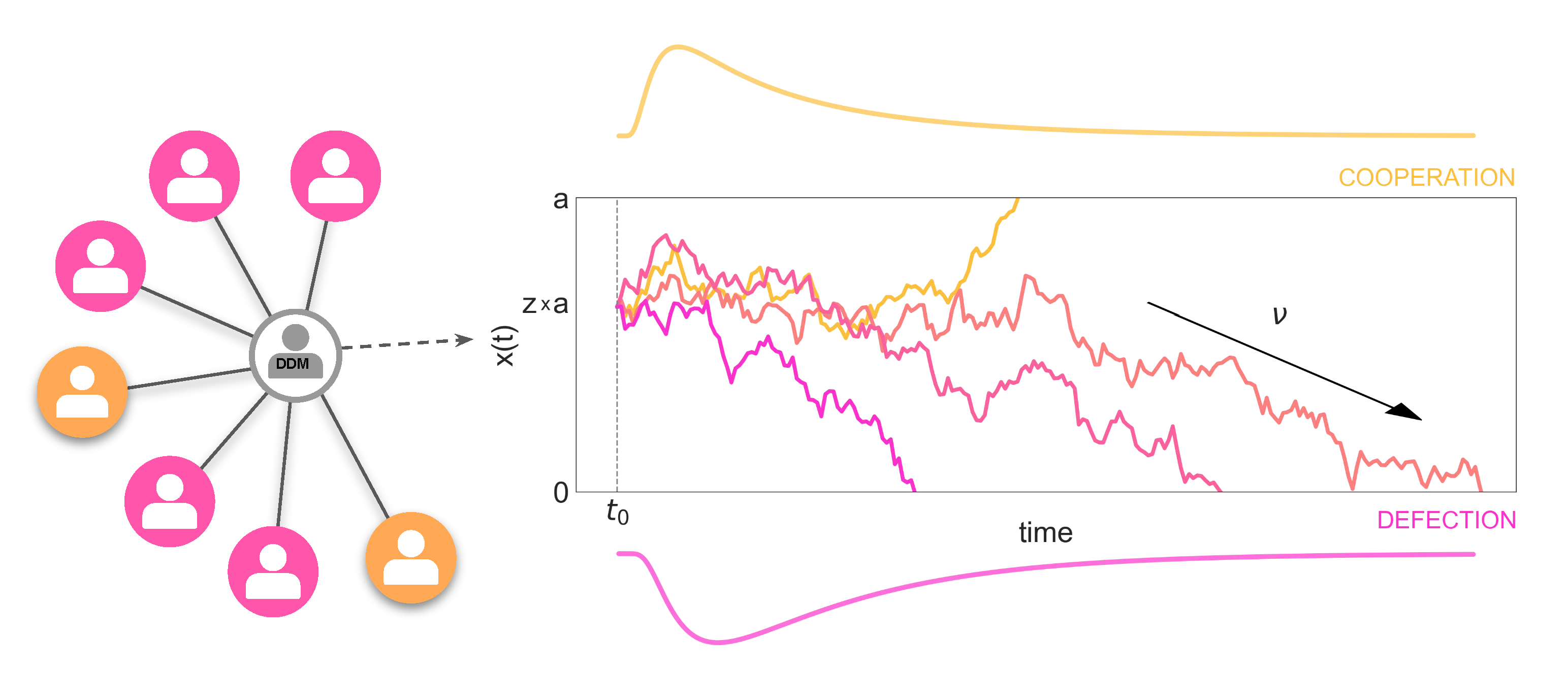}
            \caption{{\bf Illustration of DDM as a model for cooperation/defection choices}. In the context of group dynamics, an individual (here represented in the center) makes a decision following a DDM process. Starting from an initial condition $z \cdot a$, at each time step $i$, the subject will accumulate random evidence in favor of one of two alternative decisions. The $x=a$ threshold is associated with cooperation, and the $x=0$ threshold is associated with defection. Once the amount of evidence reaches one of the thresholds, the associated decision is made. The black arrow indicates the presence of a negative drift towards defection.}
            \label{diagrama-rw}
        \end{figure}

        The Drift-Diffusion Model (DDM) assumes a one-dimensional random walk behavior, representing the accumulation of information (in the form of noisy evidence) in our brain, discriminating between two alternative options \cite{ratcliff2008diffusion}. The process begins from a starting point $x(0) = z \cdot a$, where $z \in [0, \, 1]$ is indicated as the initial bias and $a$ as the decision threshold (height of the barrier), see Fig.~\ref{diagrama-rw}. At each time step, the individual gathers and processes information until one of the two boundaries is reached, $x = 0$ or $x = a$, and thus a decision is made. The continuous integration of evidence is described by $x(i)$ and is given by the equation,
        \begin{equation}
         d x  =  \nu  \; di + \sqrt{D} \; \xi(i). 
         \label{DDM-eq}
        \end{equation}
        Eq. \ref{DDM-eq} represents a one-dimensional Brownian motion characterized by: (i) a drift term with parameter $\nu$; (ii) a diffusive term composed by white Gaussian noise $\xi$, with diffusion coefficient $ \sqrt{D}$. In order to minimize the number of free parameters, in the following we consider $D=1$, assuming the dynamics to be scaled by $\sqrt{D}$. In Eq. \eqref{DDM-eq}, the letter $i$ represents the model internal time of the decision, whose final value, once a decision is made, corresponds to the response time attained.
        From a neuroscience perspective, these free parameters of the mathematical model have the following interpretation: (i) the bias $z$ is related to the prior inclination (the pre-existing ideas) the individual has about the options, (ii) the threshold or height of the barrier $a$ quantifies psychological cautiousness in response, (iii) the drift rate $\nu$ is related to the quality and velocity of the information extracted from the stimulus, and hence to the task complexity. For tasks that are easier to understand, the drift rate has a high absolute value, and responses are fast and accurate on average. Conversely, when subjects find the tasks more complex, the drift rate values are closer to zero, and responses are slower and less accurate on average. Note that the dynamic presented is the simple version of DDM, and many variations to this model have been explored \cite{ratcliff1980note, ratcliff1998modeling, moran2015optimal,edwards1965optimal, ratcliff1999connectionist}. 
        For our purposes, we consider the classic model with the addition of the non-decision time $t_0$ \cite{ratcliff1999connectionist}, given its widespread use in literature, and its higher realism and generality in describing the decision process. This parameter corresponds to the time before the actual decision process begins, namely, it is the response time component that is not due to evidence accumulation and is more related, for instance, to the time individuals take preparing themselves to start reading the headline (before they realize that the new round started) or the time used to press the button on the mouse.
        
        In the context of cooperation-defection decisions, we say that the decision to defect is made when the boundary  $x = 0$ is reached. Otherwise, if $x = a$ is reached, then the decision is to cooperate. The free parameters $z$, $a$, and $\nu$ play an important role in the process. The bias $z$ describes the individual initial tendency to cooperate or defect: if it is greater than $0.5$, this means that the individual is inclined towards cooperating, below towards defecting, and an unbiased scenario is given when $z=0.5$. The drift $\nu$ is related to the information gathering and can be analyzed in two parts: on the one hand, the module of the drift $|\nu|$ is the signal-to-noise ratio, representing the amount of evidence supporting the two alternative options (the lower the $|\nu|$, the more difficult the task). On the other hand, the sign of $\nu$ is related to the direction supporting one of the two options. When $\nu>0$ (positive), the gathered of information is tendentially in favor of cooperating, while when $\nu<0$ (negative) the evidence gathered is mostly supporting defection. The PDF of the response times in case of defective choice corresponds to the model's probability of first crossing the lower boundary $x = 0$ at instant $i$, i.e., 
        \begin{equation}\label{RT-PDF}
        \begin{split}
           P_{\rm D}(i;\nu, a, z) & = \frac{\pi}{a^2}\; exp \left(- \nu\;z\;a - \frac{\nu^2i}{2}\right)\\
                     & \times \sum^{\infty}_{k=1} \; k \; exp \left(-\frac{k^2\pi^2i} {2a^2} \right)\; sin(k \; z \; \pi),
        \end{split}
        \end{equation}
        known as the F\"urth formula for first passages \cite{feller1991introduction}. Analogously, the PDF of the response times in case of cooperative choice corresponds to the model's probability of first crossing the upper boundary $x = a$ at instant $i$ is given by $P_{\rm C}(i;\nu, a, z) = P_{\rm D}(i; - \nu, a, 1 - z)$. The expected cooperation rate is represented by the area under the curve 
        \begin{equation}
            C_{\rm C}(\nu, \, a, \, z) = \int_{0}^{\infty} P_{\rm C}(i;\nu, a, z) \, di.
            \label{expected-coop-eq}
        \end{equation}
        and an analogous expression can be derived for the expected defection rate $ C_{\rm D}(\nu, \, a, \, z)$, involving $P_{\rm D}(i;\nu, a, z)$ (see \cite{bogacz2006physics,gallotti2019quantitative} for more details). Notice that in an unbiased scenario, where $z=0.5$ and decisions only depend on the drift and the barrier length, the theoretical expression of the expected cooperation rate Eq.~(\ref{expected-coop-eq}) takes shape $C_C(\nu, \, a, \, z=0.5) = \frac{1}{1 \ + \ exp(-a \ \nu)}$.
        
        \subsubsection*{Rationality}
        
        As a measure of the extent to which responses tend to be more intuitive or rational, the authors in \cite{gallotti2019quantitative} introduce the rationality ratio parameter, defined as,
        \begin{equation}
            \mathcal{R} = \frac{|\Delta C_{rationality}|}{|\Delta C_{rationality}| + |\Delta C_{intuition}|}
            \label{eq-rationality}
        \end{equation}
        where $|\Delta C_{rationality}| = C_C(\nu, \, a, \, z=0.5) - 0.5$ is the contribution of the rationality behavior, given by the difference between the unbiased and the completely random scenarios; and $|\Delta C_{intuition}| = C_{empirical} - C_C(\nu, \, a, \, z=0.5)$ is the contribution of the intuitive bias behavior, given by the difference between the empirical value and the unbiased scenario. Hence, in the case of $\mathcal{R} \to 0$, one would expect more decisions dominated by intuition, being the bias more important, while in the opposite spectrum, $\mathcal{R} \to 1$ decisions tend to be rational, based on the gathering of information, with the drift playing a more important role in the dynamic.
        
        \subsection*{Predictive Drift-Diffusion Model}
        \label{methods-ddm-predictive}
        
        In this work, we consider a DDM as a model of the cooperation/defection decision process of an individual, and we employ a Bayesian regression approach to obtain a predictive model, capable of capturing the evolution of the DDM free parameters, based on regressors describing the interactions among the neighborhood of players. We consider a model of the form 
        \begin{equation}
            a_t = F_{a}(x_{t-1}^a, \, P_a), \qquad \nu_t = F_{\nu}(x_{t-1}^{\nu}, \, P_{\nu}), \qquad z_t = F_{z}(x_{t-1}^z, \, P_z), \qquad {t_0}_t = F_{t_0}(x_{t-1}^{{t_0}}, \, P_{t_0}) 
            \label{eq:tv-param-model-generic}
        \end{equation}
        where, for each of the free parameters $j \in \lbrace a, \, \nu , \, z, \, {t_0} \rbrace $, the vector $x_{t-1}^{j}$ contains the regressors designed to capture the effect of the interaction with other players on the DDM parameters evolution, and $P_j$ collects a set of parameters $p_i^{j}$ to be estimated, representing the relationship between the DDM parameters to be predicted and the chosen regressors. Coherently with the Bayesian regression approach, each parameter $p_i^{j}$ is modeled as a random variable with Gaussian distribution, characterized by mean $\mu(p_i^{j})$ and standard deviation $\sigma(p_i^{j})$, i.e., $p_i \sim \mathscr{N}(\mu(p_i^{j}),\sigma(p_i^{j})^2)$. 
        
        We design a simple affine model capable of approximating the DDM parameters evolution in time, thanks to regressors that measure the characteristics of the interactive behavior captured by the employed dataset. In particular, we predict the free DDM parameters at round $t$ according to
        \begin{equation}
           a_t = x_{t-1}^a \cdot P_a, \qquad \nu_t = x_{t-1}^{\nu} \cdot P_{\nu}, \qquad z_t = x_{t-1}^{z} \cdot P_{z}, \qquad {t_{0_t}} = x_{t-1}^{t_0} \cdot P_{t_{0}}.
           \label{eq:tv-param-model-linear}
        \end{equation}
        In the following subsection, we detail the design of the regressors, the parameters fitting, and both the predicting and simulating procedures.
        
        \subsubsection*{Regressors}
        
        The regressors $x_{t-1}^a$, $x_{t-1}^{\nu}$, $x_{t-1}^z$, $x_{t-1}^{t_0}$ indicated in Eqs.~\ref{eq:tv-param-model-generic}-\ref{eq:tv-param-model-linear} are defined by taking advantage of the Relative Allocation ($RA$) metric proposed by authors in \cite{montero2022fast} to measure the individual propensity to cooperate or defect and of the associated personal gains allocated to self ($a_{self}$) and to other players ($a_{others}$). 
        
        In particular, considering a MIPD with $N$ players, per each round $t>1$ and player $k$ (here indicated as the focal player), we use information gathered from the behavior observed at the previous round (i.e. the number $n_c^{t-1} \in \lbrace 0, 1, \dots, N-1 \rbrace $ of neighbors of $k$ that cooperated at round $t-1$, excluded $k$ itself), the decision $d_{t-1}$ of $k$ at round $t-1$, and define: 
        \begin{itemize}
            \item if response of $k$ at round $t-1$ is Cooperation (i.e., $d_{t-1}=1$):
            \begin{align*}
                {a}_{self}^{t-1} &= R \; \times \; (n_c^{t-1} + 1) + S \; \times \; (N - (n_c^{t-1} + 1)),\\
                {a}_{others}^{t-1} &= R \; \times \; (n_c^{t-1} + 1) + T \; \times \; (N - (n_c^{t-1} + 1)),\\
            \end{align*}
            \item if response of $k$ at round $t-1$ is Defection (i.e., $d_{t-1}=0$):
            \begin{align*}
                {a}_{self}^{t-1} &= T \; \times \; n_c^{t-1} + P \; \times \; (N - n_c^{t-1}),\\
                {a}_{others}^{t-1} &= S \; \times \; n_c^{t-1} + P \; \times \; (N - n_c^{t-1}),\\
            \end{align*}
        \end{itemize}
        We indicate as $\overline{a}_{self}^{t-1}$ and $\overline{a}_{others}^{t-1}$ the normalized version of ${a}_{self}^{t-1}$ and ${a}_{others}^{t-1}$, respectively, i.e. 
        \begin{align*}
                \overline{a}_{self}^{t-1} = \frac{{a}_{self}^{t-1}}{({a}_{self}^{t-1} +  {a}_{others}^{t-1} )}, \qquad 
                \overline{a}_{others}^{t-1} = \frac{{a}_{others}^{t-1}}{ ({a}_{self}^{t-1} + {a}_{others}^{t-1} )}
        \end{align*}
        Moreover, we consider for each round $t$
        \begin{equation}\label{param_evol}
            RA_{t-1} = \arctan \left( \frac{{a}_{others}^{t-1} }{{a}_{self}^{t-1}} \right)
        \end{equation}
        as the inclination to cooperation of the focal subject immediately before the decision at round $t$. The closer to $90^{\circ}$ is $RA_{t-1}$ (after being transformed from radians to degrees), the more selfless the individual is considered to be. Conversely, when $RA_{t-1}  \sim 0^{\circ}$, the subject is expected to adopt a more selfish behavior. When ${a}_{self}^{t-1}$ and ${a}_{others}^{t-1}$ are simultaneously equal to zero we assign $RA_{t-1} = 0$, considering that such occurrence corresponds to scenarios in which each of the neighbor players (included the focal player) defected, hence we assume the player's inclination to be towards defection.  Finally, we normalize $RA_{t-1}$ in $[-1, \, 1]$, i.e., $\overline{RA}_{t-1} = (RA_{t-1} - 45^{\circ})/45^{\circ}$.
        
        We consider that some aspects of one individual propensity to cooperation might not depend solely by the last behavior observed, but instead that choices are dependent on experience gathered on a sliding window of rounds, hence we define
        \begin{equation*}
                a^{s}_{t-1} = \frac{1}{t-T} \sum_{r = T}^{t-1} \overline{a}_{self}^r,\qquad a^{o}_{t-1} = \frac{1}{t-T} \sum_{r = T}^{t-1} \overline{a}_{others}^r,\qquad  T =  max(1, \, t-M),
        \end{equation*}
        where $M$ indicates the length of the window of ``memory" of an individual. In this initial work we assume for simplicity that all the individuals have equal memory, corresponding to $M = 5$ previous rounds. 
        In order to use them as regressors, we normalize as well the number of neighbors cooperating $n_{c}^{t-1}$, as  $\overline{n}_{c}^{t-1} = n_{c}^{t-1}/(N-1)$, and the last collected response time $RT_{t-1}$, by dividing it by the maximum value attained in the training dataset, i.e. $\overline{RT}_{t-1} = RT_{t-1} / \max_{\text{Exp1}}(RT)$, where $\max_{\text{Exp1}}(RT)$ is the maximum response time measured on the training data Experiment 1 (see section Data for details on this dataset).
        
        As additional regressors, we employ some measure of the level of experience and/or tiredness of the deciding individual, by considering the current round index $t$. We use this information in two ways, considering the characteristics of the dataset employed (see again section Data):
        \begin{itemize}
            \item we define the normalized round $\overline{t}$, dividing the index of the current round $t$ (that counts the number of rounds played by a player in the current scenario, playing with the same group of neighbors) by the maximum round $t_{max}$ experienced in training (i.e., the maximum round reached in Experiment 1, $t_{max} = 47$); this regressor can be interpreted as a measure of the amount of experience gathered both on the game and on the behavior of the current multiplayer group.
            \item We define the normalized experience $\overline{t}_{\small E}$, considering the total number of rounds ${t}_{\small E}$ of MIPD played by the individual, even if not with the same group of neighbors. It will coincide with the round $t$ on the data samples of Experiment 1, considering that Experiment 1 corresponds with the first phase of the experiment, when players are facing the game setup for the first time, while in Experiment 2 it will be obtained as ${t}_{\small E} = t_{max} + t$. We normalize again ${t}_{\small E}$ by dividing it by $t_{max}$. This regressor can be interpreted mainly as a measure of the amount of experience gathered on the game, but not on the current multiplayer group.
        \end{itemize}
        %
        The described measures are then used to define the regression vectors 
        \begin{align}
            x^{a}_{t-1} & = \begin{bmatrix} \overline{RT}_{t-1}, & d_{t-1}, & \overline{n}_c^{t-1}, &  \overline{a}_{self}^{t-1}, & \overline{t}_E,  & 1 \end{bmatrix}, \qquad 
            x^{\nu}_{t-1} = \begin{bmatrix} \overline{RA}_{t-1}, & \overline{t}, & 1 \end{bmatrix}, \qquad
            x^{z}_{t-1} = \begin{bmatrix} \overline{a}^{s}_{t-1}, & \overline{a}^{o}_{t-1},  & 1 \end{bmatrix} \nonumber\\
            & \qquad \qquad \qquad \qquad \qquad \qquad x^{t_0}_{t-1} = \begin{bmatrix} \overline{RT}_{t-1}, & d_{t-1}, & \overline{n}_c^{t-1}, & 1 \end{bmatrix}.
            \label{regressors}
        \end{align}
        thus we can rewrite Eq.~\ref{eq:tv-param-model-linear} as
        \begin{equation}
            \begin{array}{lcccccccccccc} 
            a_t & = & \overline{RT}_{t-1} \cdot p_1^{a} & + & d_{t-1} \cdot p_2^{a} & + & \overline{n}_c^{t-1} \cdot p_3^{a} & + & \overline{a}_{self}^{t-1} \cdot p_4^{a} & + & \overline{t}_E \cdot p_5^{a} & + & p_6^{a}, \vspace{0.25cm} \\
            \nu_t & = & \overline{RA}_{t-1} \cdot p_1^{\nu} & + &  \overline{t} \cdot p_2^{\nu} & + &  p_3^{\nu}, & \quad &  \quad & \quad & \quad & \quad & \quad \vspace{0.25cm} \\
            z_t & = & \overline{a}^{s}_{t-1} \cdot p_1^{z} & + & \overline{a}^{o}_{t-1}  \cdot p_2^{z} & + & p_3^{z}, & \quad &  \quad & \quad & \quad & \quad & \quad \vspace{0.25cm} \\
            {t_0}_t & = & \overline{RT}_{t-1} \cdot p_1^{t_0} & + & d_{t-1} \cdot p_2^{t_0} & + & \overline{n}_c^{t-1} \cdot p_3^{t_0} & + &  p_4^{t_0}. & \quad & \quad & \quad & \quad \vspace{0.25cm} \\
            \end{array}
            \label{model-reg}
        \end{equation}
        
        \subsubsection*{Model fitting and obtained parameters}
        
        The parameters $P_a$, $P_{\nu}$, $P_z$ and $P_{t_0}$ were fitted using a Python-based toolbox called Hierarchical Drift Diffusion Model (HDDM) \cite{wiecki2013hddm} version=0.9.8. The library performs Bayesian estimation for the DDM free parameters and allows the estimation of higher level parameters (in our case $P_a$, $P_{\nu}$, $P_{z}$, and $P_{t_0}$), describing the influence of regressors on such parameters, through the function hddm.HDDMRegressor.
        
        Conceptually, we assume that all the participants are copies of an average subject by aggregating across participants. Thus, we let our model use the regressors computed from data in Experiment 1 (training dataset) to estimate the parameter vectors $P_a$, $P_{\nu}$, $P_z$, $P_{t_0}$. The parameters obtained this way are included in Table~\ref{slope-intercept}. Fitting  $P_a$, $P_{\nu}$, $P_z$, and $P_{t_0}$ permits us to estimate the evolution of the original DDM parameters $\lbrace a_t, \, \nu_t , \, z_t, \, {t_{0}}_t \rbrace_t $ at each round $t$, which are the ones that we use in the analysis presented in section Results, jointly with Eq.~\ref{RT-PDF}-\ref{expected-coop-eq}.
        
        It is interesting to notice that some of the obtained parameters are easily interpretable and fit with the expected underlying dynamics. For instance, observing the model of the bias $z$ evolution, we can see that its evolution is directly proportional to the individual inclination towards others $\overline{a}_{t-1}^o$, and inversely proportional to the individual inclination towards self $\overline{a}_{t-1}^s$, both measured just before the decision. Moreover, the drift $\nu$, representing the individual velocity at extracting information from the stimulus, and the task complexity, is inversely proportional with respect to the growing number $\overline{t}$ of iterative games in a match. Similar observations can drown on the cautiousness $a$, which results to be inversely proportional to the normalized number of cooperators $\overline{n}_c^{t-1}$ and payoff $\overline{a}_{self}^{t-1}$ at the previous iteration, and to the length of the experience $\overline{t}_E$ collected on the game.
        
        \begin{table}
            \centering
            \begin{tabular}{ |c|c|c|c|c|}
             \hline
             DDM parameter$\TBstrut$ & Fitted Parameter$\TBstrut$ &  Regressor$\TBstrut$   & Mean$\TBstrut$ &   Standard deviation$\TBstrut$\\
             \hline
             $a (sec^{1/2})\Tstrut$     & $p_1^a \Tstrut$       &  $\overline{RT}_{t-1} \Tstrut$          & $5.512131 \Tstrut$   &   $0.244568  \Tstrut$\\
             $\quad$         & $p_2^a \TBstrut$      &  $d_{t-1} \TBstrut$                     & $-0.423748 \TBstrut$ &   $0.101119 \TBstrut$\\
             $\quad$         & $p_3^a \TBstrut$      &  $\overline{n}_c^{t-1} \TBstrut$          & $-0.528015 \TBstrut$ &   $0.165799 \TBstrut$\\
             $\quad$         & $p_4^a \TBstrut$      &  $\overline{a}_{self}^{t-1}  \TBstrut$  & $-0.08742 \TBstrut$  &   $0.130688 \TBstrut$\\
             $\quad$         & $p_5^a \TBstrut$      &  $\overline{t}_E \TBstrut$              & $-0.533712 \TBstrut$ &   $0.088916 \TBstrut$\\
             $\quad$         & $p_6^a \TBstrut$      &  $1 \text{ (intercept) }\Bstrut$        & $4.723139 \TBstrut$  &   $0.121242 \TBstrut$\\
             \hline
             $\nu (sec^{-1/2}) \Tstrut$   & $p_1^{\nu} \Tstrut$   &  $\overline{RA}_{t-1} \Tstrut $         & $0.136728 \Tstrut$   &   $0.006899 \Tstrut$\\
             $\quad$         & $p_2^{\nu} \TBstrut$  &  $\overline{t} \TBstrut$                & $-0.136411 \TBstrut$ &   $0.016796 \TBstrut$\\
             $\quad \Bstrut$ & $p_3^{\nu} \Bstrut$   &  $1 \text{ (intercept) }\Bstrut$        & $-0.086914 \Bstrut$  &   $0.010545 \Bstrut$\\
             \hline
             $z \Tstrut$     & $p_1^{z} \Tstrut$     &  $\overline{a}_{t-1}^s \Tstrut$         & $-0.056263 \Tstrut$  &   $0.021486 \Tstrut$\\
             $\quad$         & $p_2^{z} \TBstrut$    &  $\overline{a}_{t-1}^o \TBstrut$        & $0.033669 \TBstrut$  &   $0.023936 \TBstrut$\\
             $\quad \Bstrut$ & $p_3^{z} \Bstrut$     &  $1 \text{ (intercept) }\Bstrut$        & $0.568235 \Bstrut$   &   $0.019991 \Bstrut$\\
             \hline
             $t_0 (sec) \Tstrut$   & $p_1^{t_0} \Tstrut$   &  $\overline{RT}_{t-1} \Tstrut$          & $1.328126 \Tstrut$   &    $0.107053 \Tstrut$\\
             $\quad$         & $p_2^{t_0} \TBstrut$  &  $d_{t-1} \TBstrut$                     & $-0.065272 \TBstrut$ &   $0.016784 \TBstrut$\\
             $\quad$         & $p_3^{t_0} \TBstrut$  &  $\overline{n}_c^{t-1} \TBstrut$          & $0.908602 \TBstrut$  &    $0.071989 \TBstrut$\\
             $\quad$         & $p_4^{t_0} \Bstrut$   &  $1 \text{ (intercept) }\Bstrut$        & $0.054696 \Bstrut$   &   $0.018869 \Bstrut$\\
             \hline
             \end{tabular}
            \caption{Estimated values of parameters $P_a$, $P_{\nu}$, $P_z$ and $P_{t_0}$ associated with the Drift Diffusion Model \eqref{DDM-eq} with time-varying parameters described by dynamics in Eq.~\ref{model-reg}.}
            \label{slope-intercept}
        \end{table}
        
        \subsubsection*{Using the model to predict and to simulate}
        
        Once parameters $P_a = (p_i^a)_{i=1}^6$, $P_{\nu} = (p_i^{\nu})_{i=1}^3$, $P_z = (p_i^z)_{i=1}^3$, $P_{t_0} = (p_i^{t_0})_{i=1}^4$ in Table~\ref{slope-intercept} are fitted, we want to test the results obtained by using them to predict the behaviors contained in Experiment 2. In order to do so, for each round $t \, = \, 2, \, 3, \, \dots , \, 58$ and for each player $k$ included in the considered dataset we first proceed in computing the regression vectors $x_{t-1}^a(k)$, $x_{t-1}^{\nu}(k)$, $x_{t-1}^z(k)$, $x_{t-1}^{t_0}(k)$. Then we employ the values $\mu(p_i^j)$ for $j \in \lbrace a, \, \nu, \, z, \, t_0 \rbrace$ to predict the evolution of parameters $a_t(k)$, ${\nu}_t(k)$, $z_t(k)$, ${t_0}_t(k)$ associated with each individual $k$ at each round $t$, using Eq.~\ref{model-reg}. Finally, in order to represent the behavior of the average subject in Experiment 2, we consider as aggregated parameters at each round $t \, = \, 2, \, 3, \, \dots , \, 58$ of the test set
        \begin{equation}\label{param-model}
           a_t = \frac{1}{N_{\text{subj}}} \sum_{k = 1}^{N_{\text{subj}}} a_t(k)  , \qquad \nu_t = \frac{1}{N_{\text{subj}}} \sum_{k = 1}^{N_{\text{subj}}} \nu_t(k), \qquad z_t = \frac{1}{N_{\text{subj}}} \sum_{k = 1}^{N_{\text{subj}}} z_t(k), \qquad {t_{0_t}} = \frac{1}{N_{\text{subj}}} \sum_{k = 1}^{N_{\text{subj}}} t_{0_t}(k).
        \end{equation}
        The same method is applied to estimate the evolution of the parameters in simulated environments, with the sole difference that the regression vectors $x_{t-1}^a(k)$, $x_{t-1}^{\nu}(k)$, $x_{t-1}^z(k)$, $x_{t-1}^{t_0}(k)$ are not computed from data collected from a real-world experiment, as it is done to predict over the test set. In simulation instead, a DDM augmented with the fitted parameters dynamics is instantiated per each individual whose decision-making behavior is to be simulated. Such simulated individual will make actual choices based on the dynamics decribed in Eq. \eqref{DDM-eq}-\eqref{model-reg}, in the context of a set of $N_{\text{realiz}}$ simulated MIPD, playing in groups of $N$ players (either all DDM based or representing different taylored decison making strategies). The players choices up to round $t-1$ will be collected and used to compute the simulated regression vectors $x_{t-1}^a(k)$, $x_{t-1}^{\nu}(k)$, $x_{t-1}^z(k)$, $x_{t-1}^{t_0}(k)$, based on which the model would evolve in time, influencing decisions at round $t$. By averaging among realizations and individuals, we obtained the average evolution of the parameters in different simulating scenarios, described in the second part of section Results.
        
    \section*{Acknowledgements}
    
    We thank the authors of ref \cite{grujic2010social}, J. Grujić, C. Fosco, L. Araujo, J. A. Cuesta, A. Sánchez for allowing us to conduct this study on the mentioned data. This project is funded by the European Union. However, the views and opinions expressed are those of the author(s) only and do not necessarily reflect those of the European Union or the European Health and Digital Executive Agency (HaDEA). Neither the European Union nor the granting authority can be held responsible for them. Grant Agreement no. 101120763 - TANGO. LGAZ acknowledges financial support from the project `Understanding Misinformation and Science in Societal Debates' (UnMiSSeD), funded by the European Media and Information Fund.

    \bibliographystyle{unsrt}

\begin{thebibliography}{100}

        \bibitem{baron2023thinking}
        Jonathan Baron.
        \newblock {\em Thinking and deciding}.
        \newblock Cambridge University Press, 2023.
        
        \bibitem{bassett2011understanding}
        Danielle~S Bassett and Michael~S Gazzaniga.
        \newblock Understanding complexity in the human brain.
        \newblock {\em Trends in cognitive sciences}, 15(5):200--209, 2011.
        
        \bibitem{krohn2023spatiotemporal}
        Stephan Krohn, Nina von Schwanenflug, Leonhard Waschke, Amy Romanello, Martin Gell, Douglas~D Garrett, and Carsten Finke.
        \newblock A spatiotemporal complexity architecture of human brain activity.
        \newblock {\em Science Advances}, 9(5):eabq3851, 2023.
        
        \bibitem{hilbert2012toward}
        Martin Hilbert.
        \newblock Toward a synthesis of cognitive biases: how noisy information processing can bias human decision making.
        \newblock {\em Psychological bulletin}, 138(2):211, 2012.
        
        \bibitem{korteling2020cognitive}
        Johan~E Korteling and Alexander Toet.
        \newblock Cognitive biases.
        \newblock {\em Encyclopedia of behavioral neuroscience}, 2020.
        
        \bibitem{george2016affect}
        Jennifer~M George and Erik Dane.
        \newblock Affect, emotion, and decision making.
        \newblock {\em Organizational Behavior and Human Decision Processes}, 136:47--55, 2016.
        
        \bibitem{yukalov2022quantification}
        Vyacheslav~I Yukalov.
        \newblock Quantification of emotions in decision making.
        \newblock {\em Soft Computing}, 26(5):2419--2436, 2022.
        
        \bibitem{bruch2017decision}
        Elizabeth Bruch and Fred Feinberg.
        \newblock Decision-making processes in social contexts.
        \newblock {\em Annual review of sociology}, 43(1):207--227, 2017.
        
        \bibitem{danziger2011extraneous}
        Shai Danziger, Jonathan Levav, and Liora Avnaim-Pesso.
        \newblock Extraneous factors in judicial decisions.
        \newblock {\em Proceedings of the National Academy of Sciences}, 108(17):6889--6892, 2011.
        
        \bibitem{harmon1999cognitive}
        Eddie Harmon-Jones and Judson Mills.
        \newblock Cognitive dissonance.
        \newblock {\em Progress on a pivotal theory in social psychology. Washington, DC: American Psychological Association}, 1999.
        
        \bibitem{kahneman2011thinking}
        Daniel Kahneman.
        \newblock Thinking fast and slow.
        \newblock {\em Farrar, Strauss and Giroux}, 2011.
        
        \bibitem{franken2005individual}
        Ingmar~HA Franken and Peter Muris.
        \newblock Individual differences in decision-making.
        \newblock {\em Personality and Individual Differences}, 39(5):991--998, 2005.
        
        \bibitem{appelt2011decision}
        Kirstin~C Appelt, Kerry~F Milch, Michel~JJ Handgraaf, and Elke~U Weber.
        \newblock The decision making individual differences inventory and guidelines for the study of individual differences in judgment and decision-making research.
        \newblock {\em Judgment and Decision making}, 6(3):252--262, 2011.
        
        \bibitem{edwards1954theory}
        Ward Edwards.
        \newblock The theory of decision making.
        \newblock {\em Psychological bulletin}, 51(4):380, 1954.
        
        \bibitem{evans2011dual}
        Jonathan~St.B.T. Evans.
        \newblock Dual-process theories of reasoning: Contemporary issues and developmental applications.
        \newblock {\em Developmental Review}, 31(2):86--102, 2011.
        
        \bibitem{kahneman2002representativeness}
        Daniel Kahneman, Shane Frederick, et~al.
        \newblock Representativeness revisited: Attribute substitution in intuitive judgment.
        \newblock {\em Heuristics and biases: The psychology of intuitive judgment}, 49(49-81):74, 2002.
        
        \bibitem{mishra2014decision}
        Sandeep Mishra.
        \newblock Decision-making under risk: Integrating perspectives from biology, economics, and psychology.
        \newblock {\em Personality and Social Psychology Review}, 18(3):280--307, 2014.
        
        \bibitem{fellows2004cognitive}
        Lesley~K Fellows.
        \newblock The cognitive neuroscience of human decision making: a review and conceptual framework.
        \newblock {\em Behavioral and cognitive neuroscience reviews}, 3(3):159--172, 2004.
        
        \bibitem{yoon2012decision}
        Carolyn Yoon, Richard Gonzalez, Antoine Bechara, Gregory~S Berns, Alain~A Dagher, Laurette Dub{\'e}, Scott~A Huettel, Joseph~W Kable, Israel Liberzon, Hilke Plassmann, et~al.
        \newblock Decision neuroscience and consumer decision making.
        \newblock {\em Marketing letters}, 23:473--485, 2012.
        
        \bibitem{dale2015heuristics}
        Steve Dale.
        \newblock Heuristics and biases: The science of decision-making.
        \newblock {\em Business Information Review}, 32(2):93--99, 2015.
        
        \bibitem{american1966theories}
        American~Economic Association, Royal~Economic Society, and Herbert~A Simon.
        \newblock {\em Theories of decision-making in economics and behavioural science}.
        \newblock Springer, 1966.
        
        \bibitem{dube2013vaccine}
        Eve Dub{\'e}, Caroline Laberge, Maryse Guay, Paul Bramadat, R{\'e}al Roy, and Julie~A Bettinger.
        \newblock Vaccine hesitancy: an overview.
        \newblock {\em Human vaccines \& immunotherapeutics}, 9(8):1763--1773, 2013.
        
        \bibitem{benin2006qualitative}
        Andrea~L Benin, Daryl~J Wisler-Scher, Eve Colson, Eugene~D Shapiro, and Eric~S Holmboe.
        \newblock Qualitative analysis of mothers' decision-making about vaccines for infants: the importance of trust.
        \newblock {\em Pediatrics}, 117(5):1532--1541, 2006.
        
        \bibitem{ajzen1996social}
        Icek Ajzen.
        \newblock The social psychology of decision making.
        \newblock {\em Social psychology: Handbook of basic principles}, pages 297--325, 1996.
        
        \bibitem{stangor2015social}
        Charles Stangor.
        \newblock {\em Social groups in action and interaction}.
        \newblock Routledge, 2015.
        
        \bibitem{gong2017social}
        Xu~Gong and Alan~G Sanfey.
        \newblock Social rank and social cooperation: Impact of social comparison processes on cooperative decision-making.
        \newblock {\em PLoS One}, 12(4):e0175472, 2017.
        
        \bibitem{rand2013human}
        David~G Rand and Martin~A Nowak.
        \newblock Human cooperation.
        \newblock {\em Trends in cognitive sciences}, 17(8):413--425, 2013.
        
        \bibitem{keddy2012competition}
        Paul~A Keddy.
        \newblock {\em Competition}, volume~26.
        \newblock Springer Science \& Business Media, 2012.
        
        \bibitem{toma2013strategic}
        Claudia Toma, Ingrid Gilles, and Fabrizio Butera.
        \newblock Strategic use of preference confirmation in group decision making: The role of competition and dissent.
        \newblock {\em British Journal of Social Psychology}, 52(1):44--63, 2013.
        
        \bibitem{garcia2013psychology}
        Stephen~M Garcia, Avishalom Tor, and Tyrone~M Schiff.
        \newblock The psychology of competition: A social comparison perspective.
        \newblock {\em Perspectives on psychological science}, 8(6):634--650, 2013.
        
        \bibitem{batson2010altruism}
        C~Daniel Batson.
        \newblock {\em Altruism in humans}.
        \newblock Oxford University Press, 2010.
        
        \bibitem{tusche2021neurocomputational}
        Anita Tusche and Lisa~M Bas.
        \newblock Neurocomputational models of altruistic decision-making and social motives: Advances, pitfalls, and future directions.
        \newblock {\em Wiley Interdisciplinary Reviews: Cognitive Science}, 12(6):e1571, 2021.
        
        \bibitem{rossetti2024direct}
        Charlotte~SL Rossetti and Christian Hilbe.
        \newblock Direct reciprocity among humans.
        \newblock {\em Ethology}, 130(4):e13407, 2024.
        
        \bibitem{solanas2009measuring}
        Antonio Solanas, David Leiva, Vicenta Sierra, and Llu{\'\i}s Salafranca.
        \newblock Measuring and making decisions for social reciprocity.
        \newblock {\em Behavior research methods}, 41(3):742--754, 2009.
        
        \bibitem{romano2017reciprocity}
        Angelo Romano and Daniel Balliet.
        \newblock Reciprocity outperforms conformity to promote cooperation.
        \newblock {\em Psychological Science}, 28(10):1490--1502, 2017.
        
        \bibitem{constant2019regimes}
        Axel Constant, Maxwell~JD Ramstead, Samuel~PL Veissi{\`e}re, and Karl Friston.
        \newblock Regimes of expectations: an active inference model of social conformity and human decision making.
        \newblock {\em Frontiers in psychology}, 10:679, 2019.
        
        \bibitem{hertz2016influence}
        Nicholas Hertz and Eva Wiese.
        \newblock Influence of agent type and task ambiguity on conformity in social decision making.
        \newblock In {\em Proceedings of the human factors and ergonomics society annual meeting}, volume~60, pages 313--317. SAGE Publications Sage CA: Los Angeles, CA, 2016.
        
        \bibitem{aureli2000natural}
        Filippo Aureli, Frans De~Waal, et~al.
        \newblock {\em Natural conflict resolution}.
        \newblock University of California Press Berkeley, 2000.
        
        \bibitem{danesh2002has}
        Hossain~B Danesh and Roshan Danesh.
        \newblock Has conflict resolution grown up? toward a developmental model of decision making and conflict resolution.
        \newblock {\em International Journal of Peace Studies}, pages 59--76, 2002.
        
        \bibitem{kelman1990interactive}
        Herbert~C Kelman.
        \newblock Interactive problem-solving: A social-psychological approach to conflict resolution.
        \newblock In {\em Conflict: Readings in management and resolution}, pages 199--215. Springer, 1990.
        
        \bibitem{cikara2014neuroscience}
        Mina Cikara and Jay~J Van~Bavel.
        \newblock The neuroscience of intergroup relations: An integrative review.
        \newblock {\em Perspectives on Psychological Science}, 9(3):245--274, 2014.
        
        \bibitem{lee2013how}
        Victoria Lee and Lasana Harris.
        \newblock How social cognition can inform social decision making.
        \newblock {\em Frontiers in neuroscience}, 7:259, 12 2013.
        
        \bibitem{rilling2011neuroscience}
        James~K Rilling and Alan~G Sanfey.
        \newblock The neuroscience of social decision-making.
        \newblock {\em Annual review of psychology}, 62:23--48, 2011.
        
        \bibitem{sanfey2007social}
        Alan~G Sanfey.
        \newblock Social decision-making: insights from game theory and neuroscience.
        \newblock {\em Science}, 318(5850):598--602, 2007.
        
        \bibitem{boyd2006solving}
        Robert Boyd and Peter~J Richerson.
        \newblock Solving the puzzle of human cooperation.
        \newblock {\em Evolution and culture}, pages 105--132, 2006.
        
        \bibitem{kwon2019peace}
        Young-Mi Kwon and Juhwa Park.
        \newblock Peace through cooperation or peace through strength? how to achieve peace in the very intractable conflict society.
        \newblock {\em Historical Social Research/Historische Sozialforschung}, 44(4 (170):269--292, 2019.
        
        \bibitem{maas2013conflict}
        Achim Maas, Alexander Carius, and Anja Wittich.
        \newblock From conflict to cooperation? environmental cooperation as a tool for peace-building.
        \newblock In {\em Environmental security}, pages 102--120. Routledge, 2013.
        
        \bibitem{delay2019mutual}
        Etienne Delay and Cyril Piou.
        \newblock Mutual aid: When does resource scarcity favour group cooperation?
        \newblock {\em Ecological Complexity}, 40:100790, 2019.
        
        \bibitem{bouma2008trust}
        Jetske Bouma, Erwin Bulte, and Daan Van~Soest.
        \newblock Trust and cooperation: Social capital and community resource management.
        \newblock {\em Journal of environmental economics and management}, 56(2):155--166, 2008.
        
        \bibitem{fomina2018industrial}
        Alena~V Fomina, Oksana~N Berduygina, and Alexander~A Shatsky.
        \newblock Industrial cooperation and its influence on sustainable economic growth.
        \newblock {\em Entrepreneurship and Sustainability Issues}, 5(3):467--479, 2018.
        
        \bibitem{strachan2018relationship}
        Anna~Louise Strachan.
        \newblock Relationship between regional cooperation and political stability and prosperity.
        \newblock {\em K4D Helpdesk Report. Brighton, UK: Institute of Development Studies}, 2018.
        
        \bibitem{boyd2009culture}
        Robert Boyd and Peter~J Richerson.
        \newblock Culture and the evolution of human cooperation.
        \newblock {\em Philosophical Transactions of the Royal Society B: Biological Sciences}, 364(1533):3281--3288, 2009.
        
        \bibitem{johnson2013cooperation}
        David~W Johnson and Roger~T Johnson.
        \newblock Cooperation and the use of technology.
        \newblock In {\em Handbook of research on educational communications and technology}, pages 777--803. Routledge, 2013.
        
        \bibitem{van2000cooperation}
        Mark Van~Vugt, Mark Snyder, Tom~R Tyler, and Anders Biel.
        \newblock Cooperation in modern society.
        \newblock {\em Promoting the welfare of communities, states and organizations: Routledge}, 2000.
        
        \bibitem{suarez2021prevalence}
        Victor Suarez-Lledo and Javier Alvarez-Galvez.
        \newblock Prevalence of health misinformation on social media: systematic review.
        \newblock {\em Journal of medical Internet research}, 23(1):e17187, 2021.
        
        \bibitem{swire2020public}
        Briony Swire-Thompson, David Lazer, et~al.
        \newblock Public health and online misinformation: challenges and recommendations.
        \newblock {\em Annu Rev Public Health}, 41(1):433--451, 2020.
        
        \bibitem{carroll1988iterated}
        John~W Carroll.
        \newblock Iterated n-player prisoner's dilemma games.
        \newblock {\em Philosophical Studies: An International Journal for Philosophy in the Analytic Tradition}, 53(3):411--415, 1988.
        
        \bibitem{basu1977information}
        Kaushik Basu.
        \newblock Information and strategy in iterated prisoner's dilemma.
        \newblock {\em Theory and Decision}, 8(3):293, 1977.
        
        \bibitem{lee2008game}
        Daeyeol Lee.
        \newblock Game theory and neural basis of social decision making.
        \newblock {\em Nature neuroscience}, 11(4):404--409, 2008.
        
        \bibitem{rand2016cooperation}
        David~G Rand.
        \newblock Cooperation, fast and slow: Meta-analytic evidence for a theory of social heuristics and self-interested deliberation.
        \newblock {\em Psychological science}, 27(9):1192--1206, 2016.
        
        \bibitem{wood2016cooperation}
        Ruth~I Wood, Jessica~Y Kim, and Grace~R Li.
        \newblock Cooperation in rats playing the iterated prisoner's dilemma game.
        \newblock {\em Animal behaviour}, 114:27--35, 2016.
        
        \bibitem{zeng2016risk}
        Weijun Zeng, Minqiang Li, Fuzan Chen, and Guofang Nan.
        \newblock Risk consideration and cooperation in the iterated prisoner’s dilemma.
        \newblock {\em Soft Computing}, 20:567--587, 2016.
        
        \bibitem{li2016changing}
        Jiaqi Li, Chunyan Zhang, Qinglin Sun, Zengqiang Chen, and Jianlei Zhang.
        \newblock Changing the intensity of interaction based on individual behavior in the iterated prisoner’s dilemma game.
        \newblock {\em IEEE Transactions on Evolutionary Computation}, 21(4):506--517, 2016.
        
        \bibitem{raihani2011resolving}
        Nichola~J Raihani and Redouan Bshary.
        \newblock Resolving the iterated prisoner’s dilemma: theory and reality.
        \newblock {\em Journal of Evolutionary Biology}, 24(8):1628--1639, 2011.
        
        \bibitem{ishibuchi2005evolution}
        Hisao Ishibuchi and Naoki Namikawa.
        \newblock Evolution of cooperative behavior in the iterated prisoner's dilemma under random pairing in game playing.
        \newblock In {\em 2005 IEEE Congress on Evolutionary Computation}, volume~3, pages 2637--2644. IEEE, 2005.
        
        \bibitem{ratcliff2016diffusion}
        Roger Ratcliff, Philip~L Smith, Scott~D Brown, and Gail McKoon.
        \newblock Diffusion decision model: Current issues and history.
        \newblock {\em Trends in cognitive sciences}, 20(4):260--281, 2016.
        
        \bibitem{ratcliff1978theory}
        Roger Ratcliff.
        \newblock A theory of memory retrieval.
        \newblock {\em Psychological review}, 85(2):59, 1978.
        
        \bibitem{gallotti2019quantitative}
        Riccardo Gallotti and Jelena Gruji{\'c}.
        \newblock A quantitative description of the transition between intuitive altruism and rational deliberation in iterated prisoner’s dilemma experiments.
        \newblock {\em Scientific reports}, 9(1):17046, 2019.
        
        \bibitem{hutcherson2015neurocomputational}
        Cendri~A Hutcherson, Benjamin Bushong, and Antonio Rangel.
        \newblock A neurocomputational model of altruistic choice and its implications.
        \newblock {\em Neuron}, 87(2):451--462, 2015.
        
        \bibitem{krajbich2015common}
        Ian Krajbich, Todd Hare, Bj{\"o}rn Bartling, Yosuke Morishima, and Ernst Fehr.
        \newblock A common mechanism underlying food choice and social decisions.
        \newblock {\em PLoS computational biology}, 11(10):e1004371, 2015.
        
        \bibitem{andrejevic2022response}
        Milan Andrejevi{\'c}, Joshua~P White, Daniel Feuerriegel, Simon Laham, and Stefan Bode.
        \newblock Response time modelling reveals evidence for multiple, distinct sources of moral decision caution.
        \newblock {\em Cognition}, 223:105026, 2022.
        
        \bibitem{bogacz2006physics}
        Rafal Bogacz, Eric Brown, Jeff Moehlis, Philip Holmes, and Jonathan~D Cohen.
        \newblock The physics of optimal decision making: a formal analysis of models of performance in two-alternative forced-choice tasks.
        \newblock {\em Psychological review}, 113(4):700, 2006.
        
        \bibitem{ratcliff2008diffusion}
        Roger Ratcliff and Gail McKoon.
        \newblock The diffusion decision model: theory and data for two-choice decision tasks.
        \newblock {\em Neural computation}, 20(4):873--922, 2008.
        
        \bibitem{lewis2015applied}
        Colin Lewis-Beck and Michael Lewis-Beck.
        \newblock {\em Applied regression: An introduction}, volume~22.
        \newblock Sage publications, 2015.
        
        \bibitem{grujic2010social}
        Jelena Gruji{\'c}, Constanza Fosco, Lourdes Araujo, Jos{\'e}~A Cuesta, and Angel S{\'a}nchez.
        \newblock Social experiments in the mesoscale: Humans playing a spatial prisoner's dilemma.
        \newblock {\em PloS one}, 5(11):e13749, 2010.
        
        \bibitem{rand2011dynamic}
        David~G Rand, Samuel Arbesman, and Nicholas~A Christakis.
        \newblock Dynamic social networks promote cooperation in experiments with humans.
        \newblock {\em Proceedings of the National Academy of Sciences}, 108(48):19193--19198, 2011.
        
        \bibitem{traulsen2010human}
        Arne Traulsen, Dirk Semmann, Ralf~D Sommerfeld, Hans-J{\"u}rgen Krambeck, and Manfred Milinski.
        \newblock Human strategy updating in evolutionary games.
        \newblock {\em Proceedings of the National Academy of Sciences}, 107(7):2962--2966, 2010.
        
        \bibitem{mao2017resilient}
        Andrew Mao, Lili Dworkin, Siddharth Suri, and Duncan~J Watts.
        \newblock Resilient cooperators stabilize long-run cooperation in the finitely repeated prisoner’s dilemma.
        \newblock {\em Nature communications}, 8(1):13800, 2017.
        
        \bibitem{balliet2011reward}
        Daniel Balliet, Laetitia~B Mulder, and Paul~AM Van~Lange.
        \newblock Reward, punishment, and cooperation: a meta-analysis.
        \newblock {\em Psychological bulletin}, 137(4):594, 2011.
        
        \bibitem{evans2019cooperation}
        Anthony~M Evans and David~G Rand.
        \newblock Cooperation and decision time.
        \newblock {\em Current opinion in psychology}, 26:67--71, 2019.
        
        \bibitem{rand2014social}
        David~G Rand, Alexander Peysakhovich, Gordon~T Kraft-Todd, George~E Newman, Owen Wurzbacher, Martin~A Nowak, and Joshua~D Greene.
        \newblock Social heuristics shape intuitive cooperation.
        \newblock {\em Nature communications}, 5(1):3677, 2014.
        
        \bibitem{bear2016intuition}
        Adam Bear and David~G Rand.
        \newblock Intuition, deliberation, and the evolution of cooperation.
        \newblock {\em Proceedings of the National Academy of Sciences}, 113(4):936--941, 2016.
        
        \bibitem{rand2012spontaneous}
        David~G Rand, Joshua~D Greene, and Martin~A Nowak.
        \newblock Spontaneous giving and calculated greed.
        \newblock {\em Nature}, 489(7416):427--430, 2012.
        
        \bibitem{everett2017deliberation}
        Jim~AC Everett, Zach Ingbretsen, Fiery Cushman, and Mina Cikara.
        \newblock Deliberation erodes cooperative behavior—even towards competitive out-groups, even when using a control condition, and even when eliminating selection bias.
        \newblock {\em Journal of Experimental Social Psychology}, 73:76--81, 2017.
        
        \bibitem{nay2016predicting}
        John~J Nay and Yevgeniy Vorobeychik.
        \newblock Predicting human cooperation.
        \newblock {\em PloS one}, 11(5):e0155656, 2016.
        
        \bibitem{embrey2018cooperation}
        Matthew Embrey, Guillaume~R Fr{\'e}chette, and Sevgi Yuksel.
        \newblock Cooperation in the finitely repeated prisoner’s dilemma.
        \newblock {\em The Quarterly Journal of Economics}, 133(1):509--551, 2018.
        
        \bibitem{guilfoos2016predicting}
        Todd Guilfoos and Andreas~Duus Pape.
        \newblock Predicting human cooperation in the prisoner’s dilemma using case-based decision theory.
        \newblock {\em Theory and Decision}, 80:1--32, 2016.
        
        \bibitem{lin2019split}
        Baihan Lin, Djallel Bouneffouf, and Guillermo Cecchi.
        \newblock Split q learning: reinforcement learning with two-stream rewards.
        \newblock {\em arXiv preprint arXiv:1906.12350}, 2019.
        
        \bibitem{lin2019story}
        Baihan Lin, Guillermo Cecchi, Djallel Bouneffouf, Jenna Reinen, and Irina Rish.
        \newblock A story of two streams: Reinforcement learning models from human behavior and neuropsychiatry.
        \newblock {\em arXiv preprint arXiv:1906.11286}, 2019.
        
        \bibitem{lin2020unified}
        Baihan Lin, Guillermo Cecchi, Djallel Bouneffouf, Jenna Reinen, and Irina Rish.
        \newblock Unified models of human behavioral agents in bandits, contextual bandits and rl.
        \newblock {\em arXiv preprint arXiv:2005.04544}, 2020.
        
        \bibitem{lin2021models}
        Baihan Lin, Guillermo Cecchi, Djallel Bouneffouf, Jenna Reinen, and Irina Rish.
        \newblock Models of human behavioral agents in bandits, contextual bandits and rl.
        \newblock In {\em Human Brain and Artificial Intelligence: Second International Workshop, HBAI 2020, Held in Conjunction with IJCAI-PRICAI 2020, Yokohama, Japan, January 7, 2021, Revised Selected Papers 2}, pages 14--33. Springer, 2021.
        
        \bibitem{lin2022online}
        Baihan Lin, Djallel Bouneffouf, and Guillermo Cecchi.
        \newblock Online learning in iterated prisoner’s dilemma to mimic human behavior.
        \newblock In {\em Pacific rim international conference on artificial intelligence}, pages 134--147. Springer, 2022.
        
        \bibitem{forstmann2016sequential}
        Birte~U Forstmann, Roger Ratcliff, and E-J Wagenmakers.
        \newblock Sequential sampling models in cognitive neuroscience: Advantages, applications, and extensions.
        \newblock {\em Annual review of psychology}, 67:641--666, 2016.
        
        \bibitem{pardo2019mechanistic}
        Jose~L Pardo-Vazquez, Juan~R Casti{\~n}eiras-de Saa, Mafalda Valente, Iris Dami{\~a}o, Tiago Costa, M~In{\^e}s Vicente, Andr{\'e}~G Mendon{\c{c}}a, Zachary~F Mainen, and Alfonso Renart.
        \newblock The mechanistic foundation of weber’s law.
        \newblock {\em Nature neuroscience}, 22(9):1493--1502, 2019.
        
        \bibitem{lin2022predicting}
        Baihan Lin, Djallel Bouneffouf, and Guillermo Cecchi.
        \newblock Predicting human decision making in psychological tasks with recurrent neural networks.
        \newblock {\em PloS one}, 17(5):e0267907, 2022.
        
        \bibitem{mckee2023scaffolding}
        Kevin~R McKee, Andrea Tacchetti, Michiel~A Bakker, Jan Balaguer, Lucy Campbell-Gillingham, Richard Everett, and Matthew Botvinick.
        \newblock Scaffolding cooperation in human groups with deep reinforcement learning.
        \newblock {\em Nature Human Behaviour}, pages 1--10, 2023.
        
        \bibitem{walsh2002analyzing}
        William~E Walsh, Rajarshi Das, Gerald Tesauro, and Jeffrey~O Kephart.
        \newblock Analyzing complex strategic interactions in multi-agent systems.
        \newblock In {\em AAAI-02 Workshop on Game-Theoretic and Decision-Theoretic Agents}, pages 109--118, 2002.
        
        \bibitem{wellman2006methods}
        Michael~P Wellman.
        \newblock Methods for empirical game-theoretic analysis.
        \newblock In {\em AAAI}, volume 980, pages 1552--1556, 2006.
        
        \bibitem{leibo2017multi}
        Joel~Z Leibo, Vinicius Zambaldi, Marc Lanctot, Janusz Marecki, and Thore Graepel.
        \newblock Multi-agent reinforcement learning in sequential social dilemmas.
        \newblock {\em arXiv preprint arXiv:1702.03037}, 2017.
        
        \bibitem{hughes2018inequity}
        Edward Hughes, Joel~Z Leibo, Matthew Phillips, Karl Tuyls, Edgar Due{\~n}ez-Guzman, Antonio Garc{\'\i}a~Casta{\~n}eda, Iain Dunning, Tina Zhu, Kevin McKee, Raphael Koster, et~al.
        \newblock Inequity aversion improves cooperation in intertemporal social dilemmas.
        \newblock {\em Advances in neural information processing systems}, 31, 2018.
        
        \bibitem{schelling1973hockey}
        Thomas~C Schelling.
        \newblock Hockey helmets, concealed weapons, and daylight saving: A study of binary choices with externalities.
        \newblock {\em Journal of Conflict resolution}, 17(3):381--428, 1973.
        
        \bibitem{perolat2017multi}
        Julien Perolat, Joel~Z Leibo, Vinicius Zambaldi, Charles Beattie, Karl Tuyls, and Thore Graepel.
        \newblock A multi-agent reinforcement learning model of common-pool resource appropriation.
        \newblock {\em Advances in neural information processing systems}, 30, 2017.
        
        \bibitem{ratcliff1980note}
        Roger Ratcliff.
        \newblock A note on modeling accumulation of information when the rate of accumulation changes over time.
        \newblock {\em Journal of Mathematical Psychology}, 1980.
        
        \bibitem{ratcliff1998modeling}
        Roger Ratcliff and Jeffrey~N Rouder.
        \newblock Modeling response times for two-choice decisions.
        \newblock {\em Psychological science}, 9(5):347--356, 1998.
        
        \bibitem{moran2015optimal}
        Rani Moran.
        \newblock Optimal decision making in heterogeneous and biased environments.
        \newblock {\em Psychonomic bulletin \& review}, 22:38--53, 2015.
        
        \bibitem{edwards1965optimal}
        Ward Edwards.
        \newblock Optimal strategies for seeking information: Models for statistics, choice reaction times, and human information processing.
        \newblock {\em Journal of Mathematical Psychology}, 2(2):312--329, 1965.
        
        \bibitem{ratcliff1999connectionist}
        Roger Ratcliff, Trisha Van~Zandt, and Gail McKoon.
        \newblock Connectionist and diffusion models of reaction time.
        \newblock {\em Psychological review}, 106(2):261, 1999.
        
        \bibitem{feller1991introduction}
        William Feller.
        \newblock {\em An introduction to probability theory and its applications}, volume~81.
        \newblock John Wiley \& Sons, 1991.
        
        \bibitem{montero2022fast}
        Eladio Montero-Porras, Tom Lenaerts, Riccardo Gallotti, and Jelena Grujic.
        \newblock Fast deliberation is related to unconditional behaviour in iterated prisoners’ dilemma experiments.
        \newblock {\em Scientific Reports}, 12(1):20287, 2022.
        
        \bibitem{wiecki2013hddm}
        Thomas~V Wiecki, Imri Sofer, and Michael~J Frank.
        \newblock Hddm: Hierarchical bayesian estimation of the drift-diffusion model in python.
        \newblock {\em Frontiers in neuroinformatics}, page~14, 2013.
        
        \end{thebibliography}

    \section*{Supplementary Information}
    \label{additional-material}
        \setcounter{figure}{0}
        \section*{Rationality regimes over time}
        \begin{figure}[H]
            \centering
            \includegraphics[width=\textwidth]{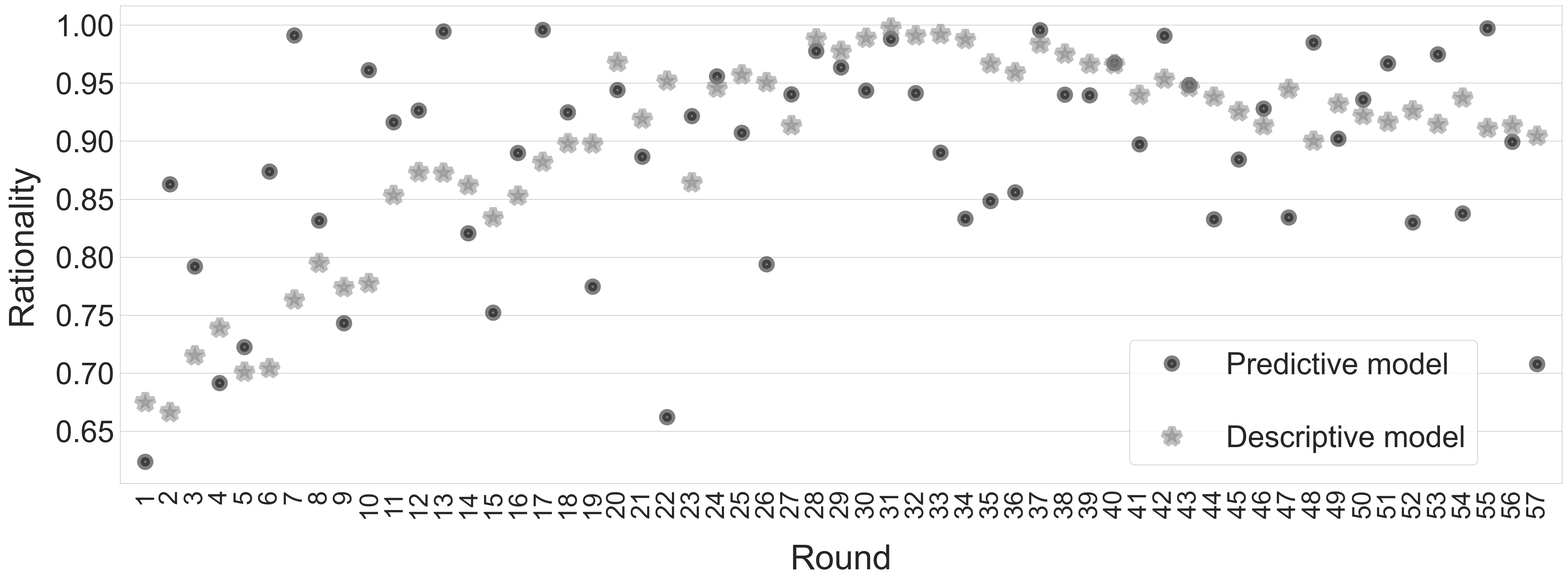}
            \caption{Rationality values attained on the test dataset by using Eq.(~\ref{eq-rationality}) employing the parameters directly fitted at each round via Bayesian Regression (gray), and by employing the parameters predicted by our model (black).}
            \label{fig:rationality}
        \end{figure}

        \section*{Predicted response time PDF at each round (test set)} 
        \begin{figure}[H]
             \centering
             \begin{subfigure}[b]{0.45\textwidth}
                 \centering
                 \includegraphics[width=\textwidth]{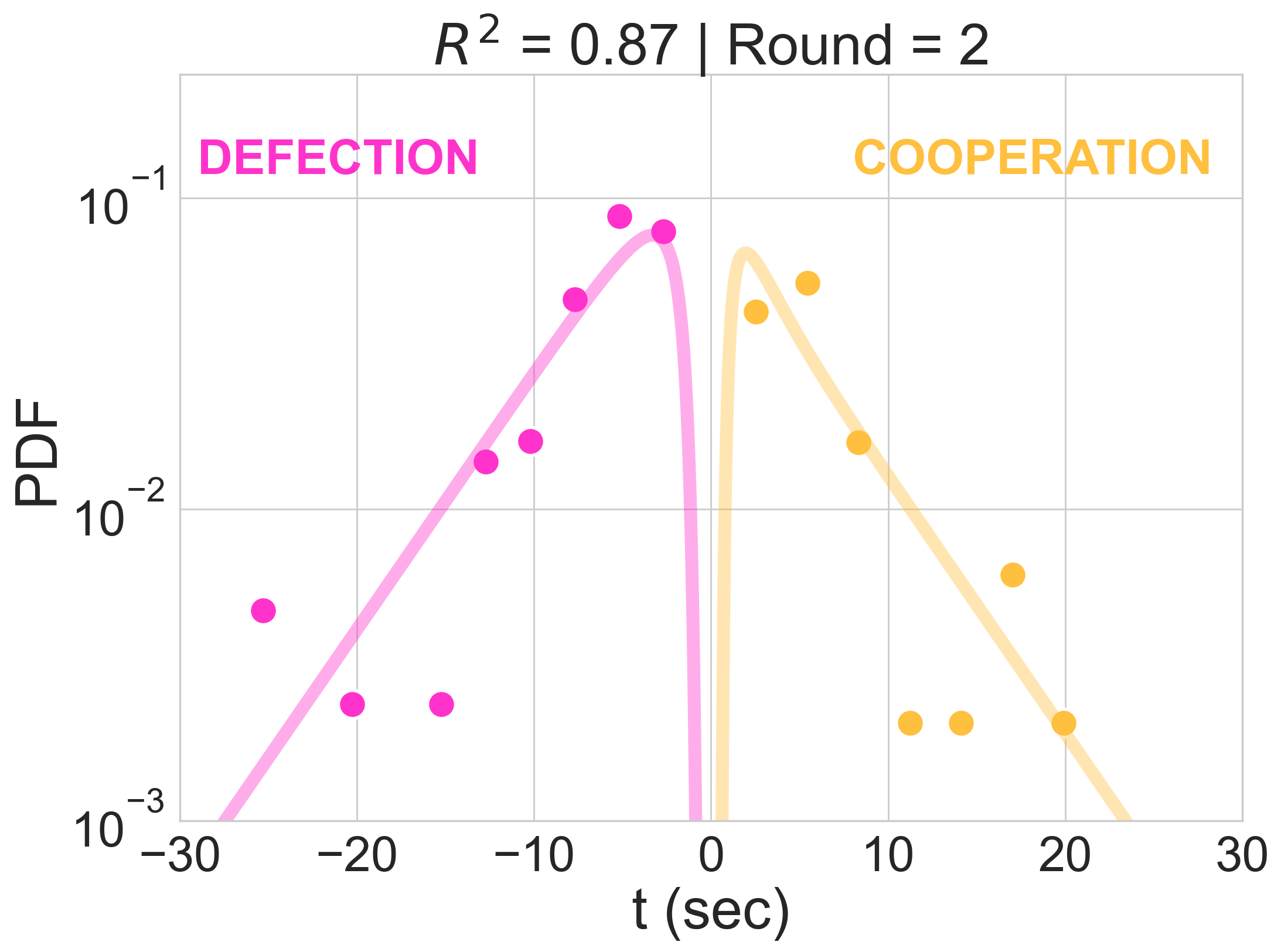}
                 \caption{}
             \end{subfigure}
             \begin{subfigure}[b]{0.45\textwidth}
                 \centering
                 \includegraphics[width=\textwidth]{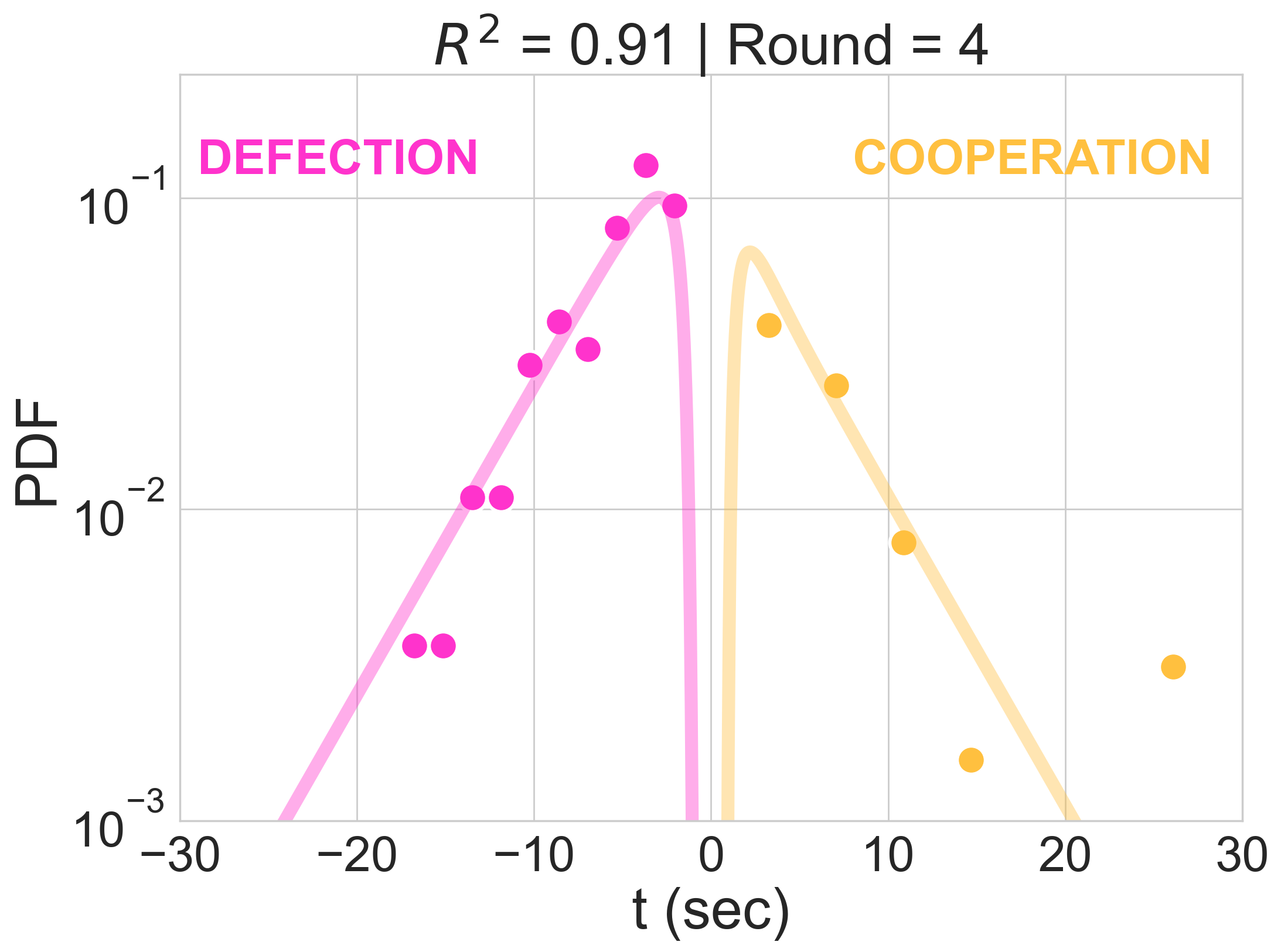}
                 \caption{}
             \end{subfigure}
             \begin{subfigure}[b]{0.45\textwidth}
                 \centering
                 \includegraphics[width=\textwidth]{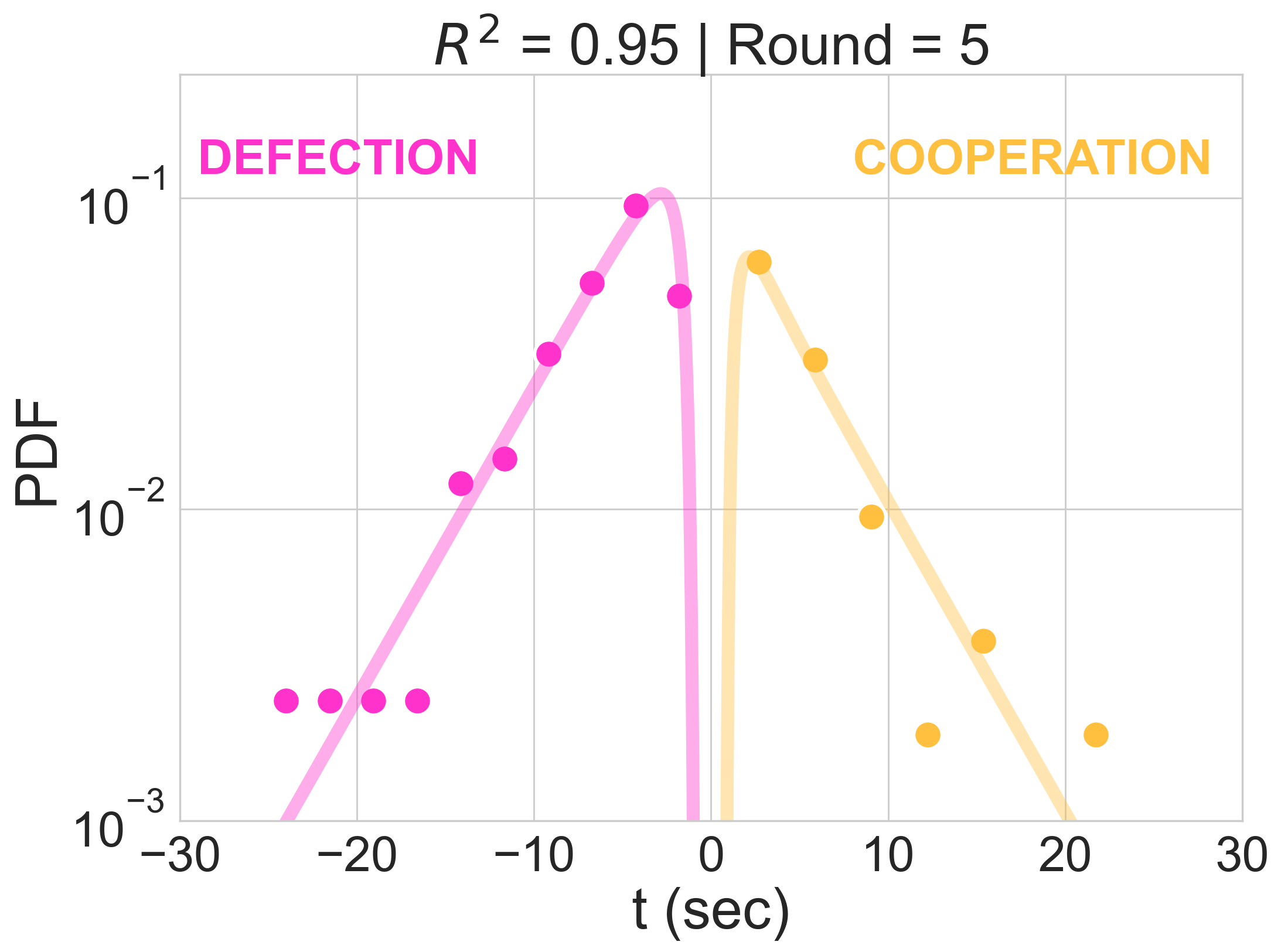}
                 \caption{}
             \end{subfigure}
             \begin{subfigure}[b]{0.45\textwidth}
                 \centering
                 \includegraphics[width=\textwidth]{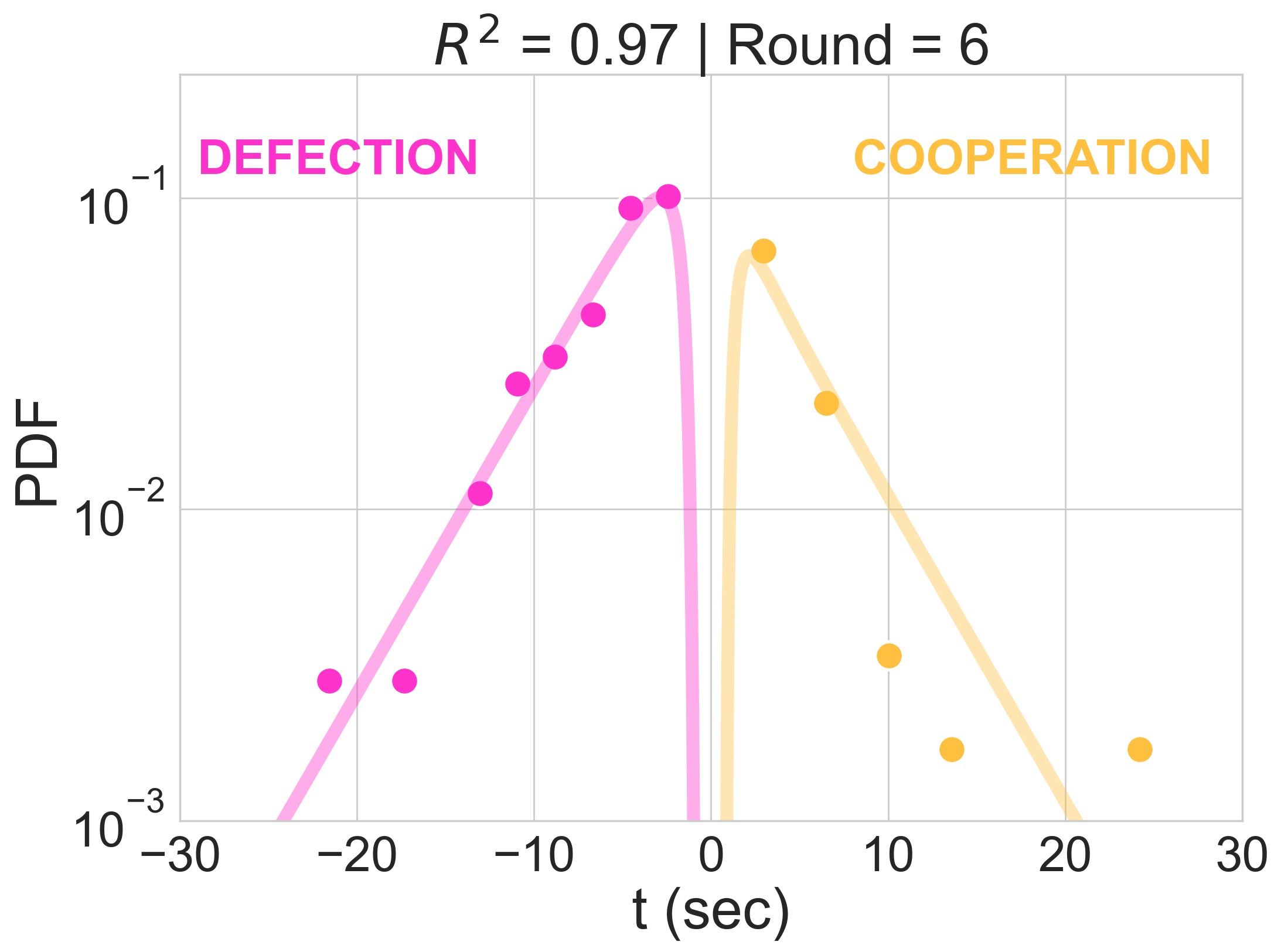}
                 \caption{}
             \end{subfigure}
             \begin{subfigure}[b]{0.45\textwidth}
                 \centering
                 \includegraphics[width=\textwidth]{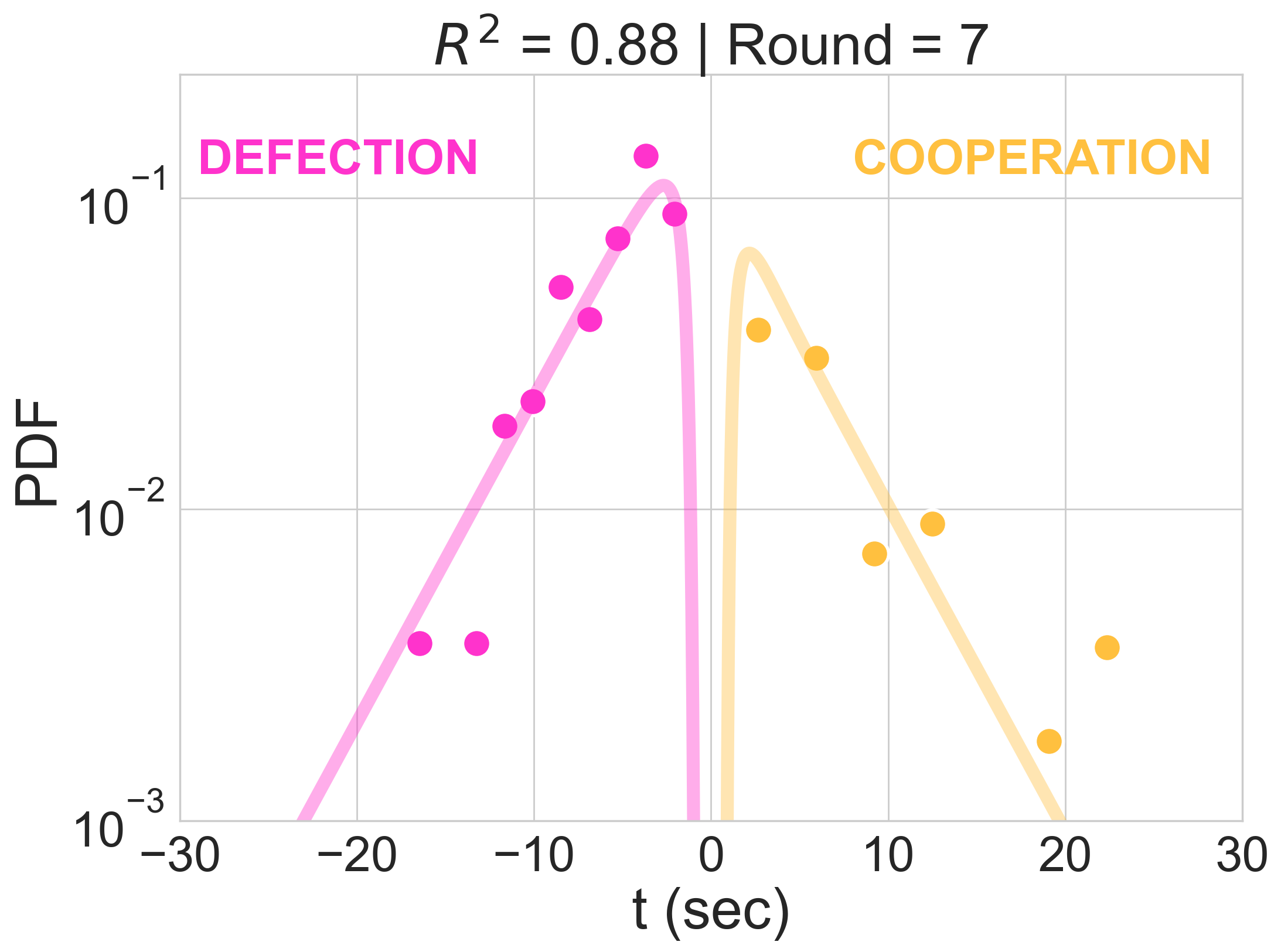}
                 \caption{}
             \end{subfigure}
             \begin{subfigure}[b]{0.45\textwidth}
                 \centering
                 \includegraphics[width=\textwidth]{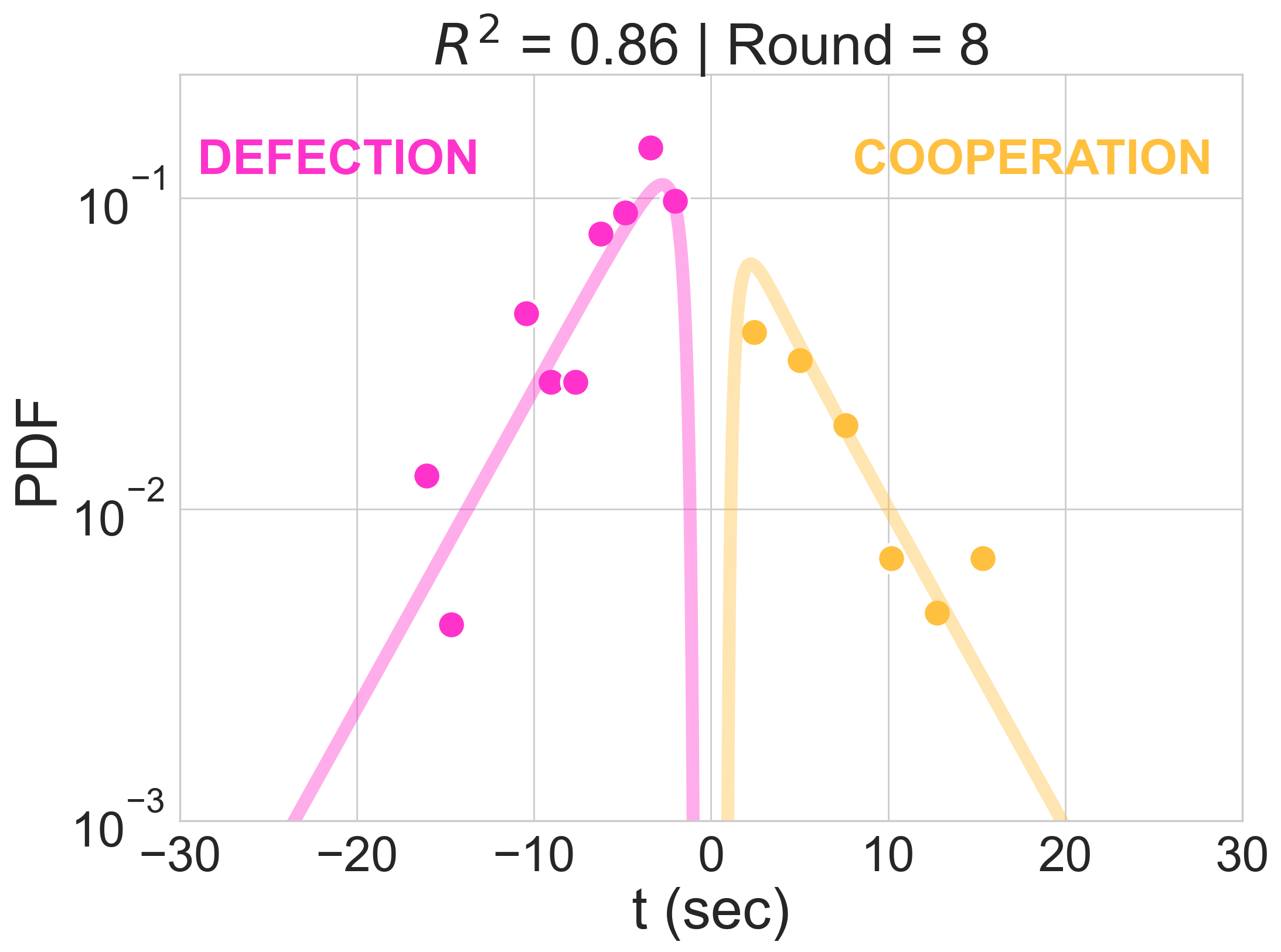}
                 \caption{}
             \end{subfigure}
             \caption{{\bf Accuracy over the testing set.} Response times PDFs for rounds $2$,$4-8$ (results on round $3$ are included in Fig.\ref{PDF-test}, in section Results of the main paper). Results are obtained using i) data (dots), and ii) our predictive model (lines). The left side (pink) corresponds to defection responses, while the right side (orange) corresponds to cooperation responses.}
        \label{SI_PDF1}
        \end{figure}
        \begin{figure}[H]
             \centering
             \begin{subfigure}[b]{0.45\textwidth}
                 \centering
                 \includegraphics[width=\textwidth]{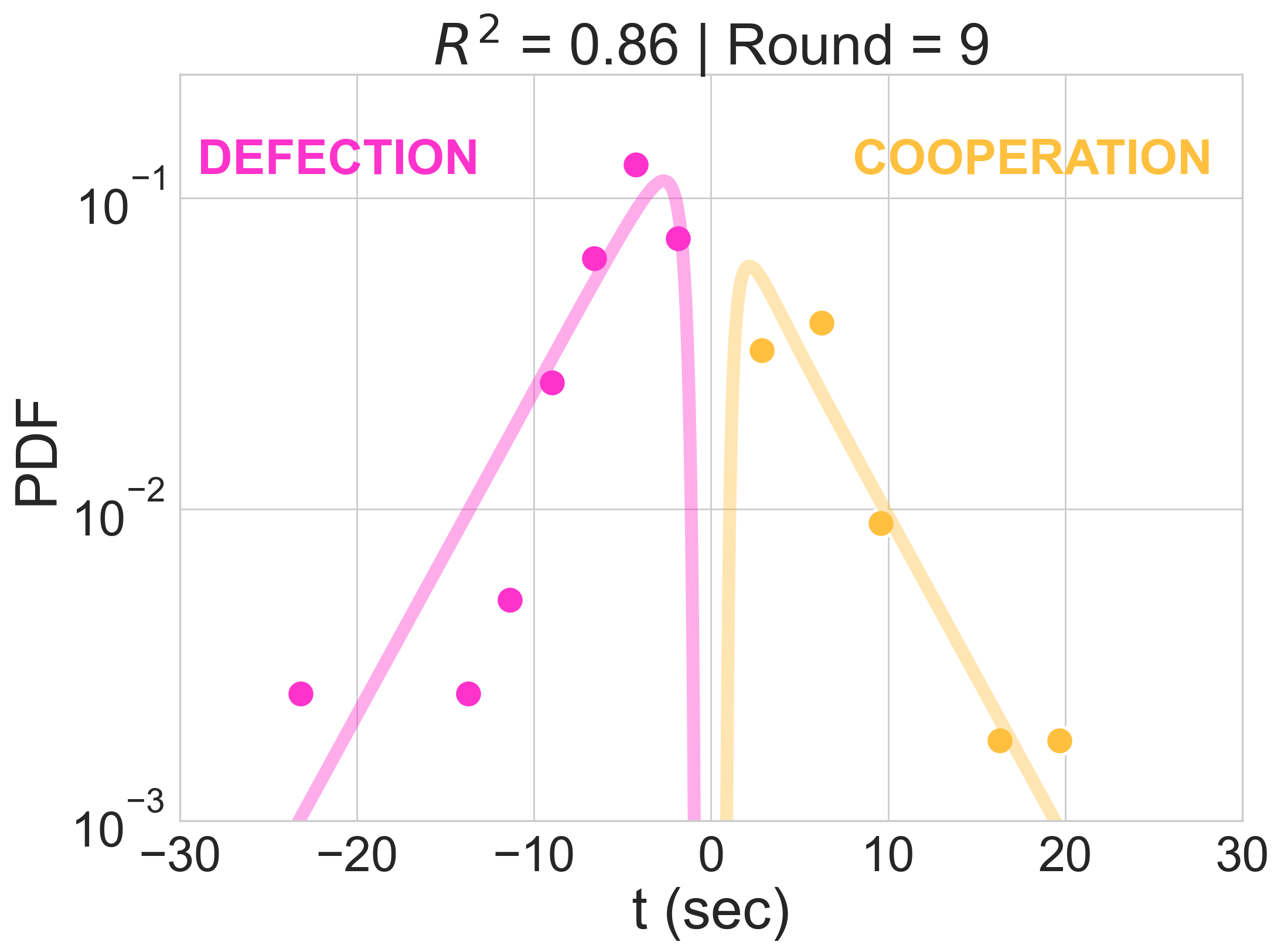}
                 \caption{}
             \end{subfigure}
             \begin{subfigure}[b]{0.45\textwidth}
                 \centering
                 \includegraphics[width=\textwidth]{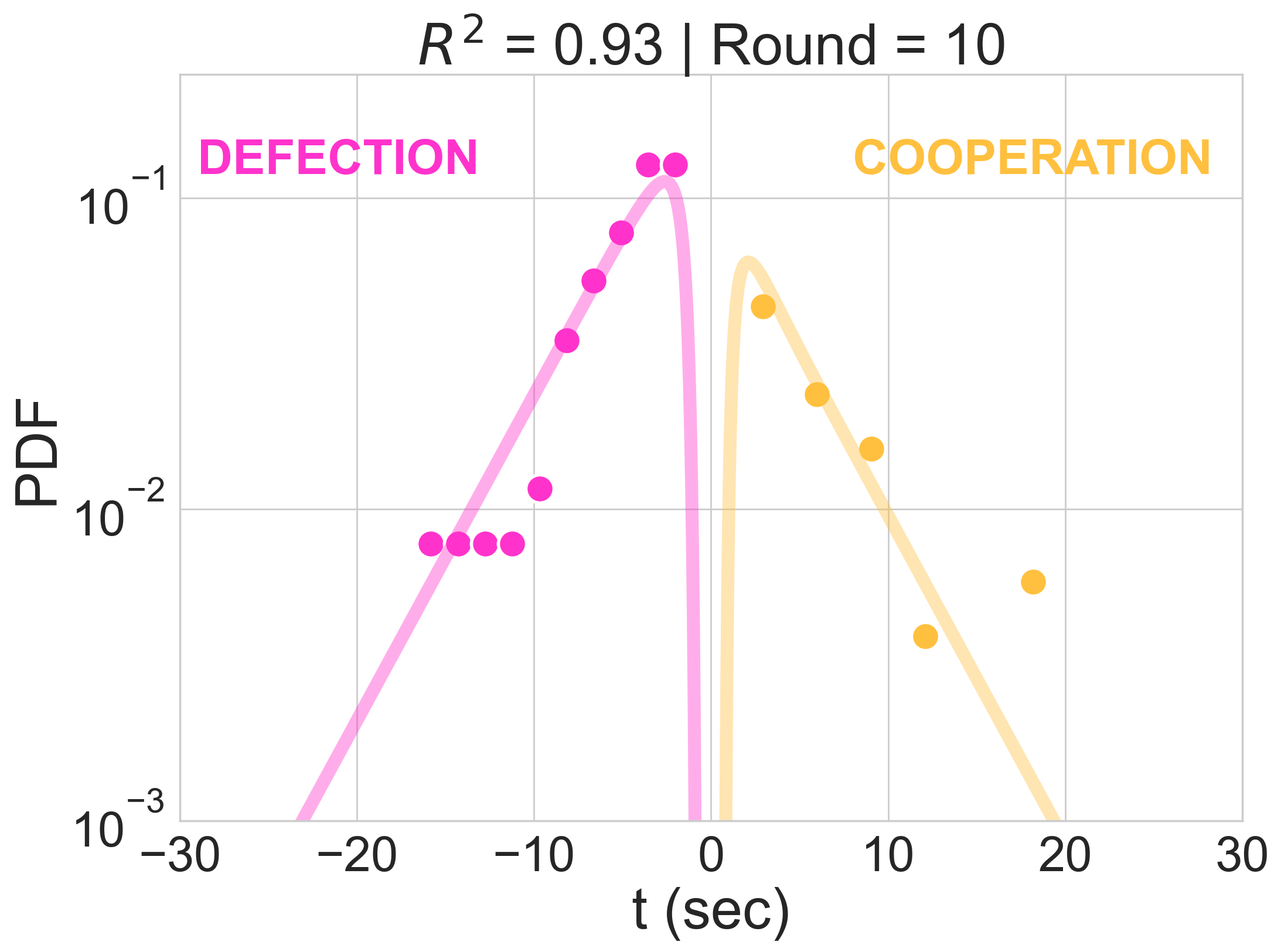}
                 \caption{}
             \end{subfigure}
             \begin{subfigure}[b]{0.45\textwidth}
                 \centering
                 \includegraphics[width=\textwidth]{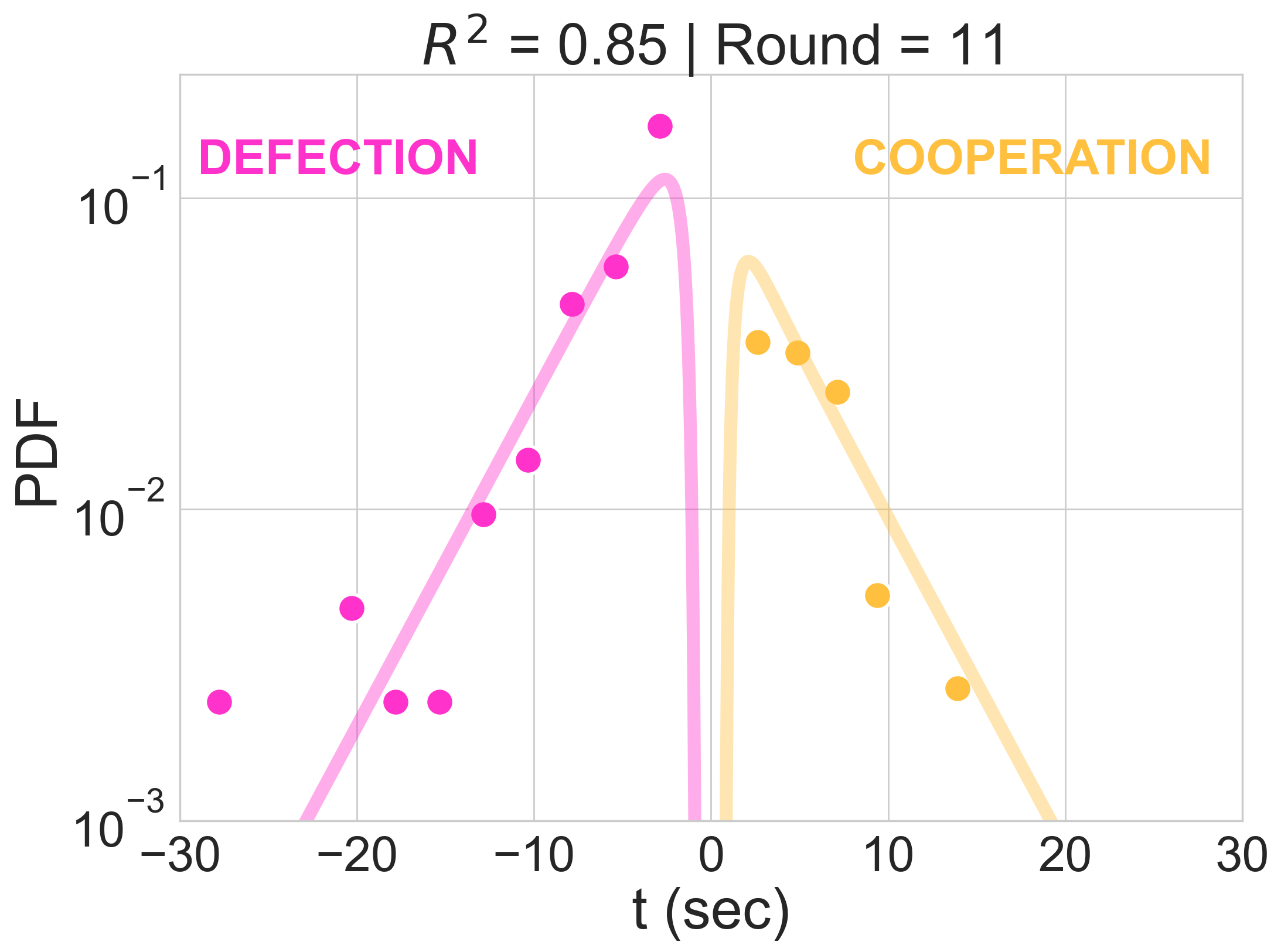}
                 \caption{}
             \end{subfigure}
             \begin{subfigure}[b]{0.45\textwidth}
                 \centering
                 \includegraphics[width=\textwidth]{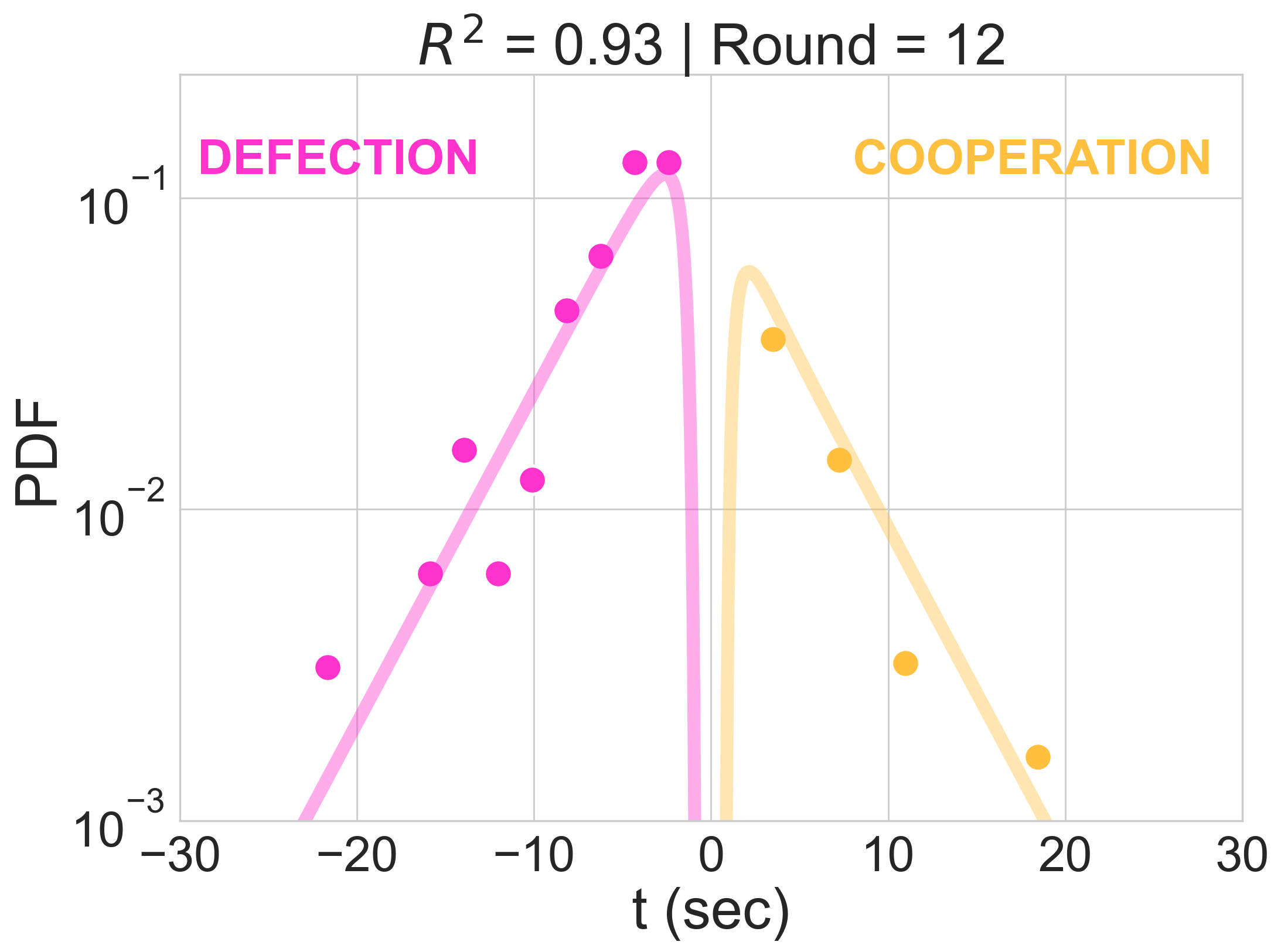}
                 \caption{}
             \end{subfigure}
             \begin{subfigure}[b]{0.45\textwidth}
                 \centering
                 \includegraphics[width=\textwidth]{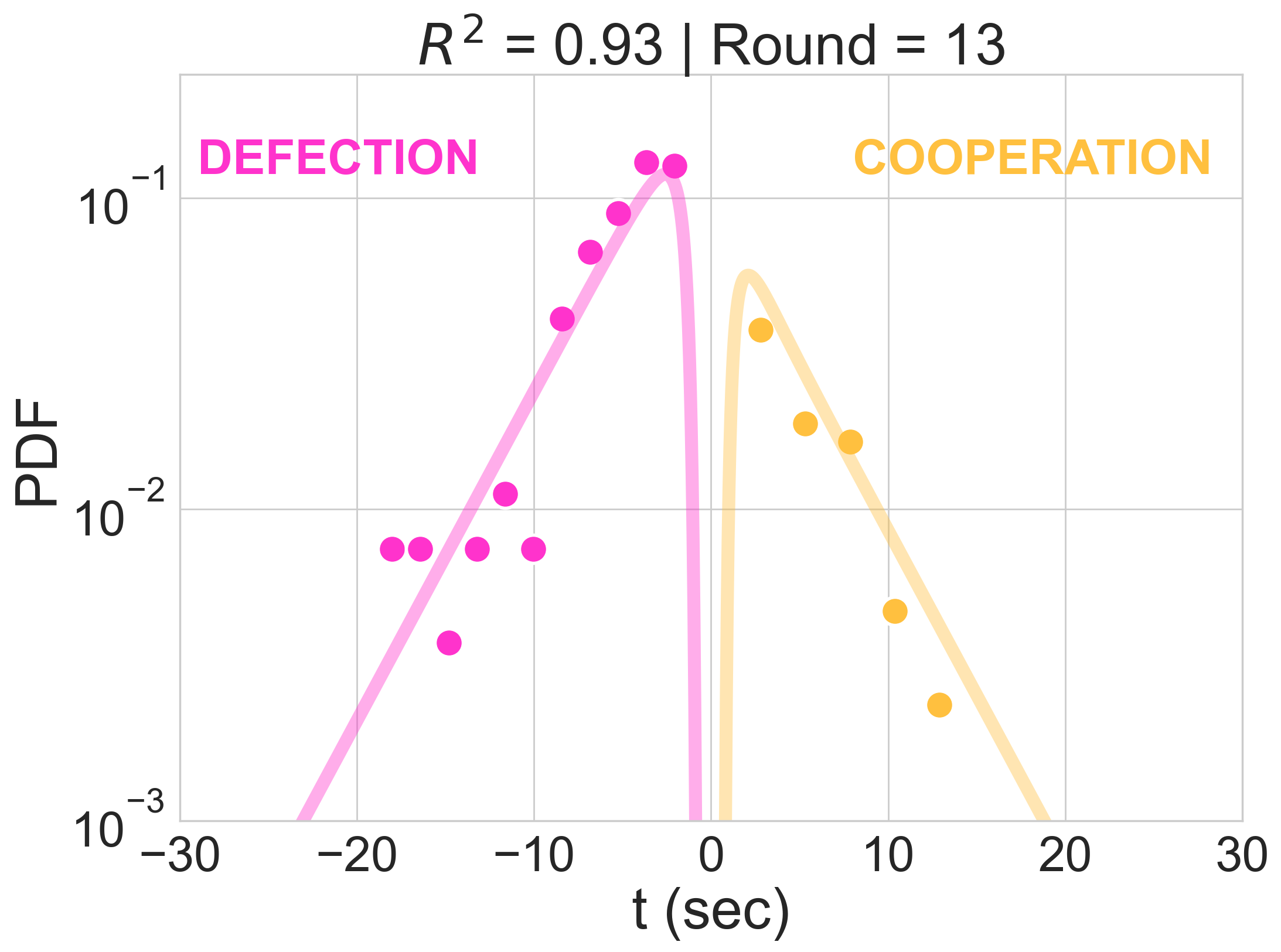}
                 \caption{}
             \end{subfigure}
             \begin{subfigure}[b]{0.45\textwidth}
                 \centering
                 \includegraphics[width=\textwidth]{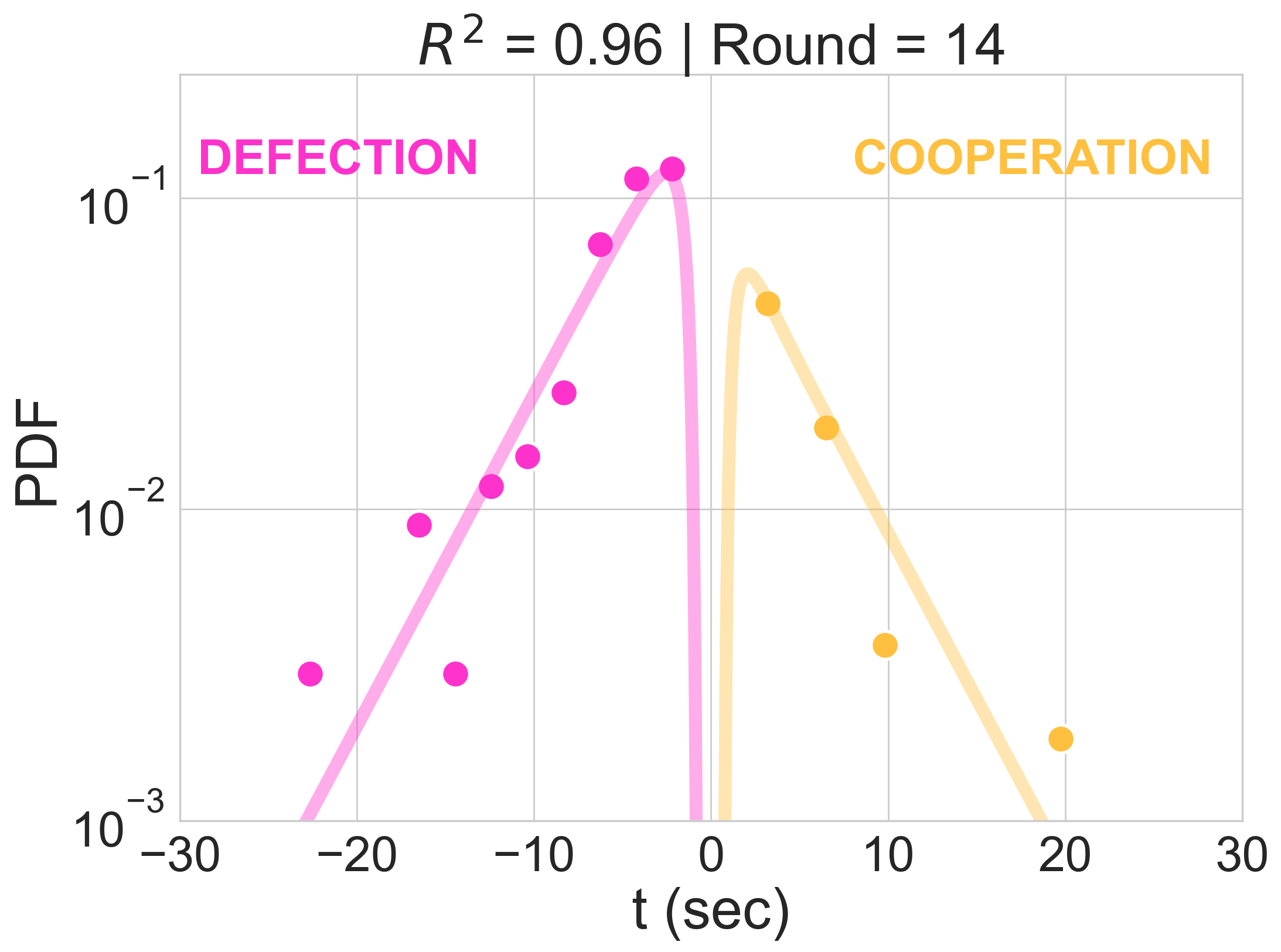}
                 \caption{}
             \end{subfigure}
             \caption{{\bf Accuracy over the testing set.} Response times PDFs for rounds $9-14$. Results are obtained using i) data (dots), and ii) our predictive model (lines). The left side (pink) corresponds to defection responses, while the right side (orange) corresponds to cooperation responses.}
        \label{SI_PDF2}
        \end{figure}
        \begin{figure}[H]
             \centering
             \begin{subfigure}[b]{0.45\textwidth}
                 \centering
                 \includegraphics[width=\textwidth]{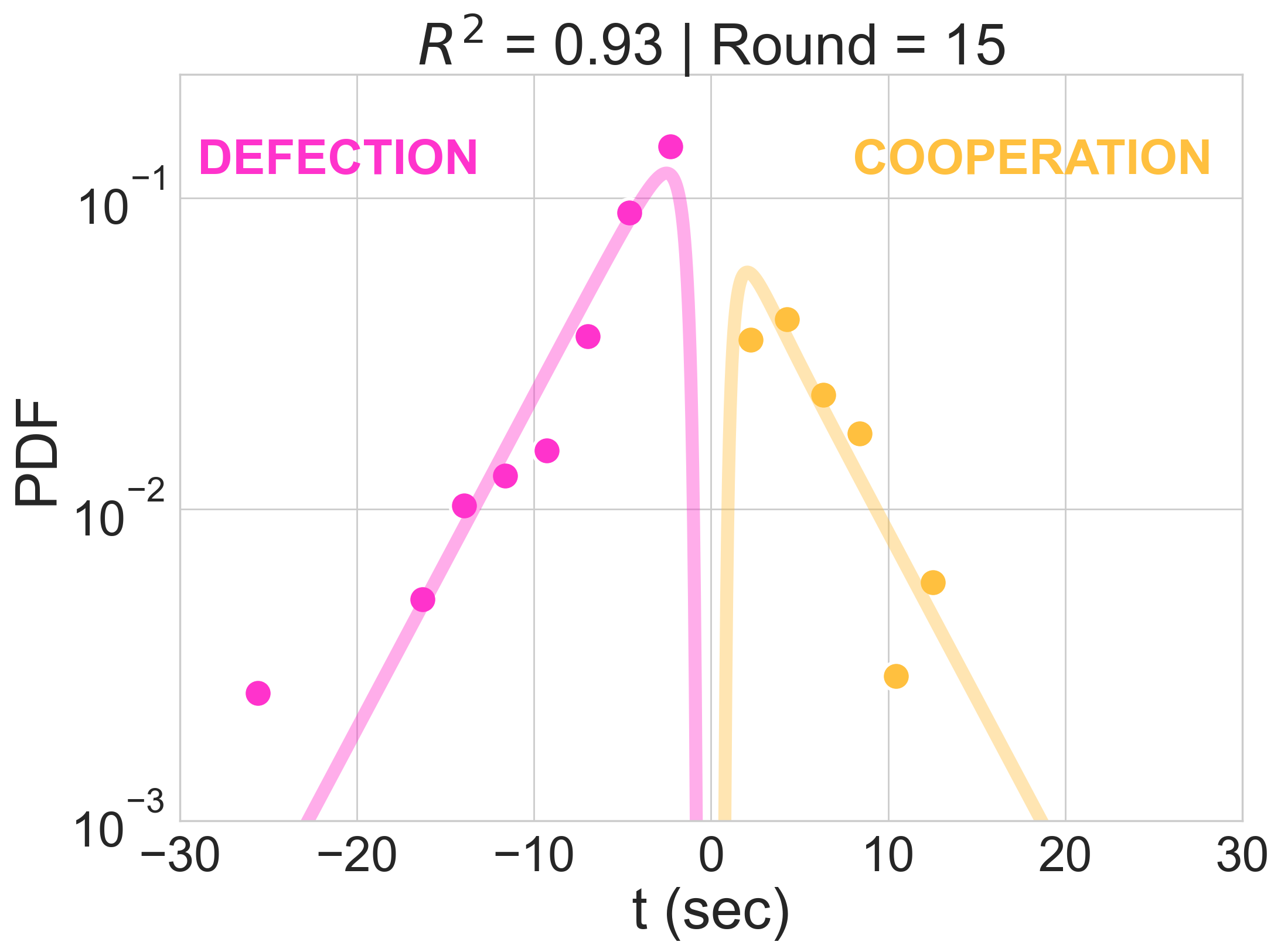}
                 \caption{}
             \end{subfigure}
             \begin{subfigure}[b]{0.45\textwidth}
                 \centering
                 \includegraphics[width=\textwidth]{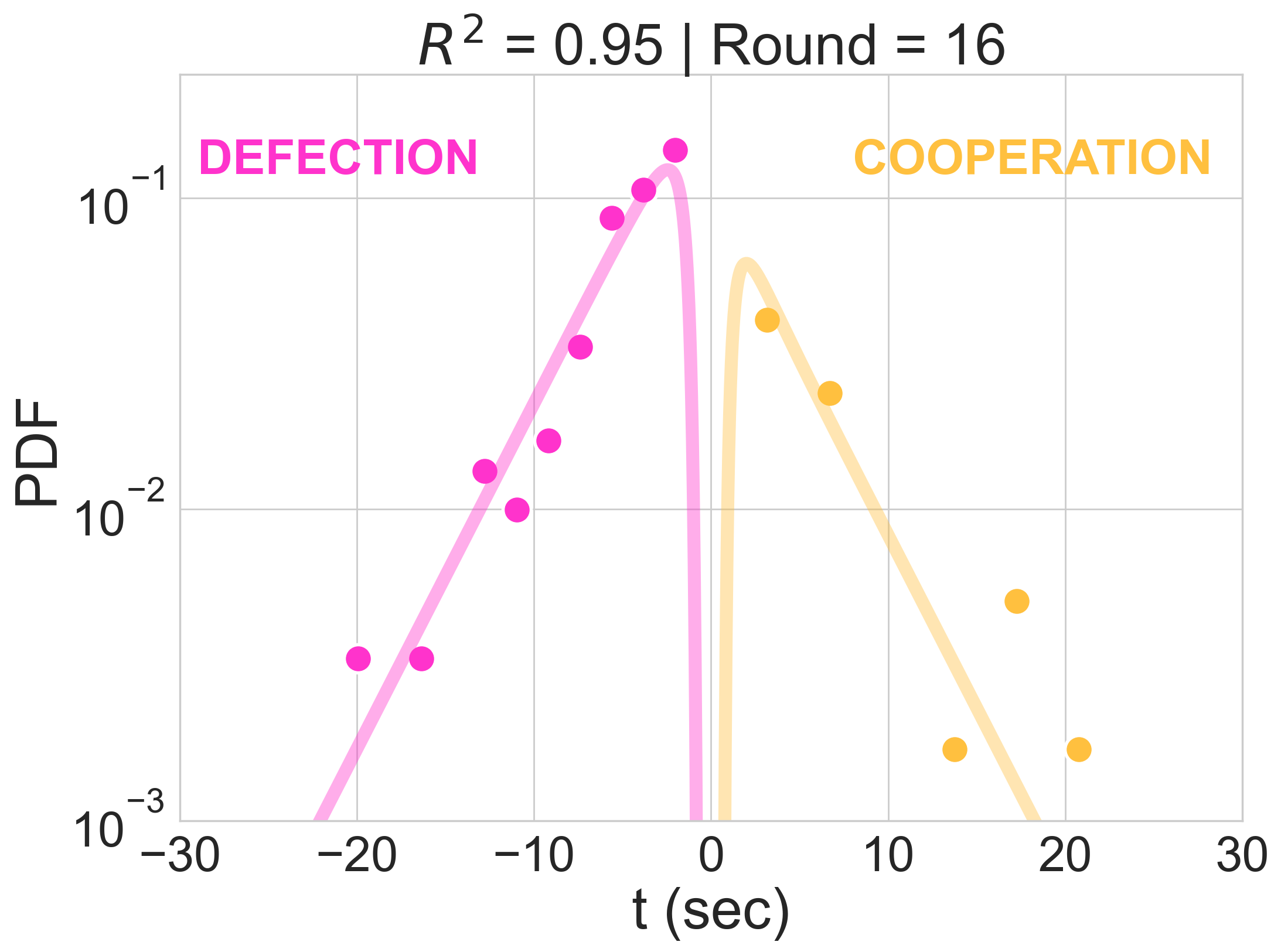}
                 \caption{}
             \end{subfigure}
             \begin{subfigure}[b]{0.45\textwidth}
                 \centering
                 \includegraphics[width=\textwidth]{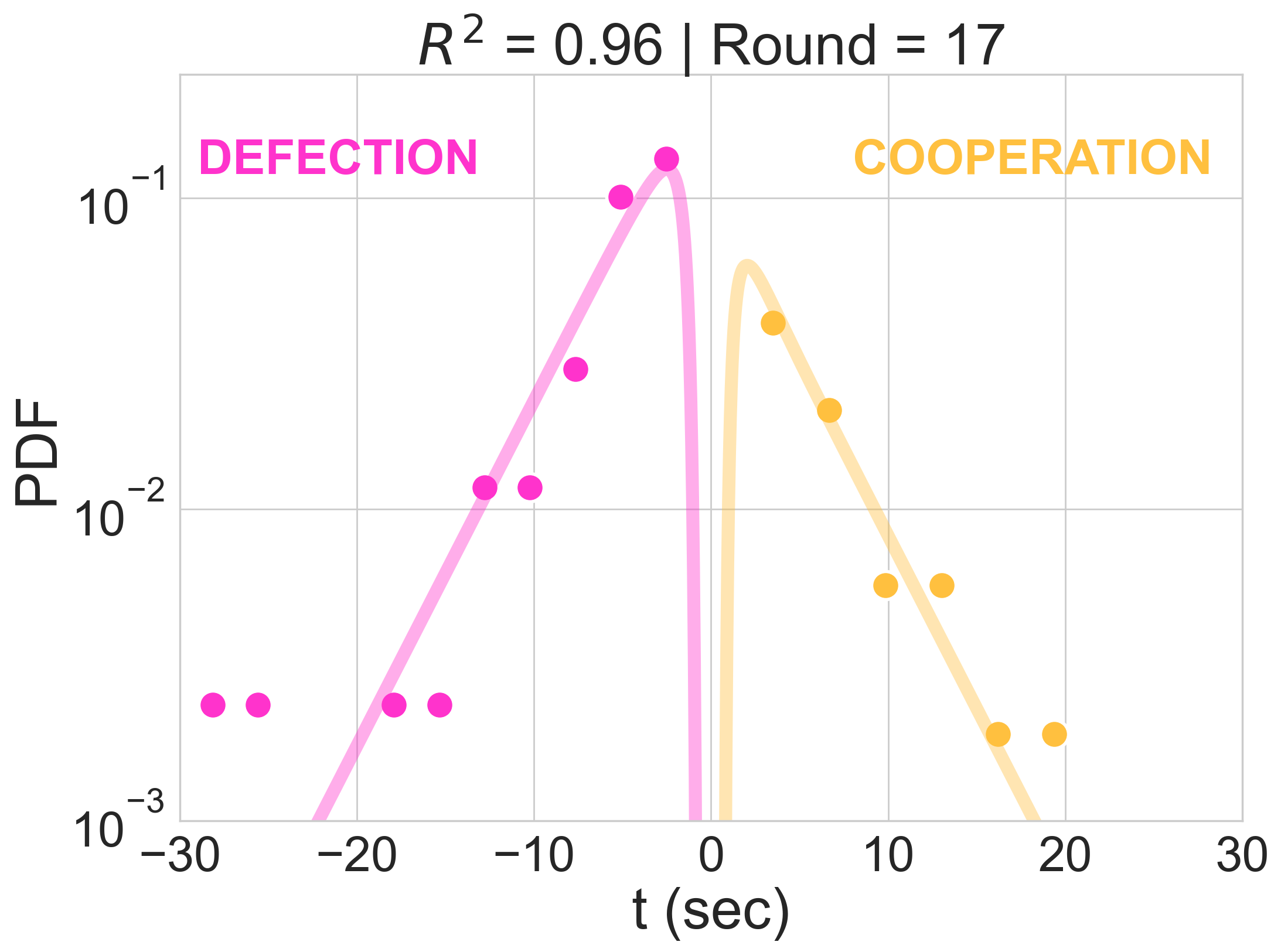}
                 \caption{}
             \end{subfigure}
             \begin{subfigure}[b]{0.45\textwidth}
                 \centering
                 \includegraphics[width=\textwidth]{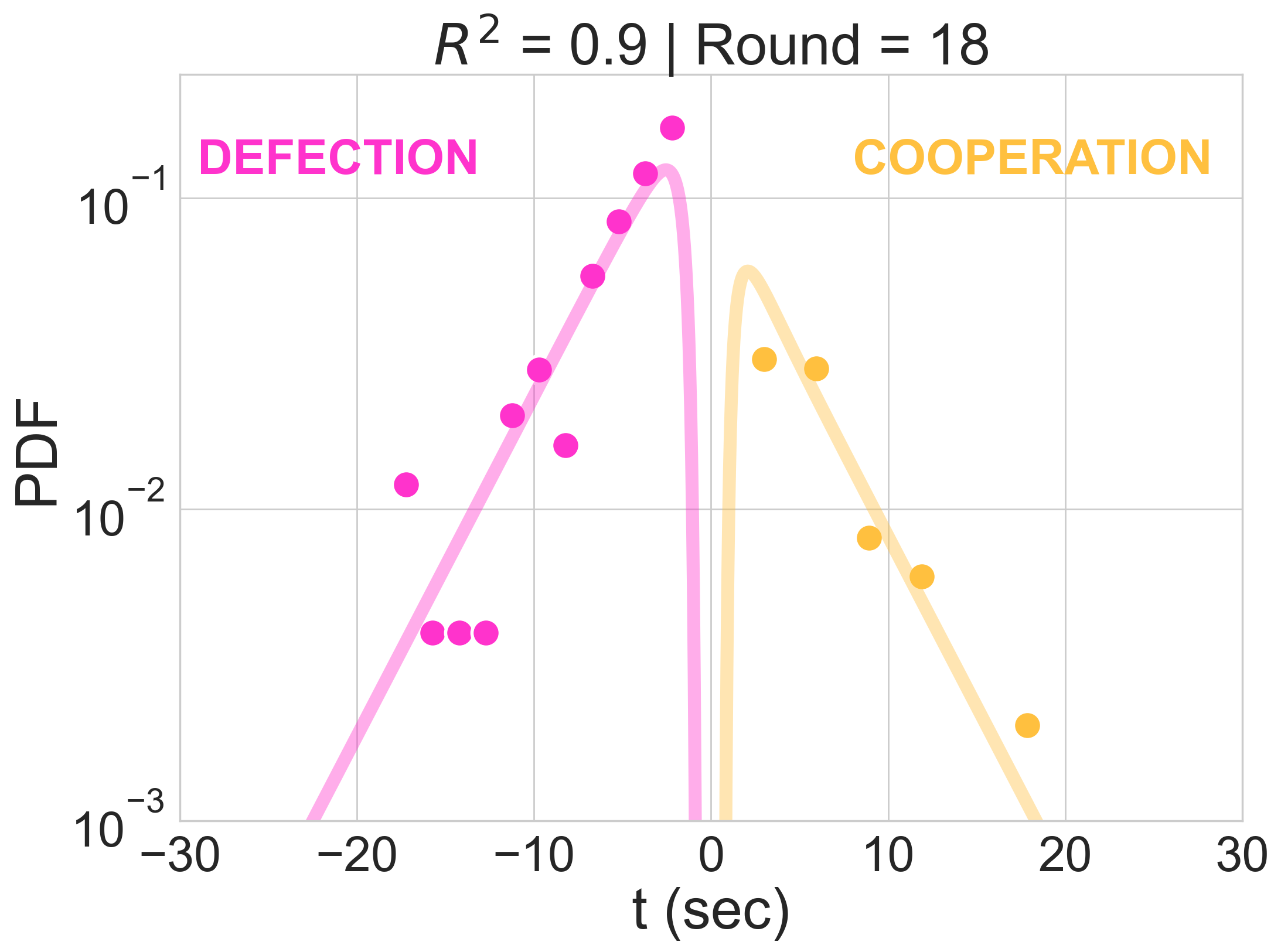}
                 \caption{}
             \end{subfigure}
             \begin{subfigure}[b]{0.45\textwidth}
                 \centering
                 \includegraphics[width=\textwidth]{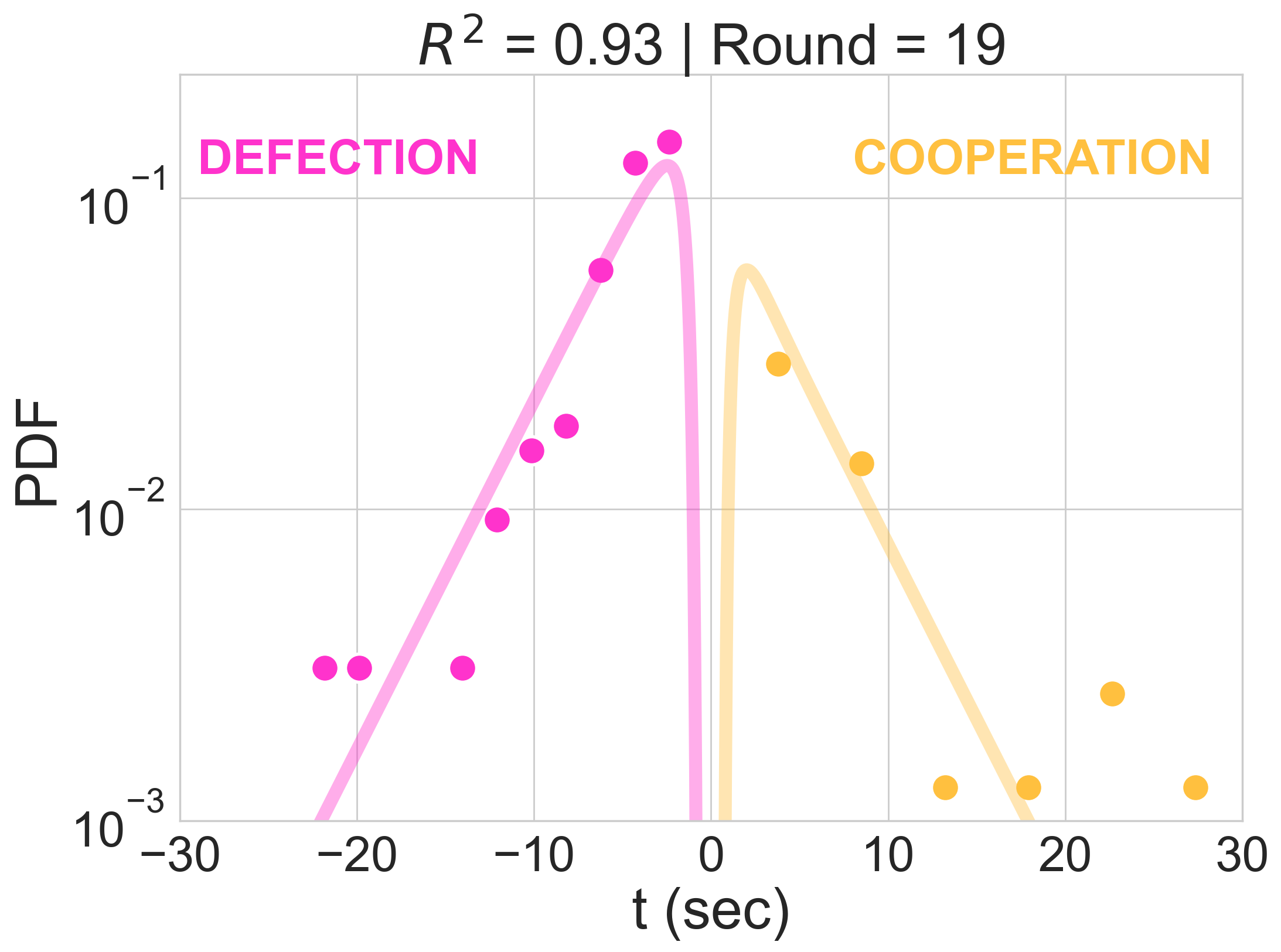}
                 \caption{}
             \end{subfigure}
             \begin{subfigure}[b]{0.45\textwidth}
                 \centering
                 \includegraphics[width=\textwidth]{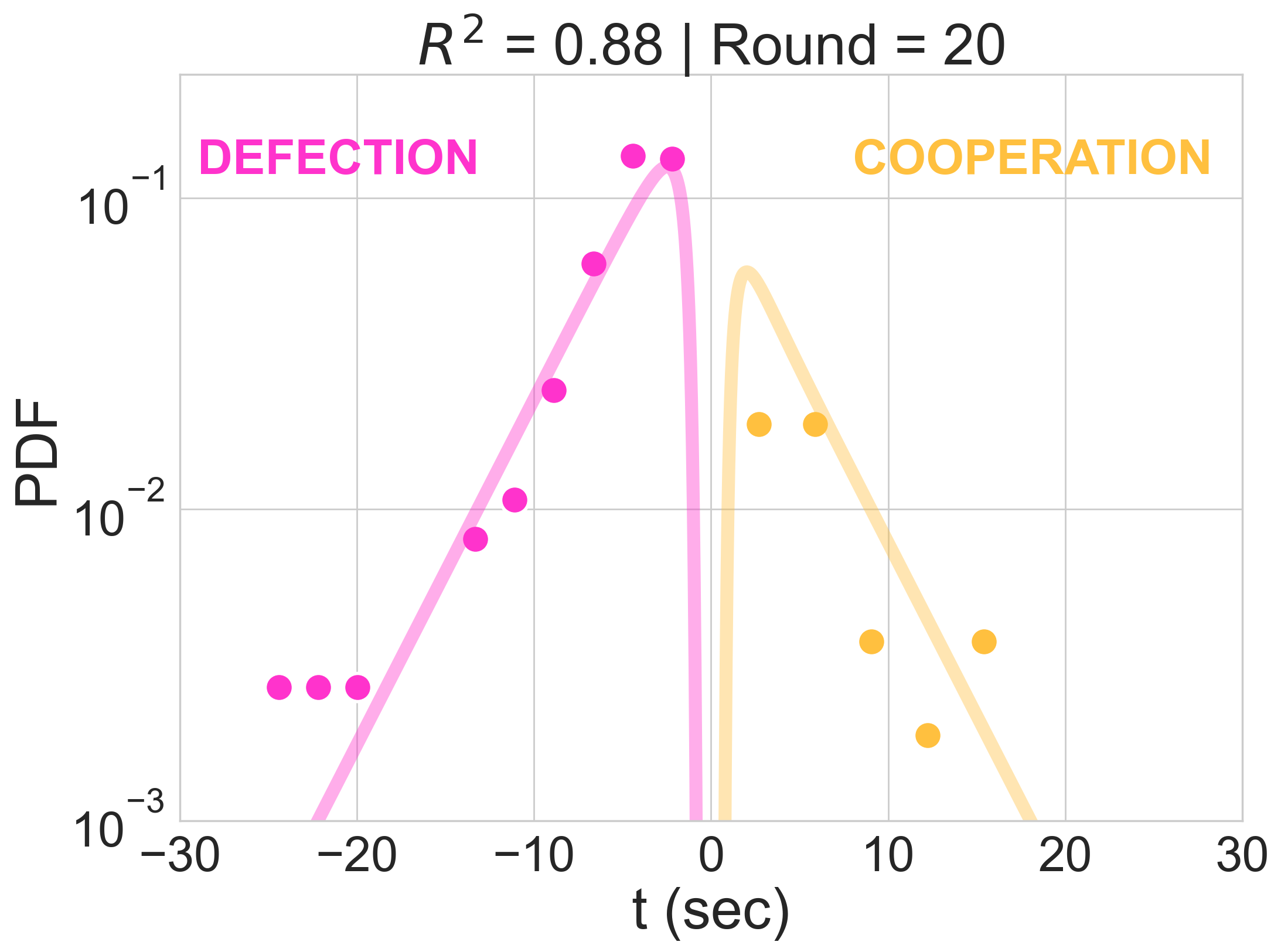}
                 \caption{}
             \end{subfigure}
             \caption{{\bf Accuracy over the testing set.} Response times PDFs for rounds $15-20$ (results on round $3$ are included in Fig.\ref{PDF-test}, in section Results of the main paper). Results are obtained using i) data (dots), and ii) our predictive model (lines). The left side (pink) corresponds to defection responses, while the right side (orange) corresponds to cooperation responses.}
        \label{SI_PDF3}
        \end{figure}
        \begin{figure}[H]
             \centering
             \begin{subfigure}[b]{0.45\textwidth}
                 \centering
                 \includegraphics[width=\textwidth]{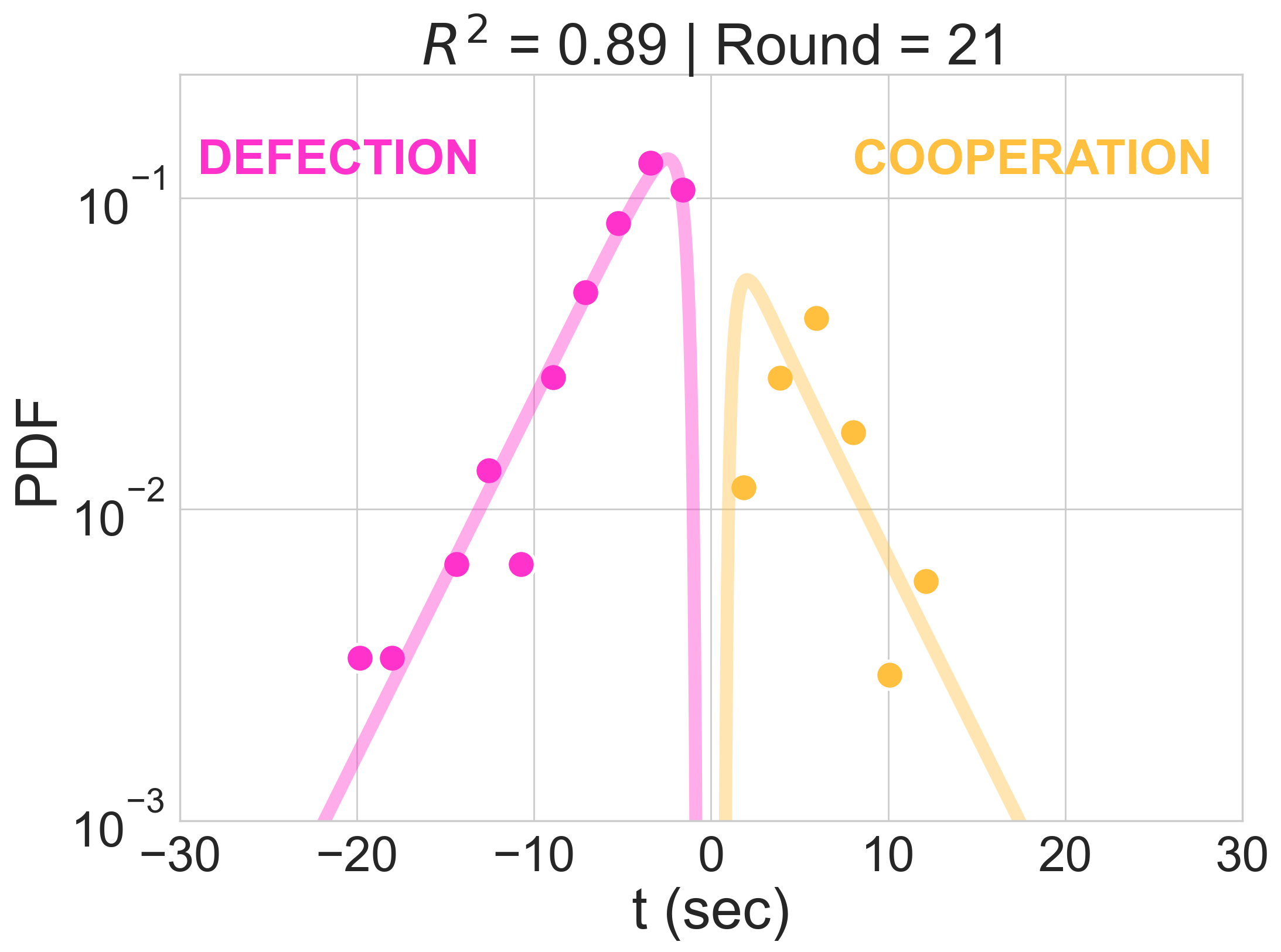}
                 \caption{}
             \end{subfigure}
             \begin{subfigure}[b]{0.45\textwidth}
                 \centering
                 \includegraphics[width=\textwidth]{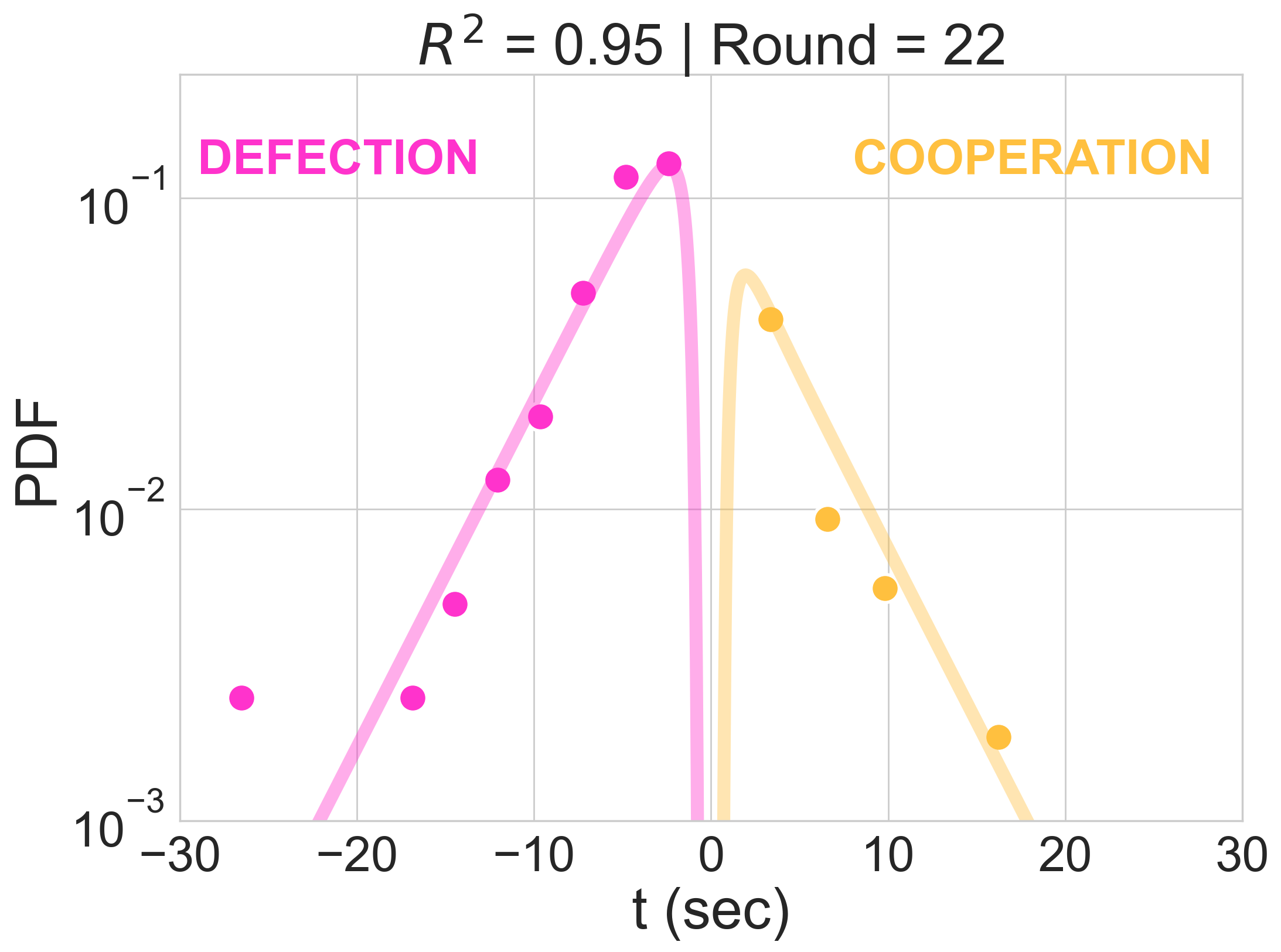}
                 \caption{}
             \end{subfigure}
             \begin{subfigure}[b]{0.45\textwidth}
                 \centering
                 \includegraphics[width=\textwidth]{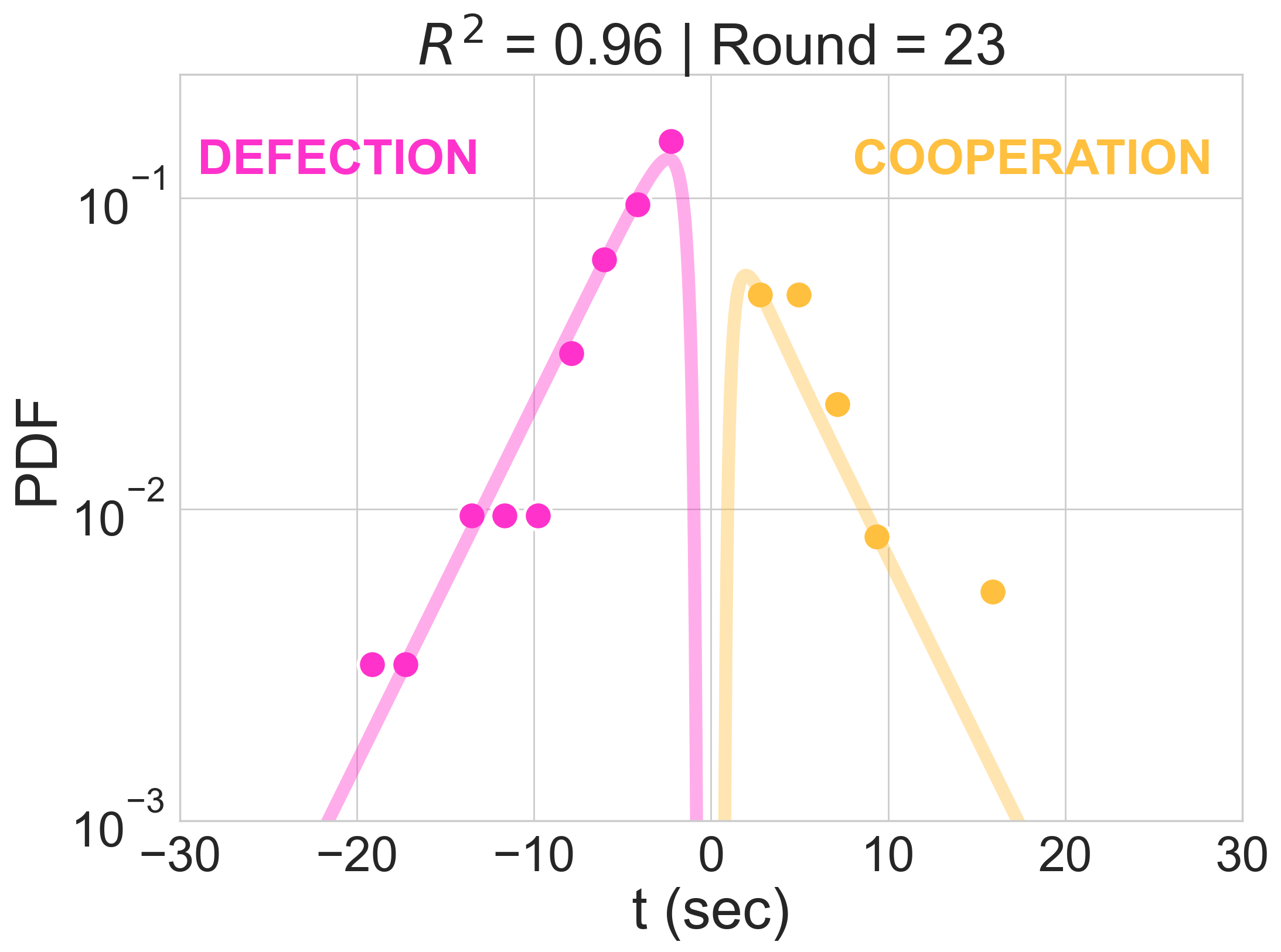}
                 \caption{}
             \end{subfigure}
             \begin{subfigure}[b]{0.45\textwidth}
                 \centering
                 \includegraphics[width=\textwidth]{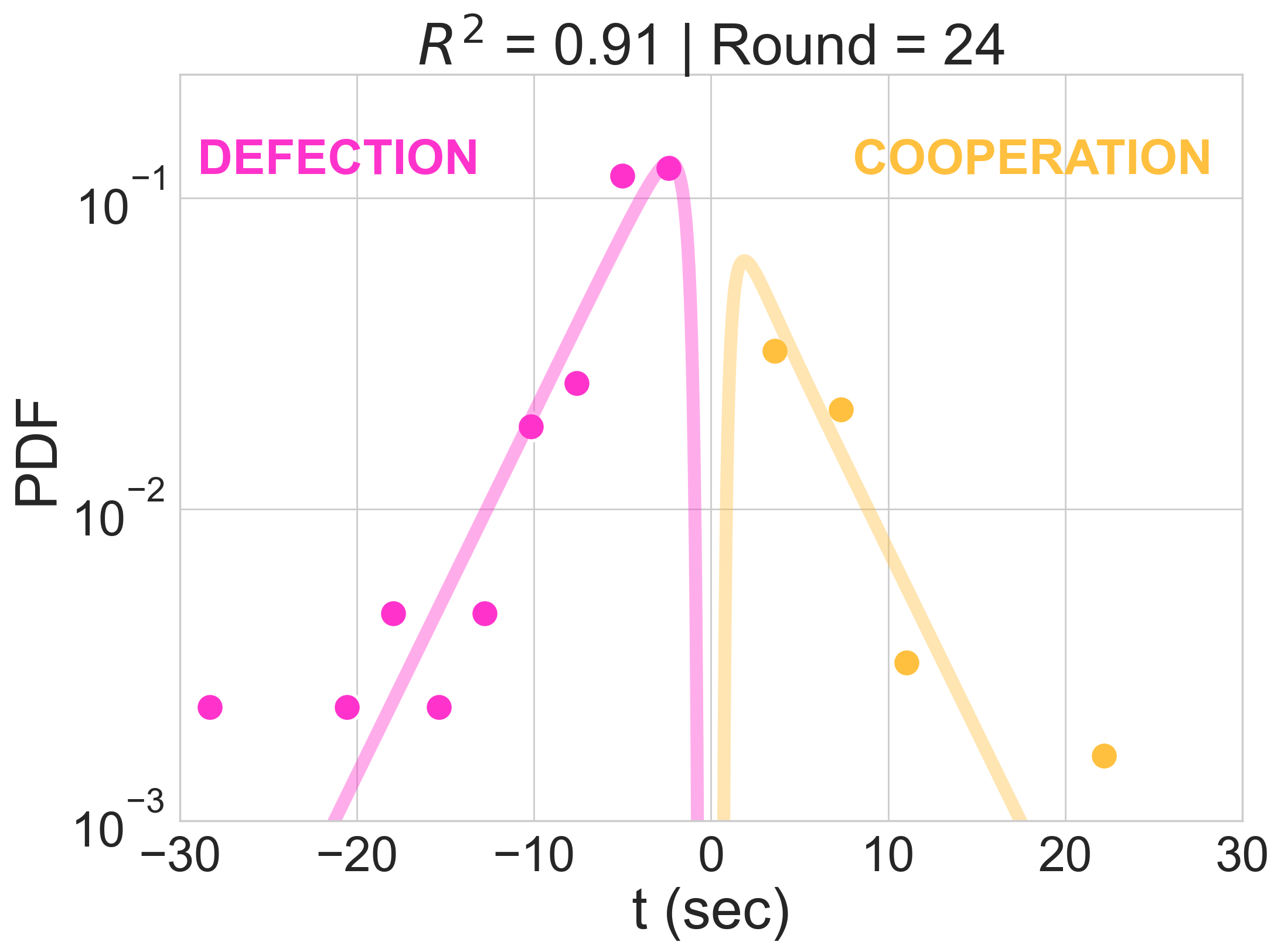}
                 \caption{}
             \end{subfigure}
             \begin{subfigure}[b]{0.45\textwidth}
                 \centering
                 \includegraphics[width=\textwidth]{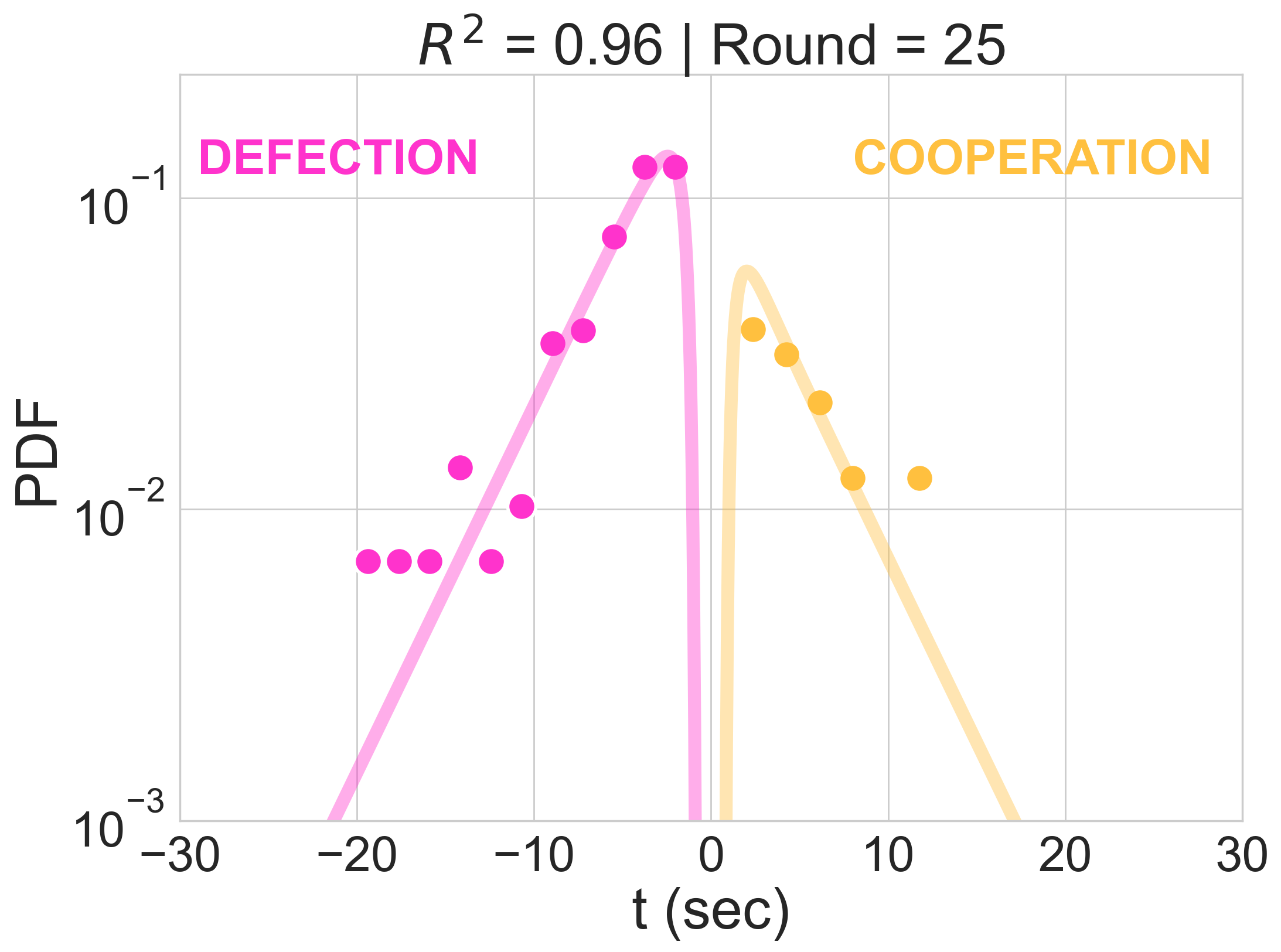}
                 \caption{}
             \end{subfigure}
             \begin{subfigure}[b]{0.45\textwidth}
                 \centering
                 \includegraphics[width=\textwidth]{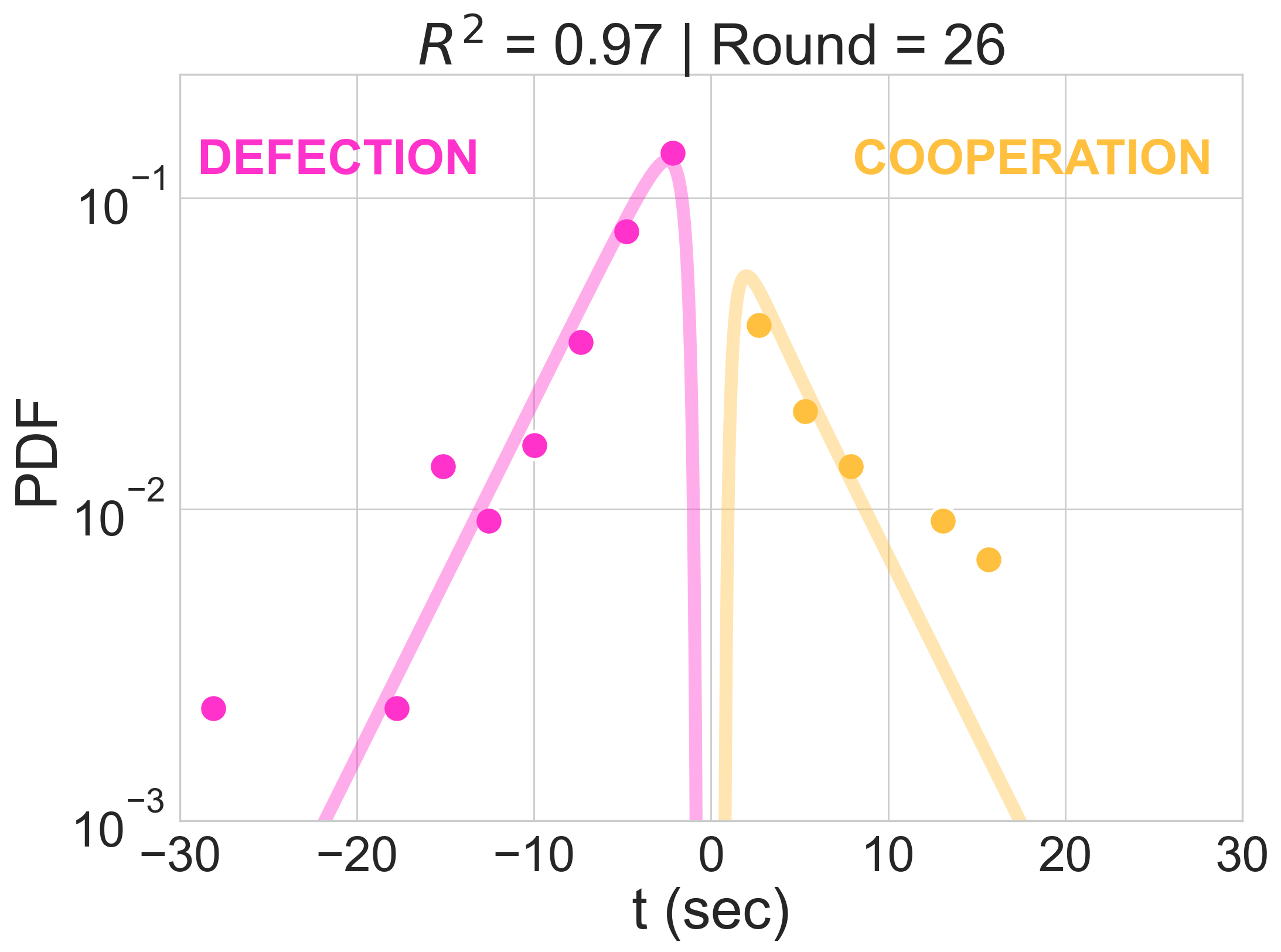}
                 \caption{}
             \end{subfigure}
             \caption{{\bf Accuracy over the testing set.} Response times PDFs for rounds $21-26$. Results are obtained using i) data (dots), and ii) our predictive model (lines). The left side (pink) corresponds to defection responses, while the right side (orange) corresponds to cooperation responses.}
             \label{SI_PDF4}
        \end{figure}
        \begin{figure}[H]
             \centering
             \begin{subfigure}[b]{0.45\textwidth}
                 \centering
                 \includegraphics[width=\textwidth]{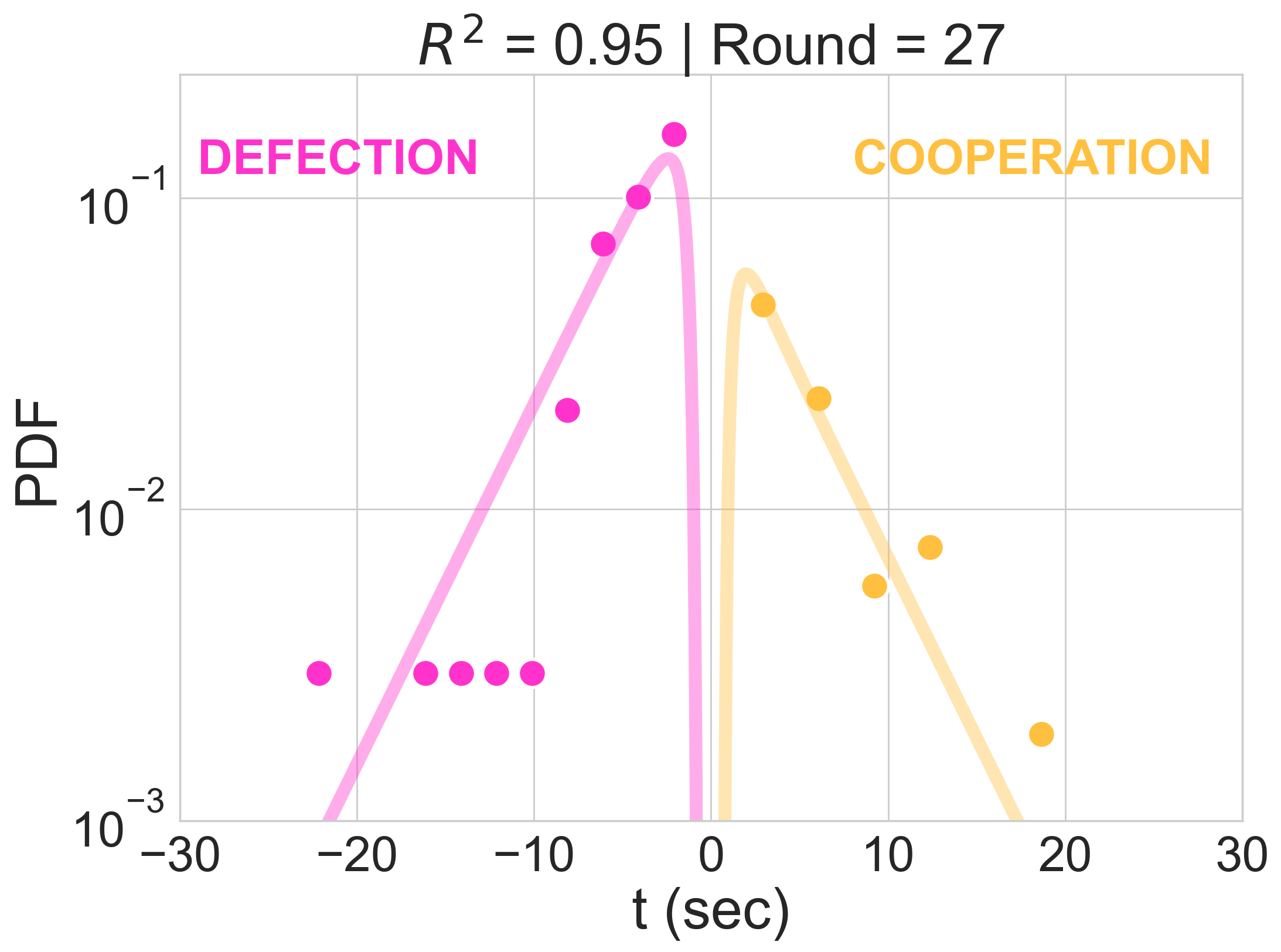}
                 \caption{}
             \end{subfigure}
             \begin{subfigure}[b]{0.45\textwidth}
                 \centering
                 \includegraphics[width=\textwidth]{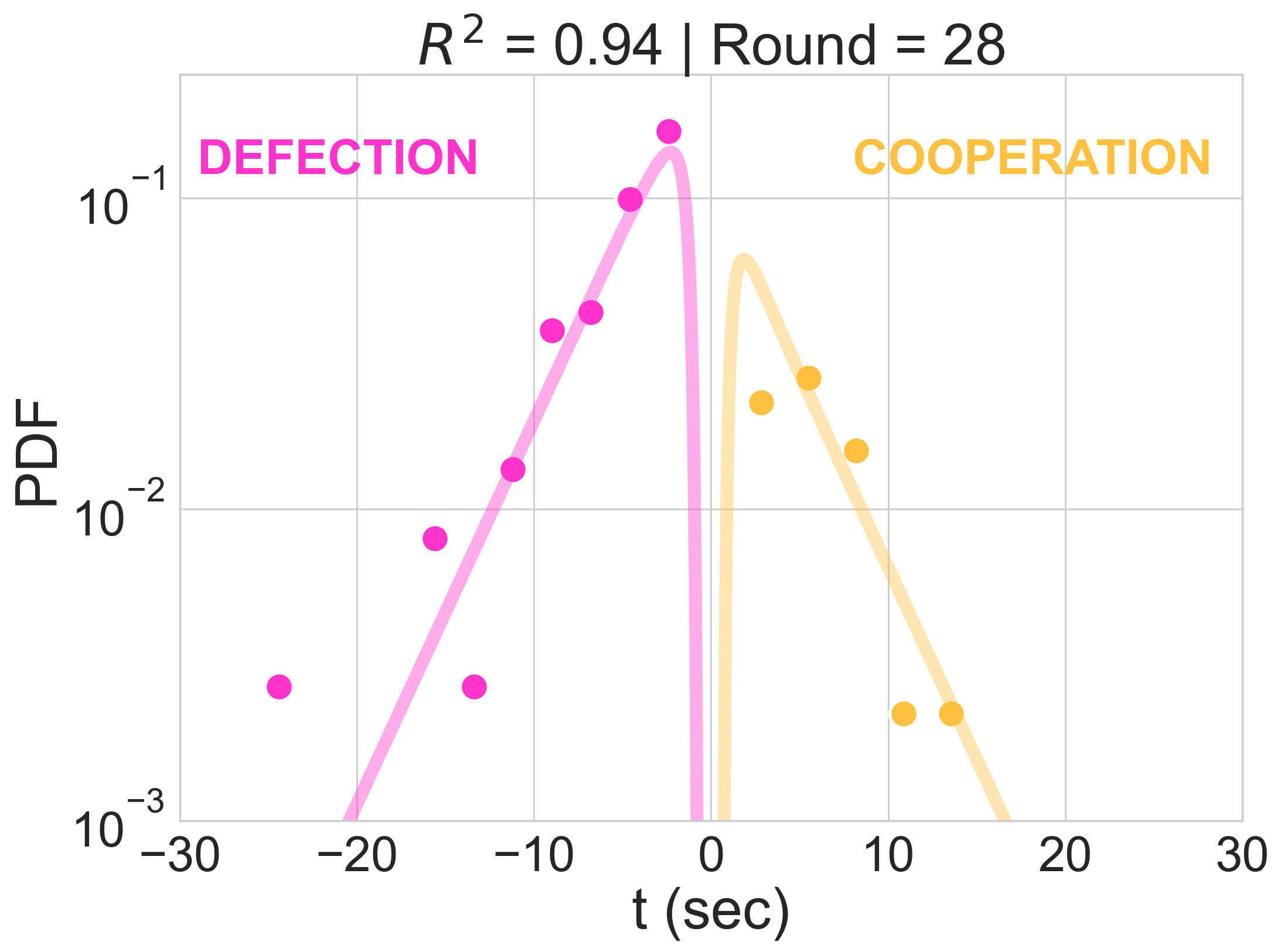}
                 \caption{}
             \end{subfigure}
             \begin{subfigure}[b]{0.45\textwidth}
                 \centering
                 \includegraphics[width=\textwidth]{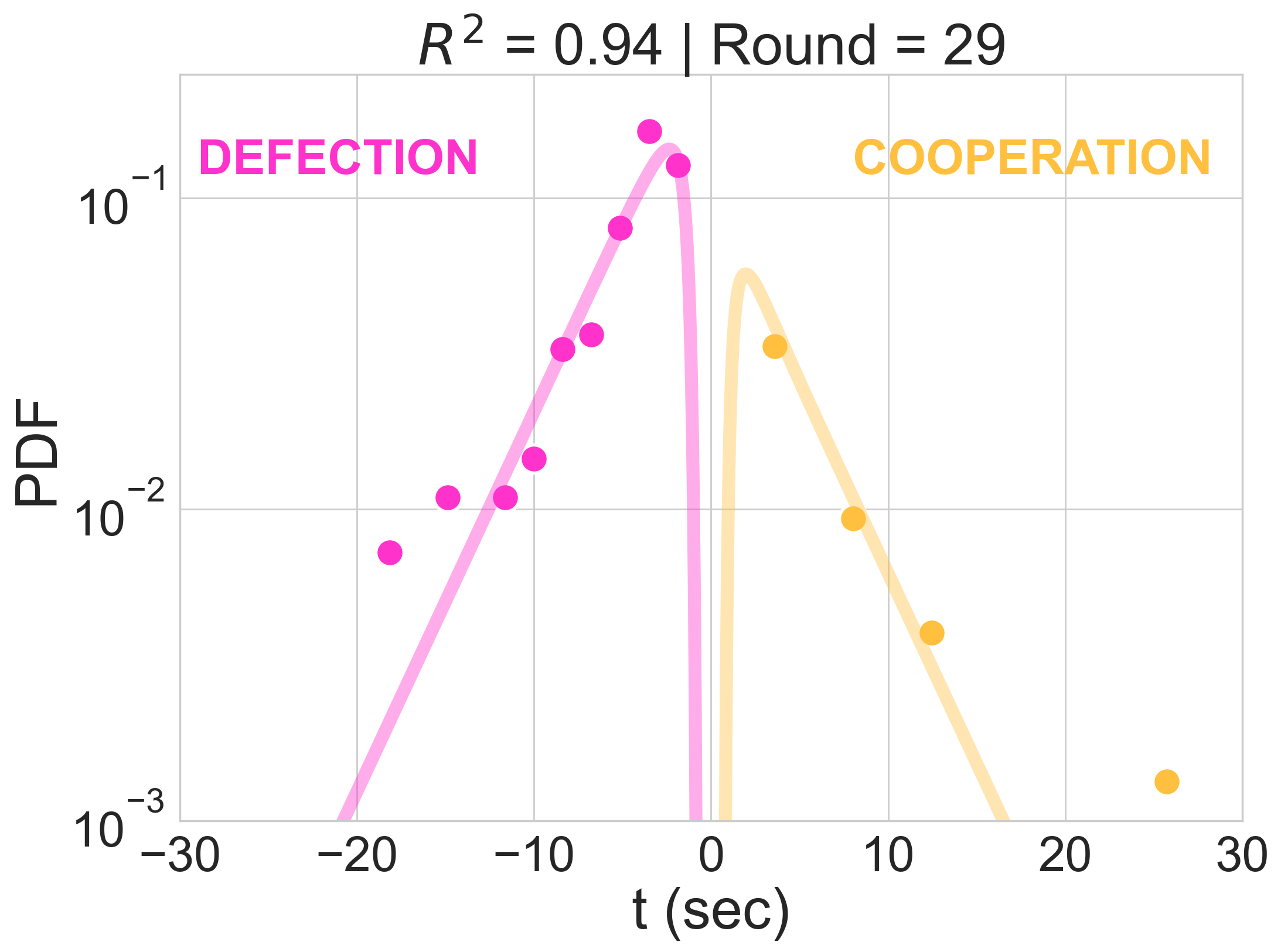}
                 \caption{}
             \end{subfigure}
             \begin{subfigure}[b]{0.45\textwidth}
                 \centering
                 \includegraphics[width=\textwidth]{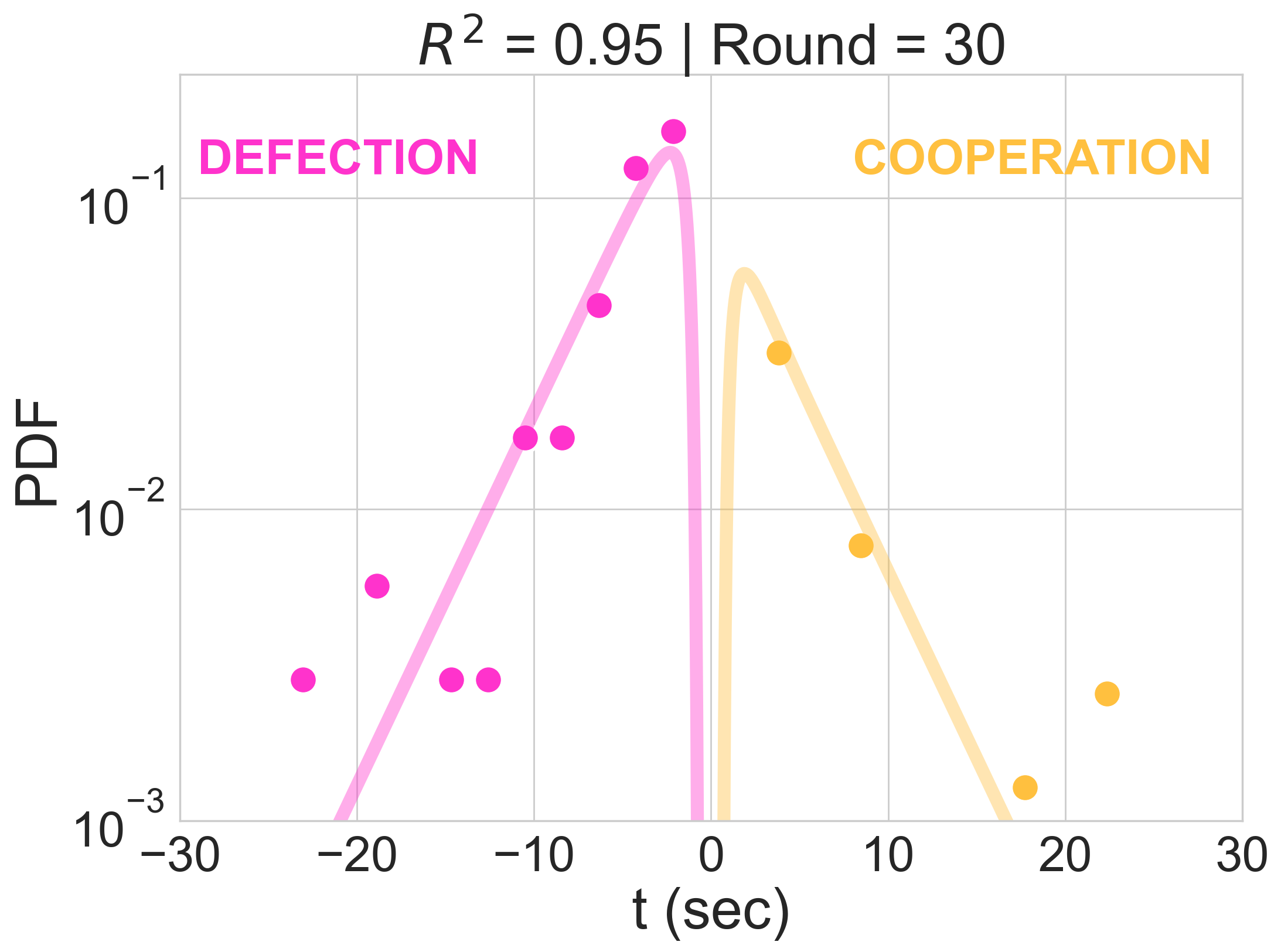}
                 \caption{}
             \end{subfigure}
             \begin{subfigure}[b]{0.45\textwidth}
                 \centering
                 \includegraphics[width=\textwidth]{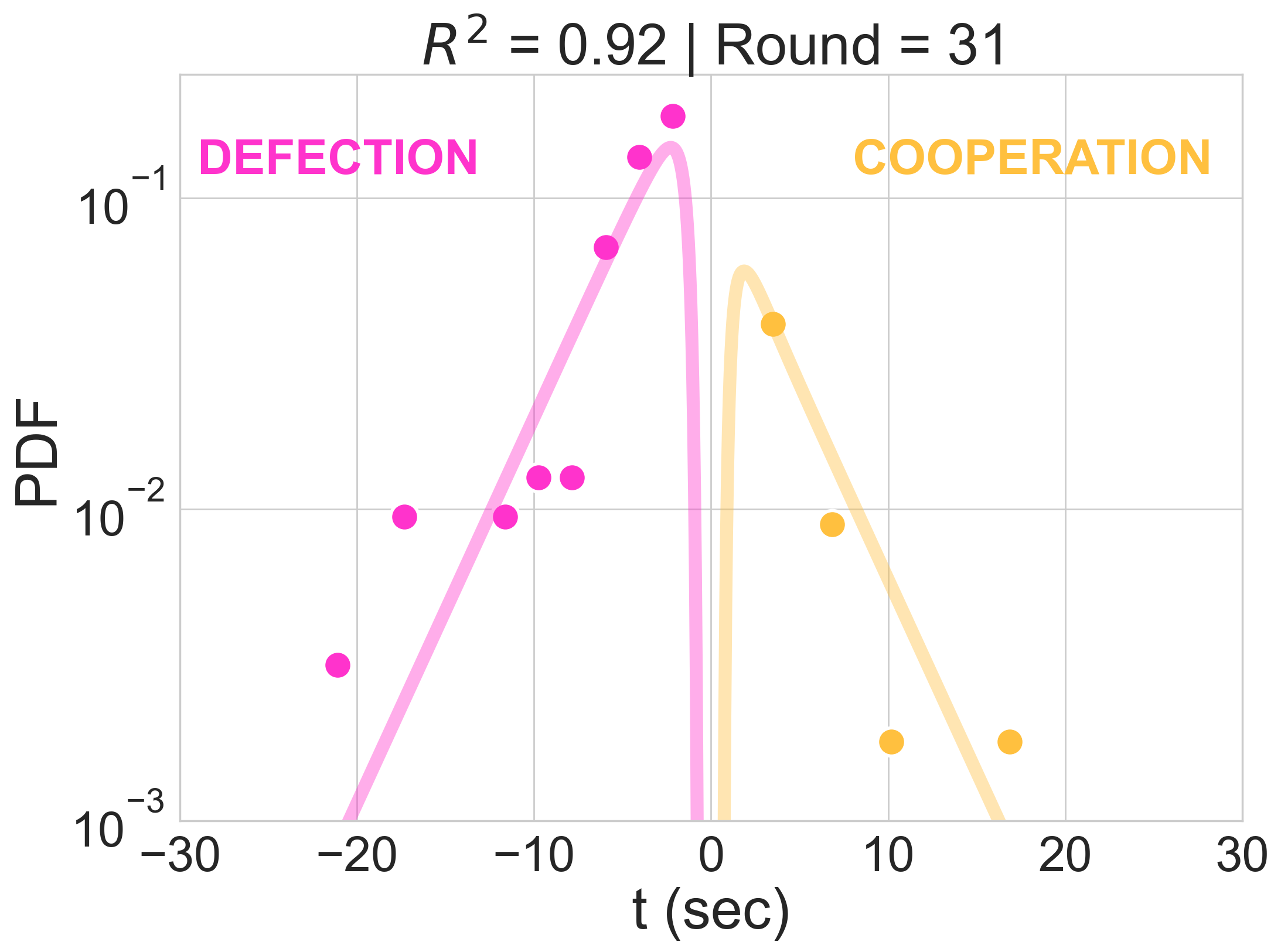}
                 \caption{}
             \end{subfigure}
             \begin{subfigure}[b]{0.45\textwidth}
                 \centering
                 \includegraphics[width=\textwidth]{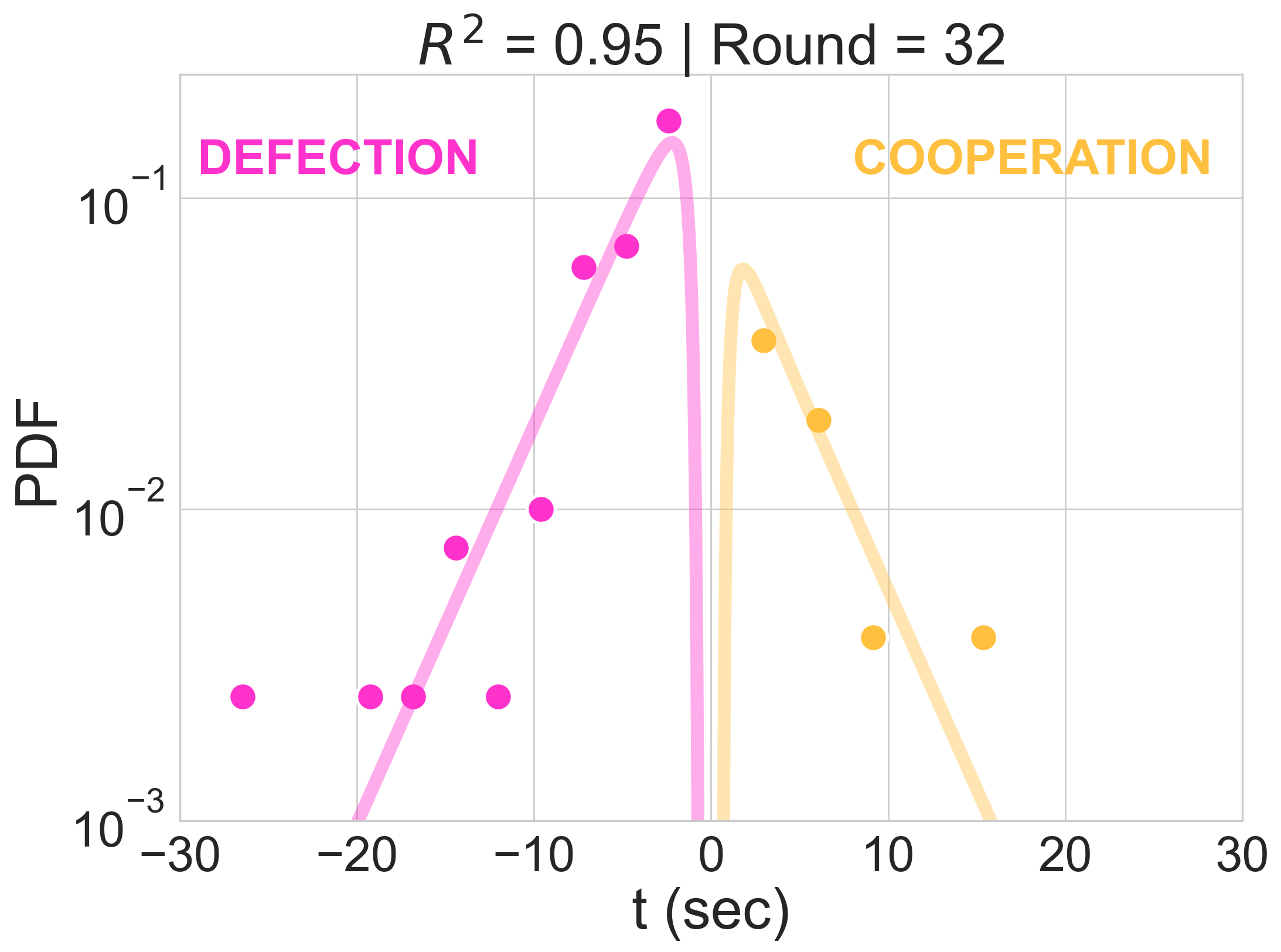}
                 \caption{}
             \end{subfigure}
             \caption{{\bf Accuracy over the testing set.} Response times PDFs for rounds $27-32$. Results are obtained using i) data (dots), and ii) our predictive model (lines). The left side (pink) corresponds to defection responses, while the right side (orange) corresponds to cooperation responses.}
             \label{SI_PDF5}
        \end{figure}
        \begin{figure}[H]
             \centering
             \begin{subfigure}[b]{0.45\textwidth}
                 \centering
                 \includegraphics[width=\textwidth]{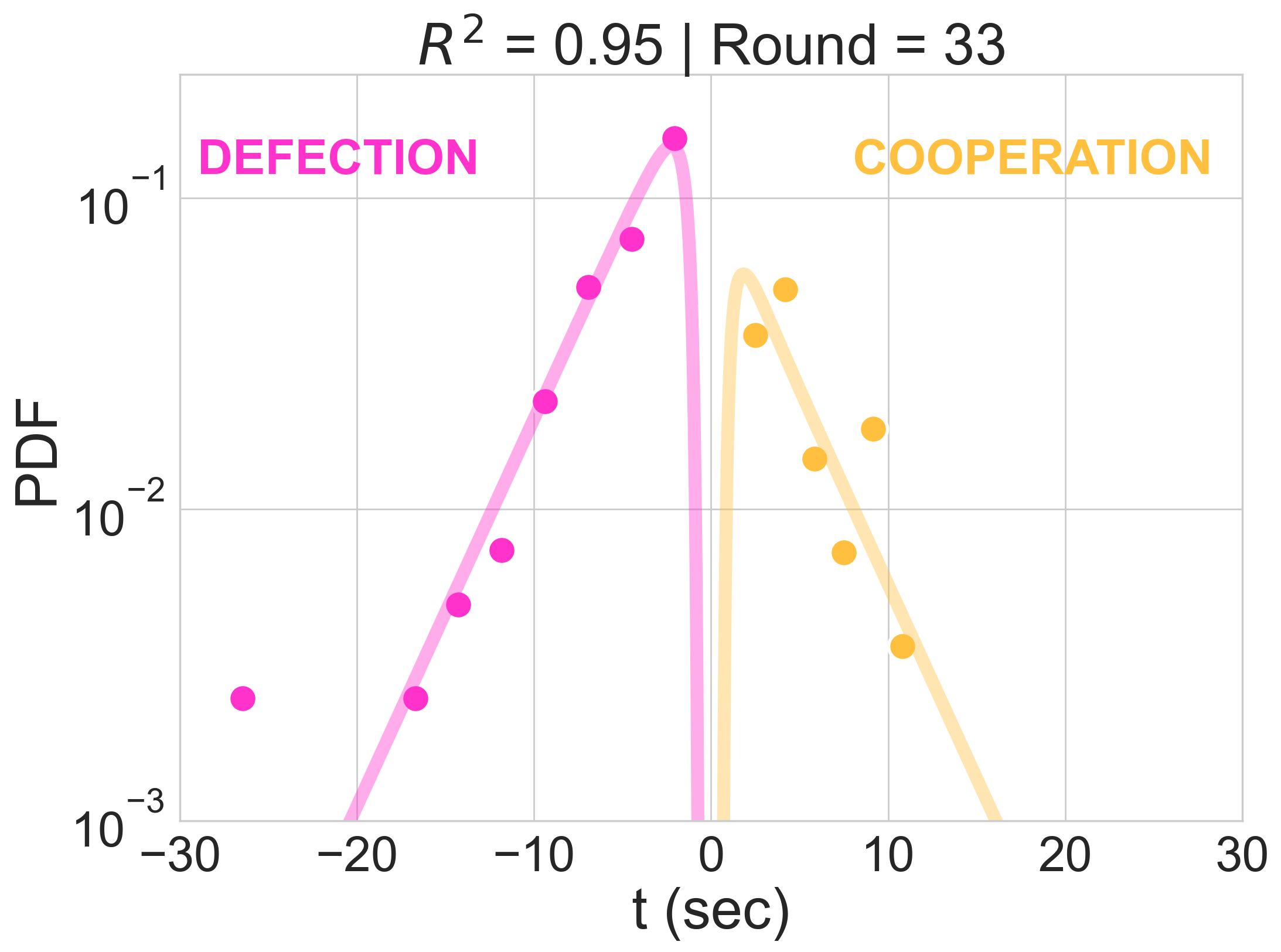}
                 \caption{}
             \end{subfigure}
             \begin{subfigure}[b]{0.45\textwidth}
                 \centering
                 \includegraphics[width=\textwidth]{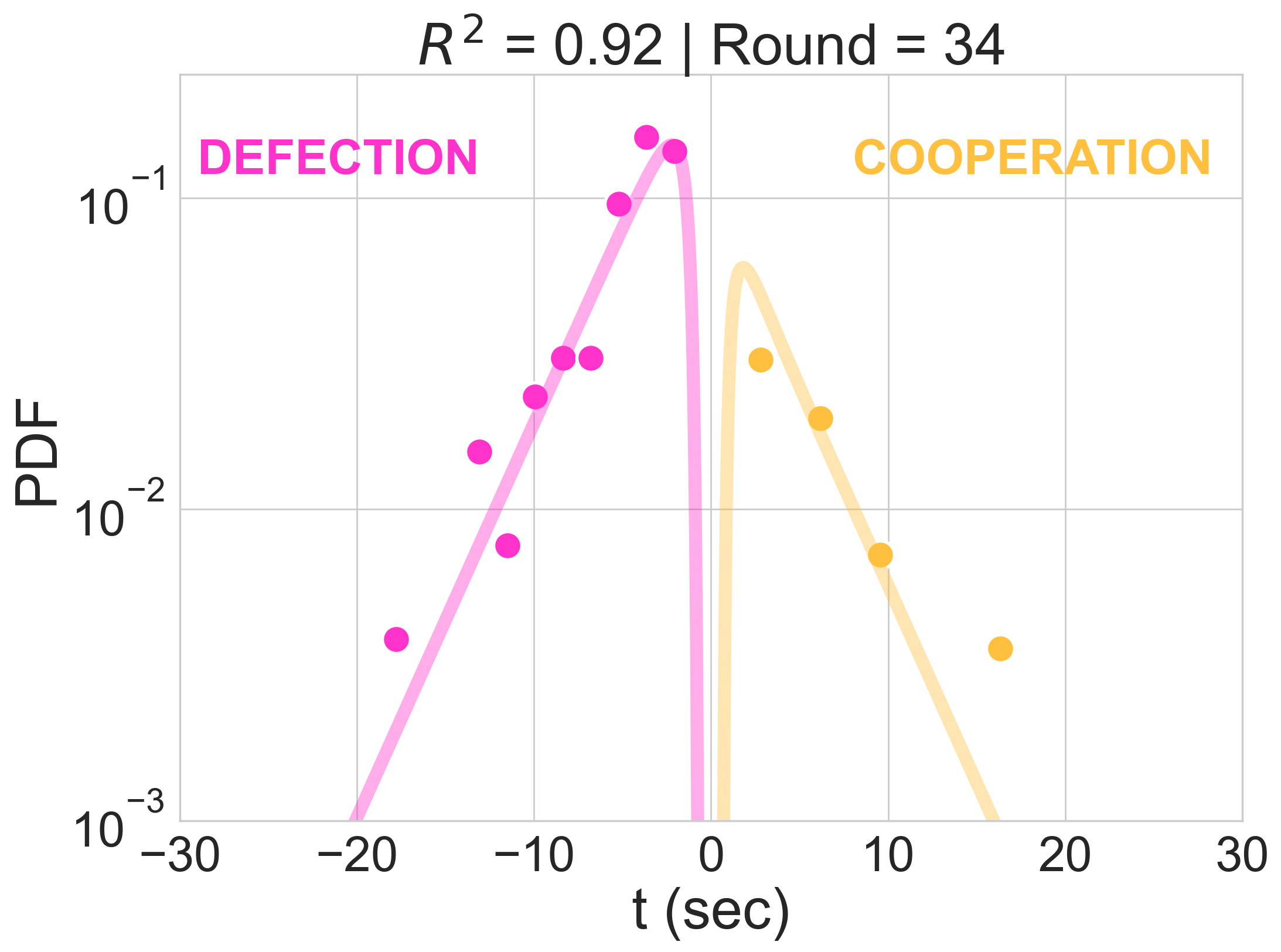}
                 \caption{}
             \end{subfigure}
             \begin{subfigure}[b]{0.45\textwidth}
                 \centering
                 \includegraphics[width=\textwidth]{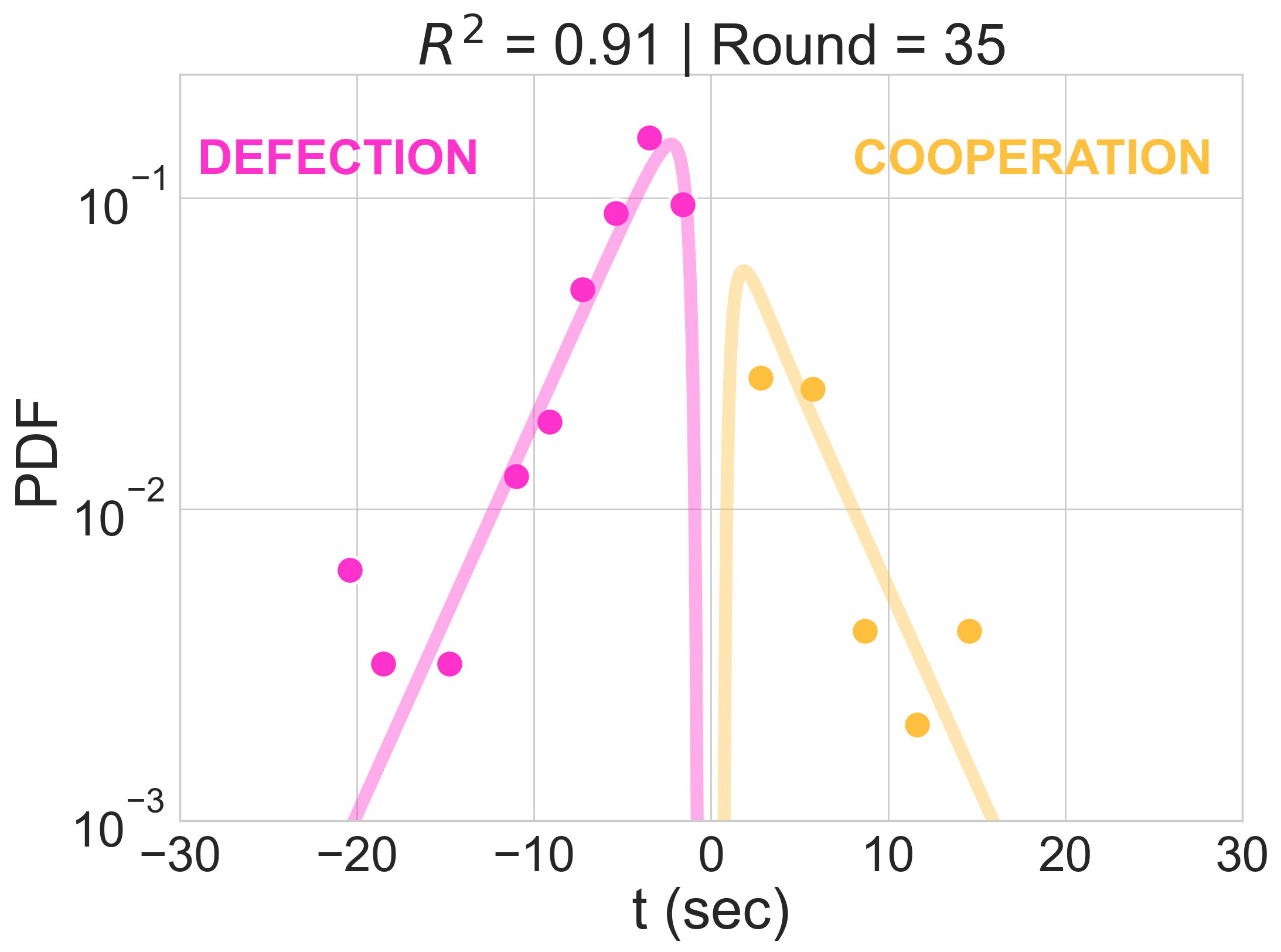}
                 \caption{}
             \end{subfigure}
             \begin{subfigure}[b]{0.45\textwidth}
                 \centering
                 \includegraphics[width=\textwidth]{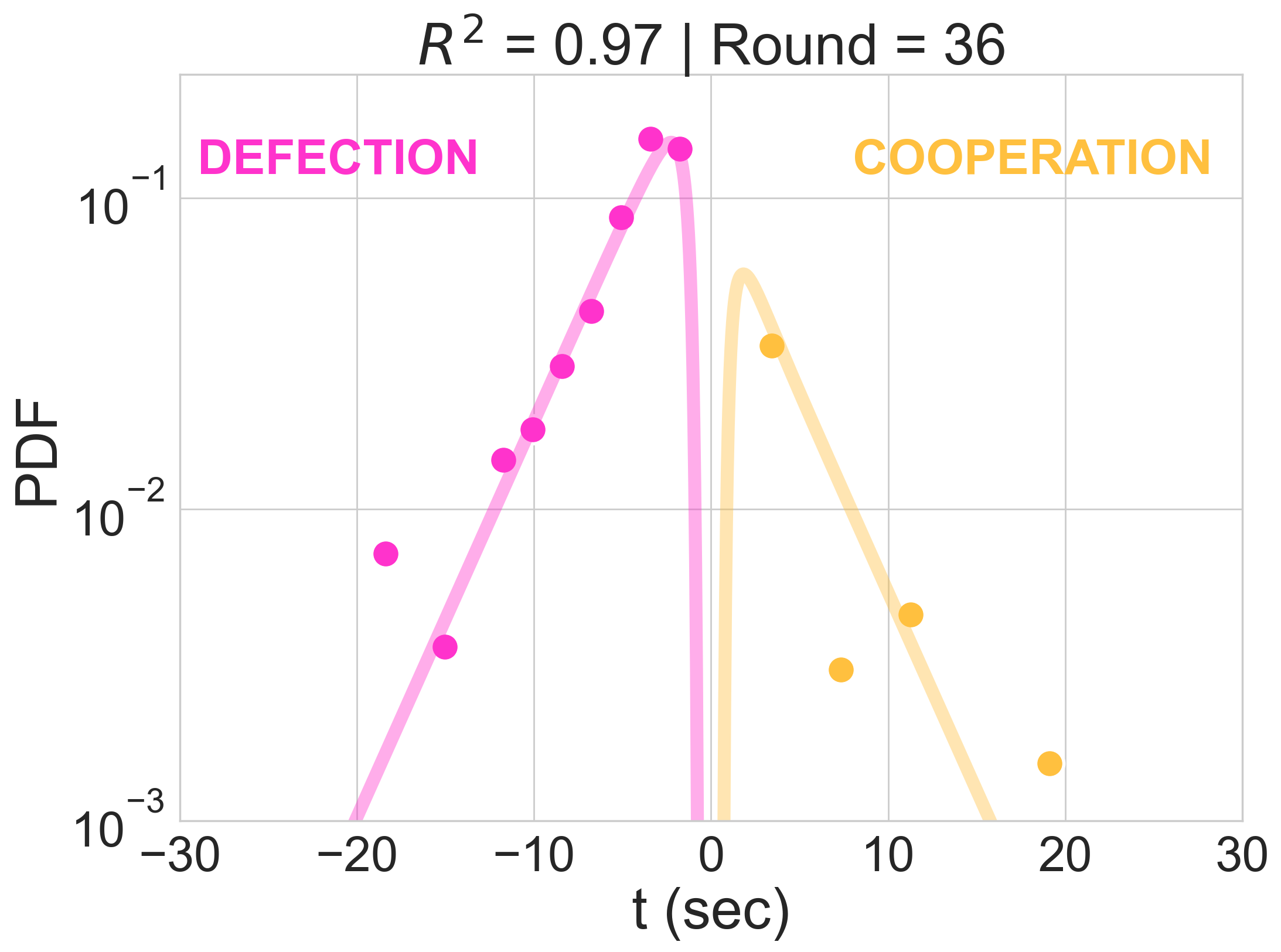}
                 \caption{}
             \end{subfigure}
             \begin{subfigure}[b]{0.45\textwidth}
                 \centering
                 \includegraphics[width=\textwidth]{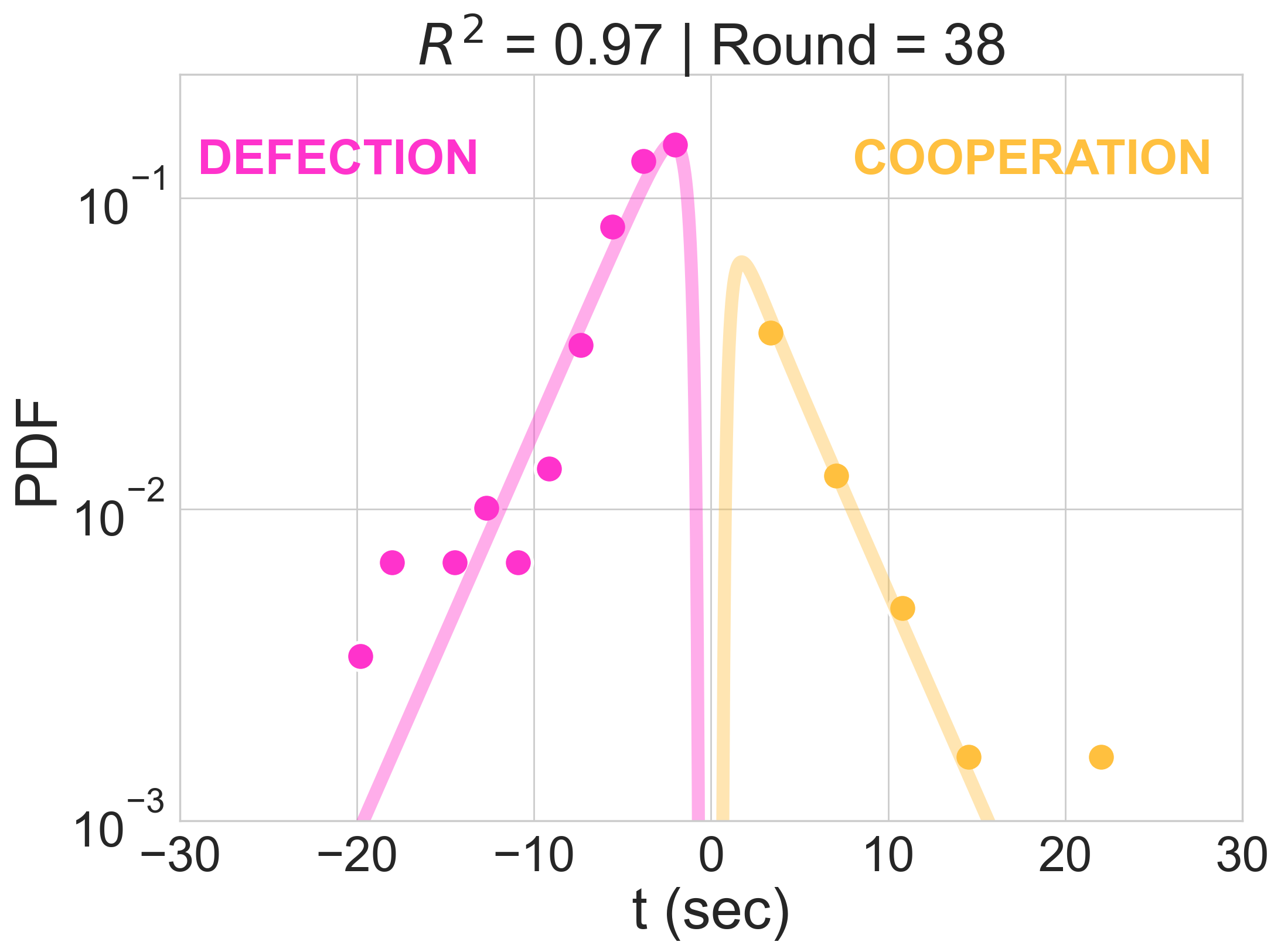}
                 \caption{}
             \end{subfigure}
             \begin{subfigure}[b]{0.45\textwidth}
                 \centering
                 \includegraphics[width=\textwidth]{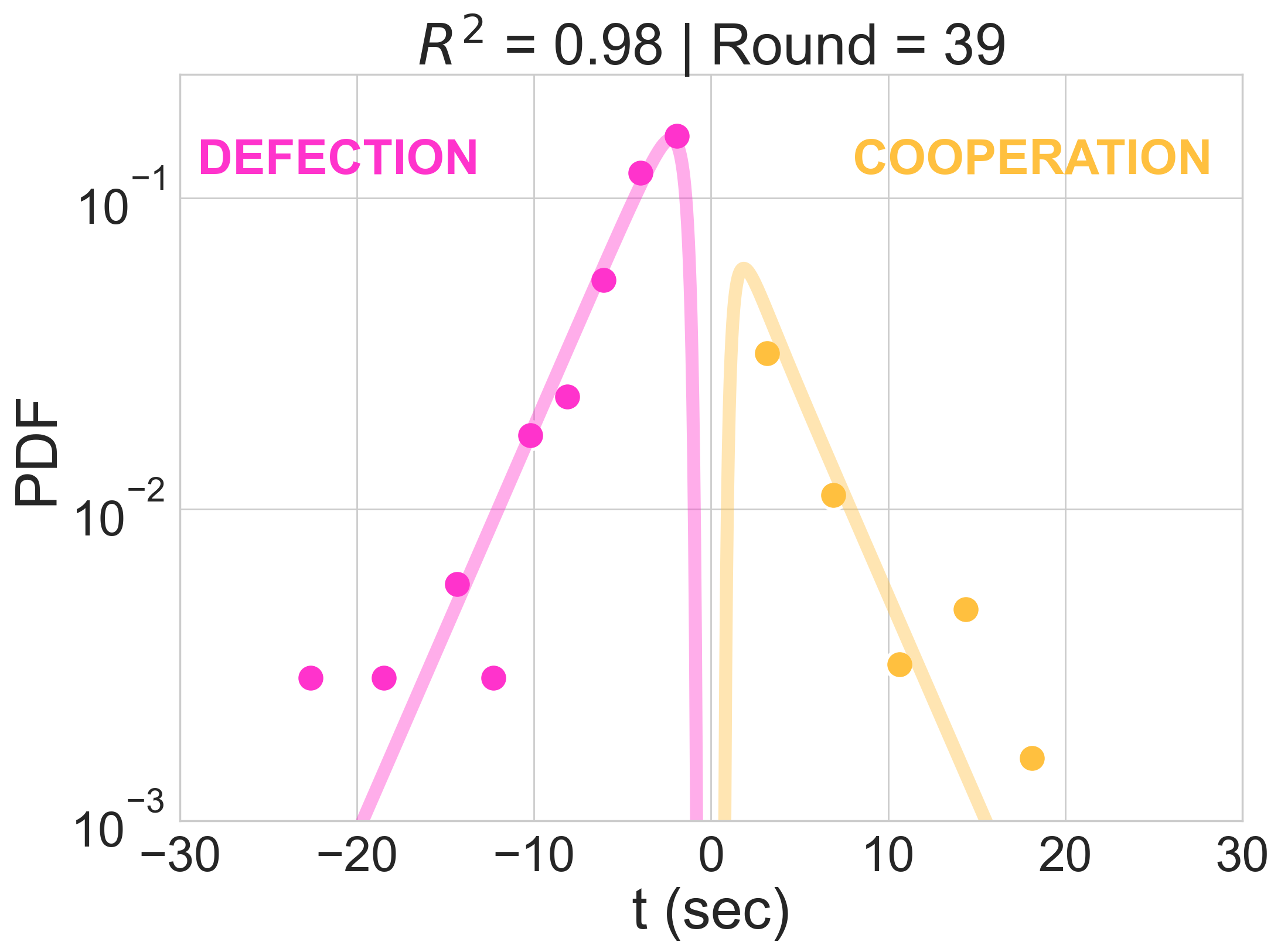}
                 \caption{}
             \end{subfigure}
             \caption{{\bf Accuracy over the testing set.} Response times PDFs for rounds $33-36$,$38-39$ (results on round $37$ are included in Fig.\ref{PDF-test}, in section Results of the main paper). Results are obtained using i) data (dots), and ii) our predictive model (lines). The left side (pink) corresponds to defection responses, while the right side (orange) corresponds to cooperation responses.}
             \label{SI_PDF6}
        \end{figure}
        \begin{figure}[H]
             \centering
             \begin{subfigure}[b]{0.45\textwidth}
                 \centering
                 \includegraphics[width=\textwidth]{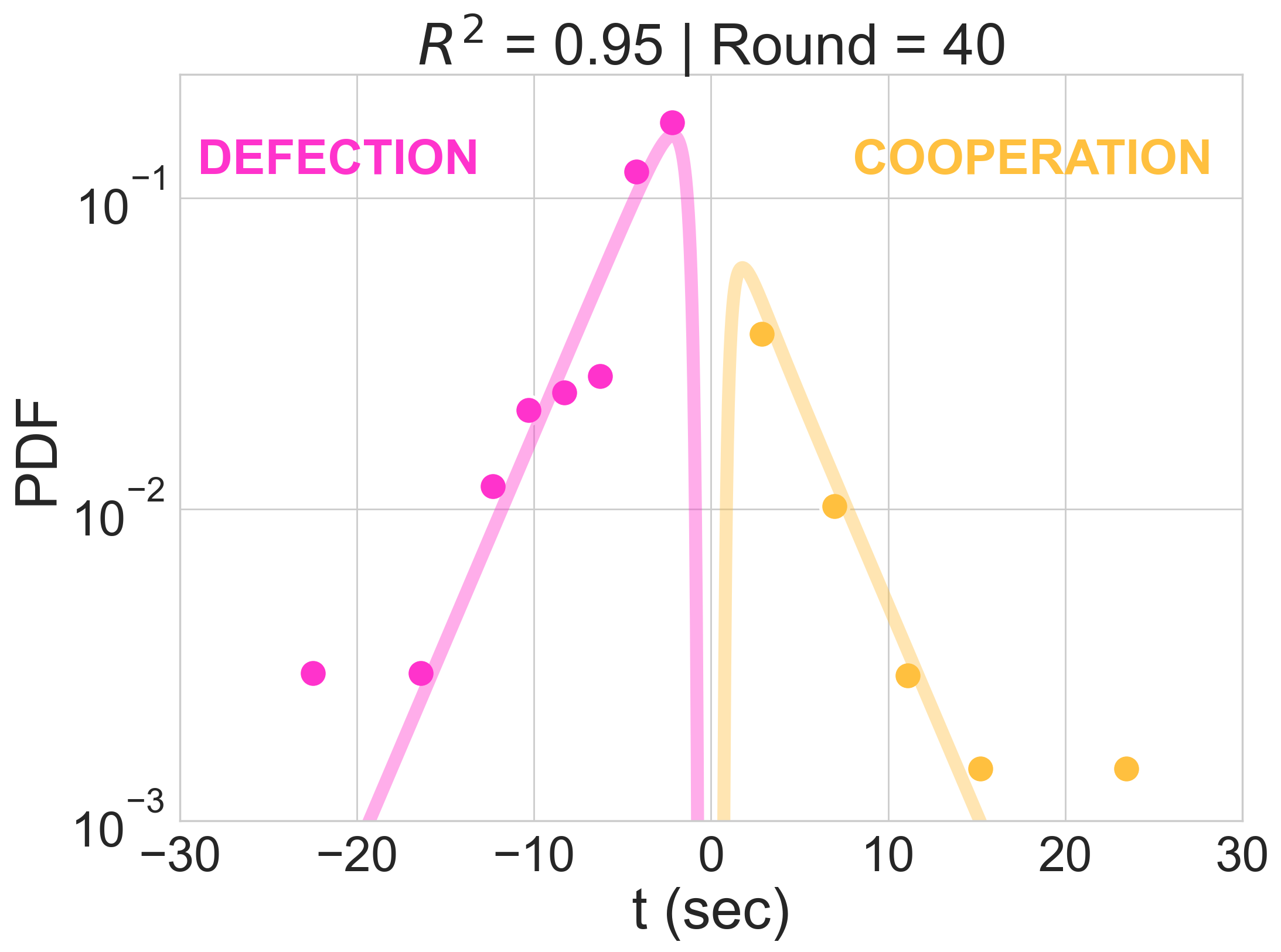}
                 \caption{}
             \end{subfigure}
             \begin{subfigure}[b]{0.45\textwidth}
                 \centering
                 \includegraphics[width=\textwidth]{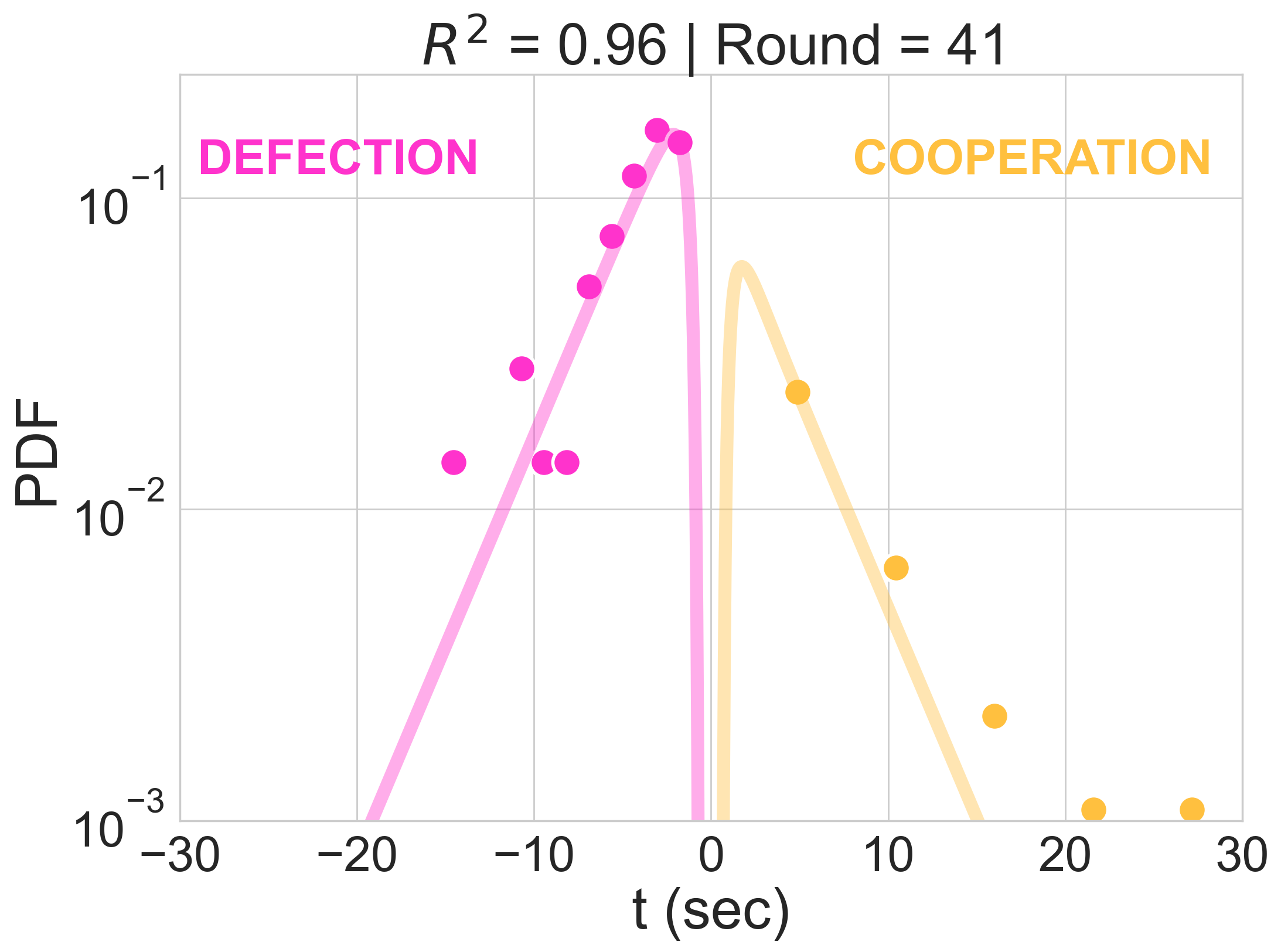}
                 \caption{}
             \end{subfigure}
             \begin{subfigure}[b]{0.45\textwidth}
                 \centering
                 \includegraphics[width=\textwidth]{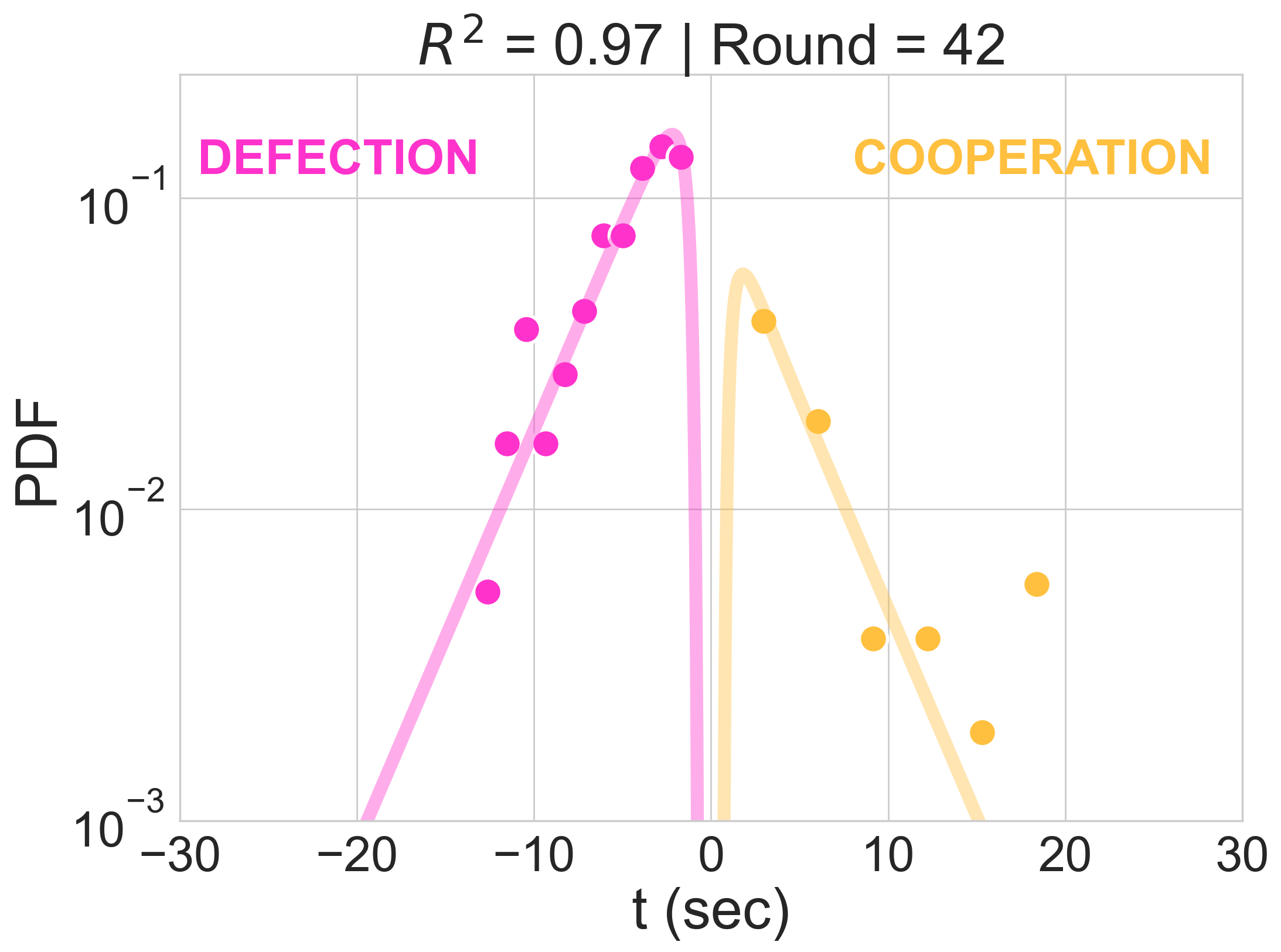}
                 \caption{}
             \end{subfigure}
             \begin{subfigure}[b]{0.45\textwidth}
                 \centering
                 \includegraphics[width=\textwidth]{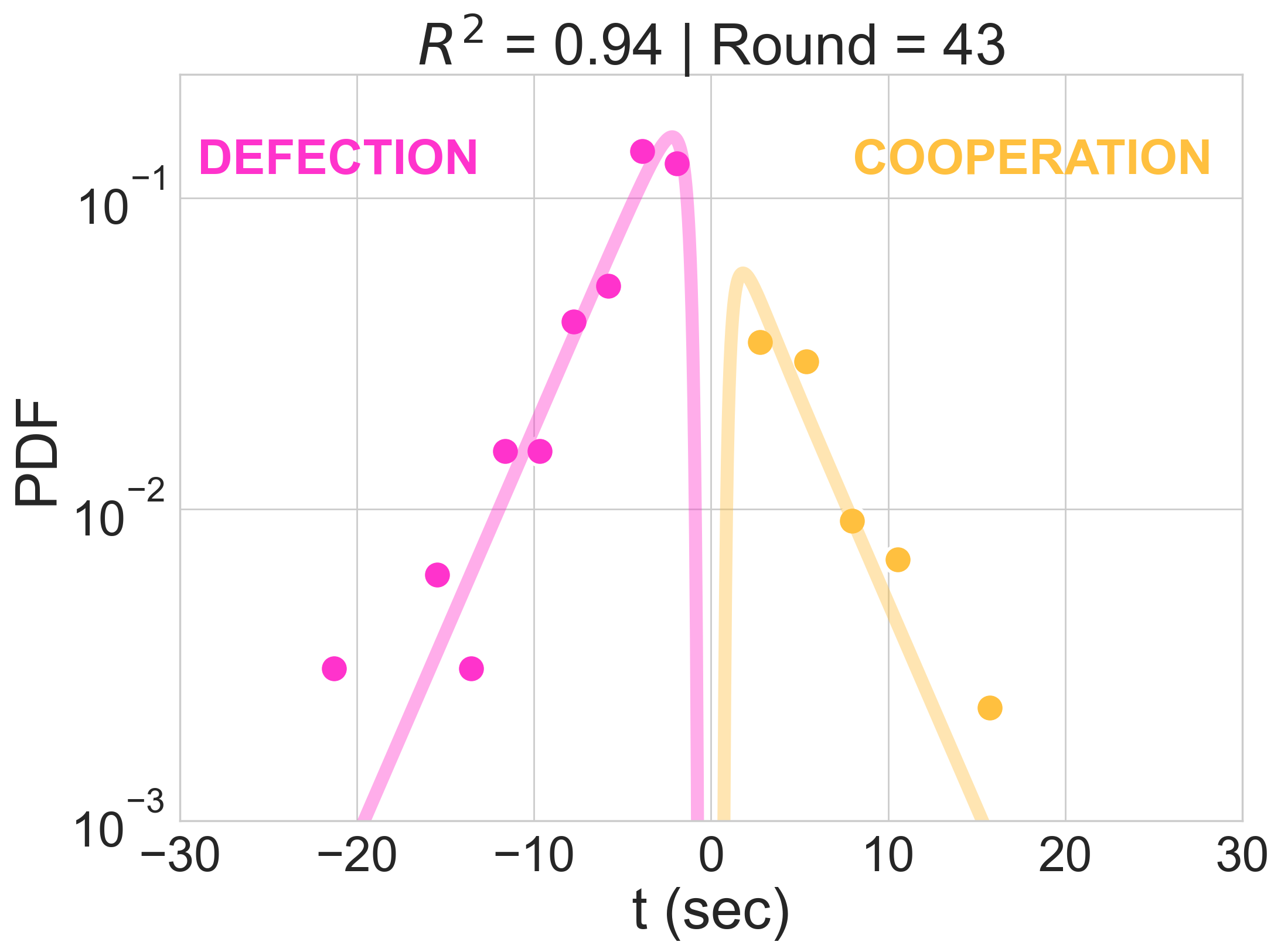}
                 \caption{}
             \end{subfigure}
             \begin{subfigure}[b]{0.45\textwidth}
                 \centering
                 \includegraphics[width=\textwidth]{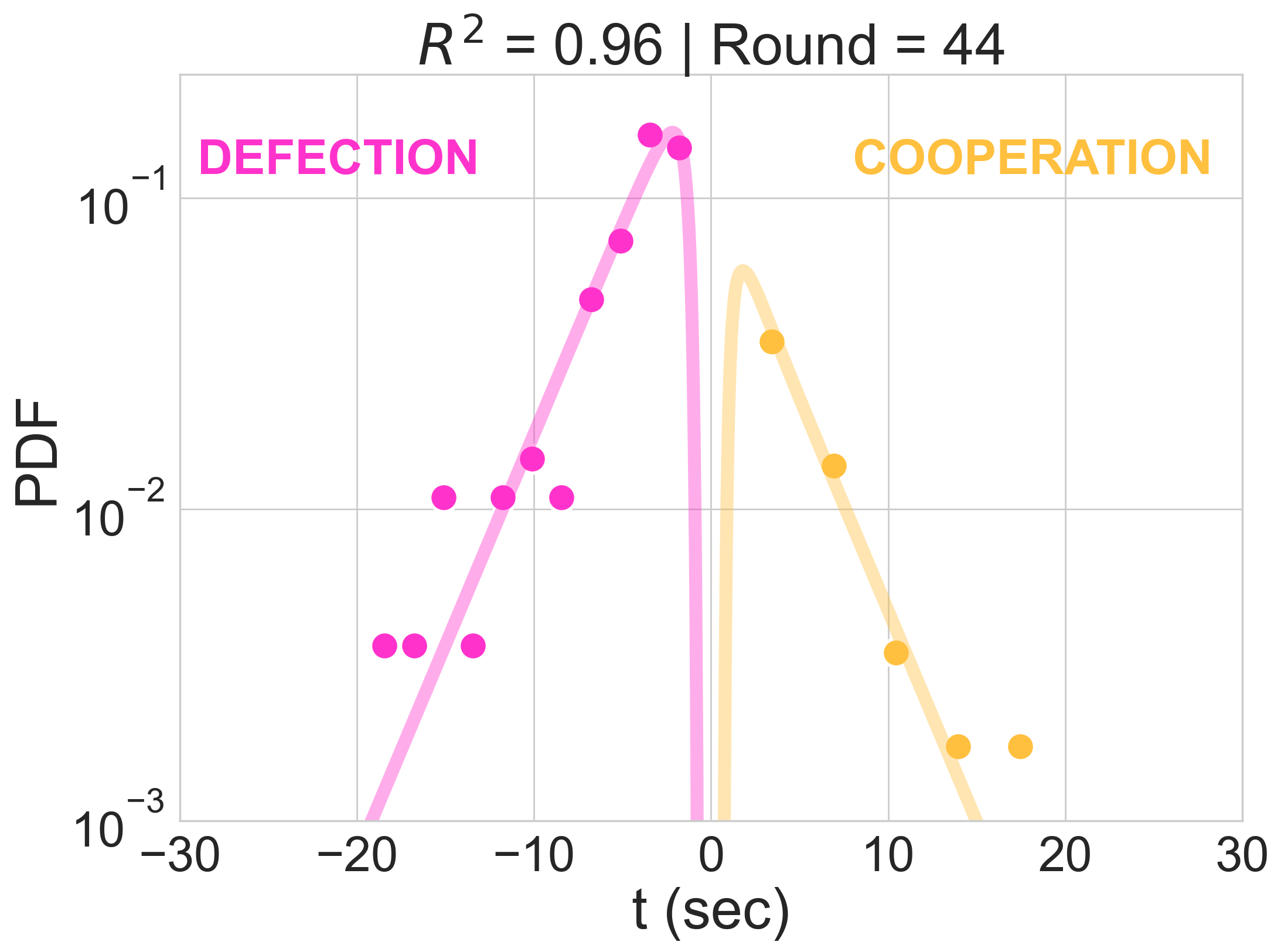}
                 \caption{}
             \end{subfigure}
             \begin{subfigure}[b]{0.45\textwidth}
                 \centering
                 \includegraphics[width=\textwidth]{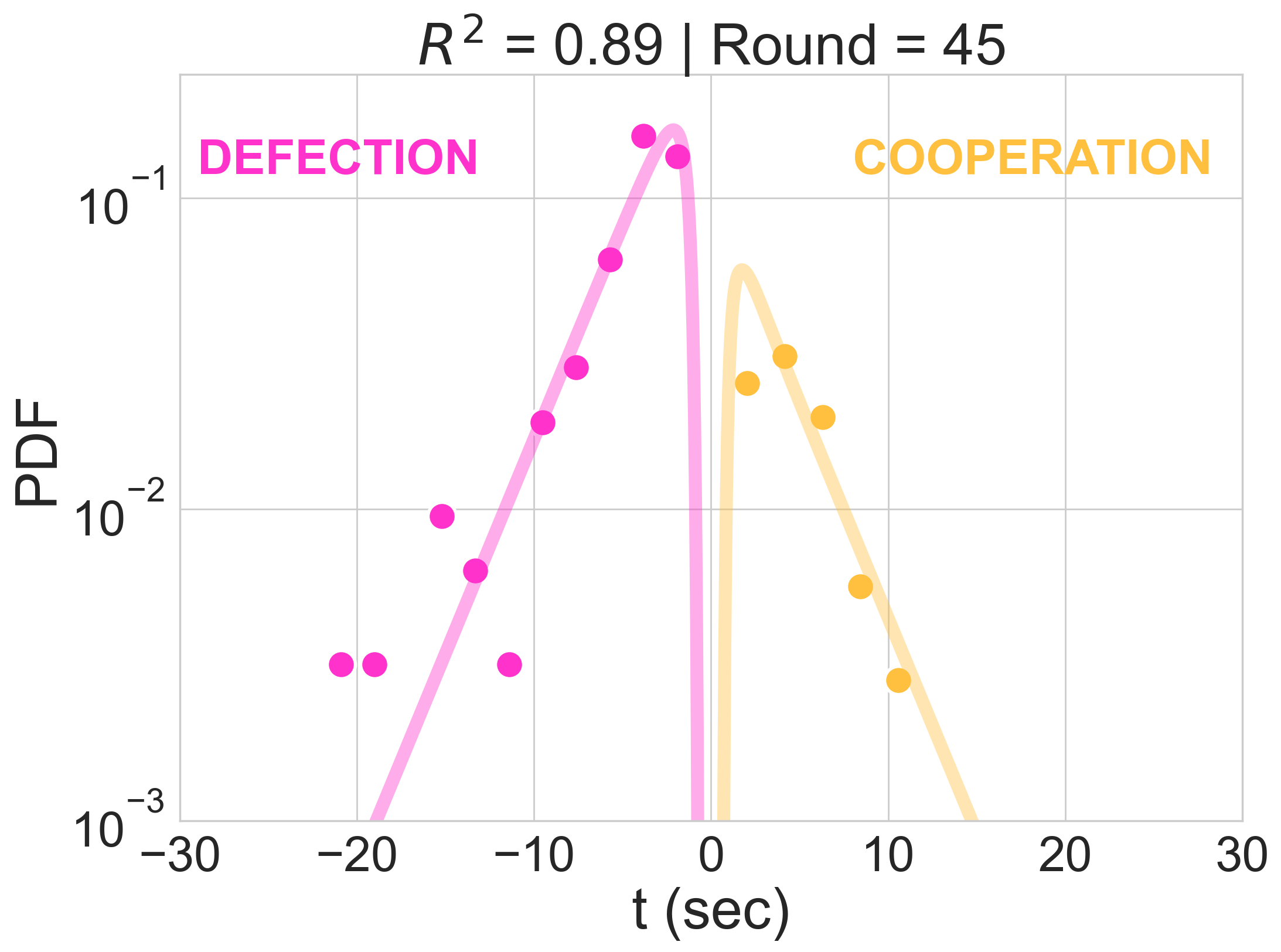}
                 \caption{}
             \end{subfigure}
             \caption{{\bf Accuracy over the testing set.} Response times PDFs for rounds $40-45$. Results are obtained using i) data (dots), and ii) our predictive model (lines). The left side (pink) corresponds to defection responses, while the right side (orange) corresponds to cooperation responses.}
             \label{SI_PDF7}
        \end{figure}
        \begin{figure}[H]
             \centering
             \begin{subfigure}[b]{0.45\textwidth}
                 \centering
                 \includegraphics[width=\textwidth]{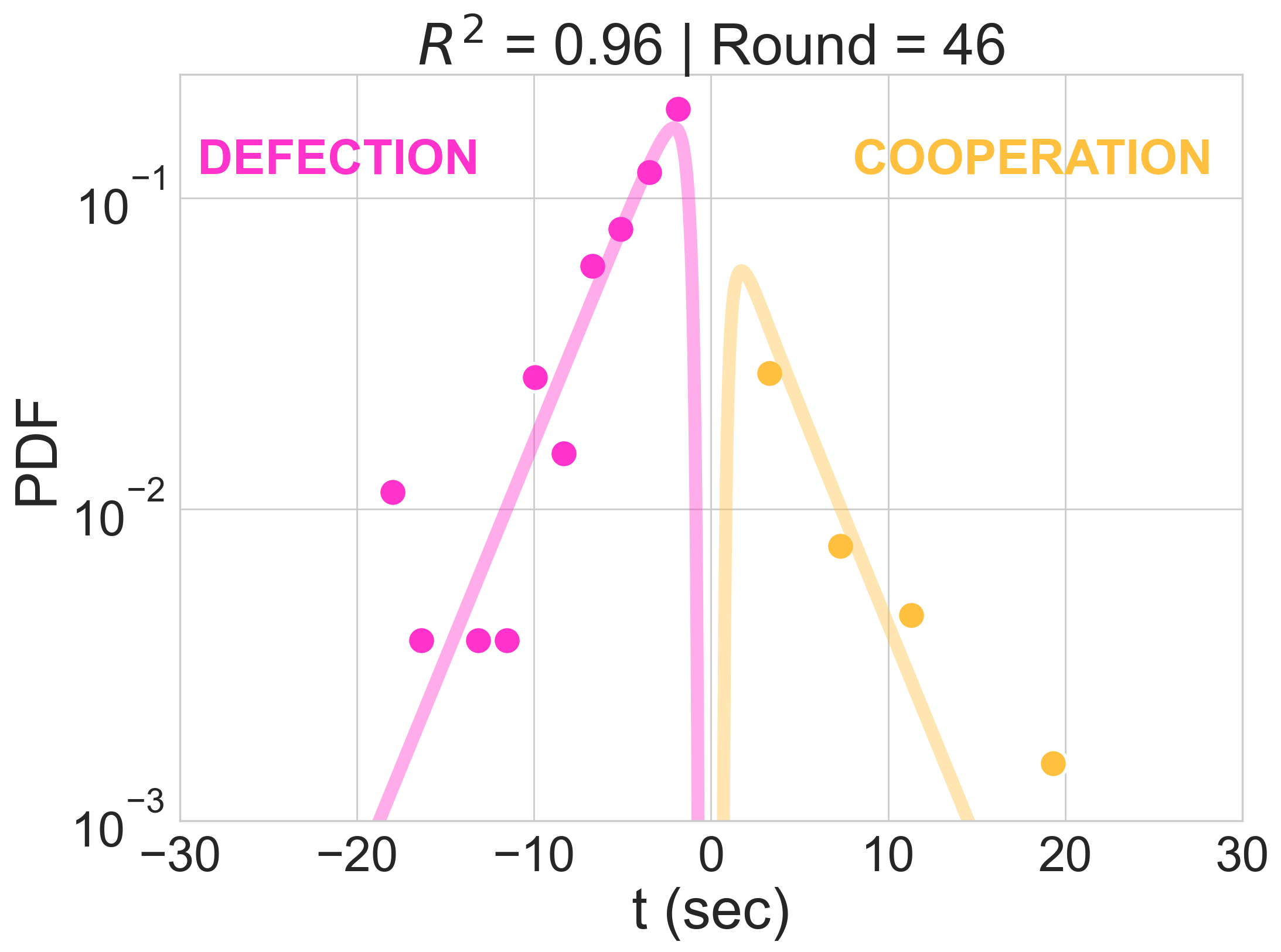}
                 \caption{}
             \end{subfigure}
             \begin{subfigure}[b]{0.45\textwidth}
                 \centering
                 \includegraphics[width=\textwidth]{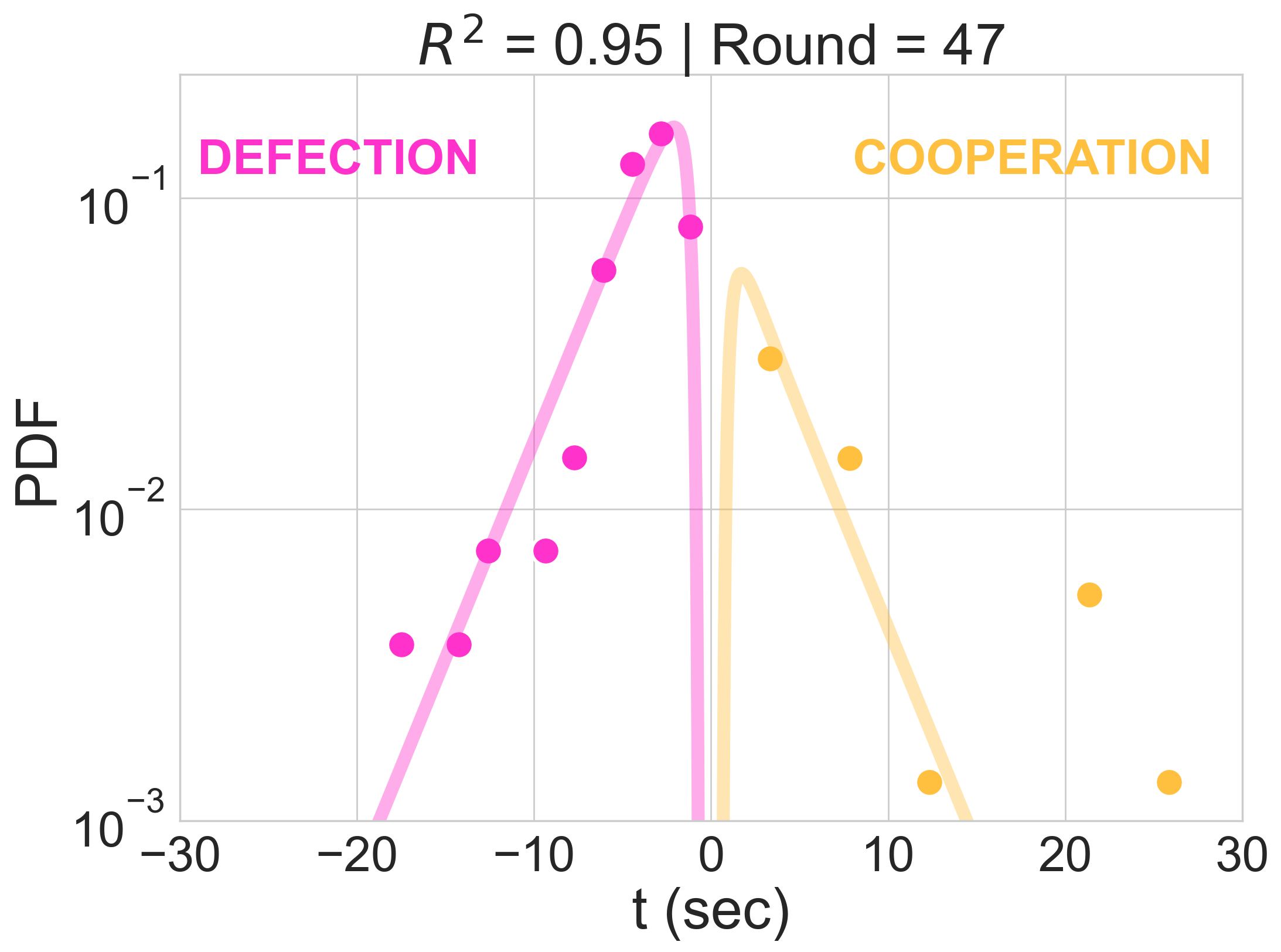}
                 \caption{}
             \end{subfigure}
             \begin{subfigure}[b]{0.45\textwidth}
                 \centering
                 \includegraphics[width=\textwidth]{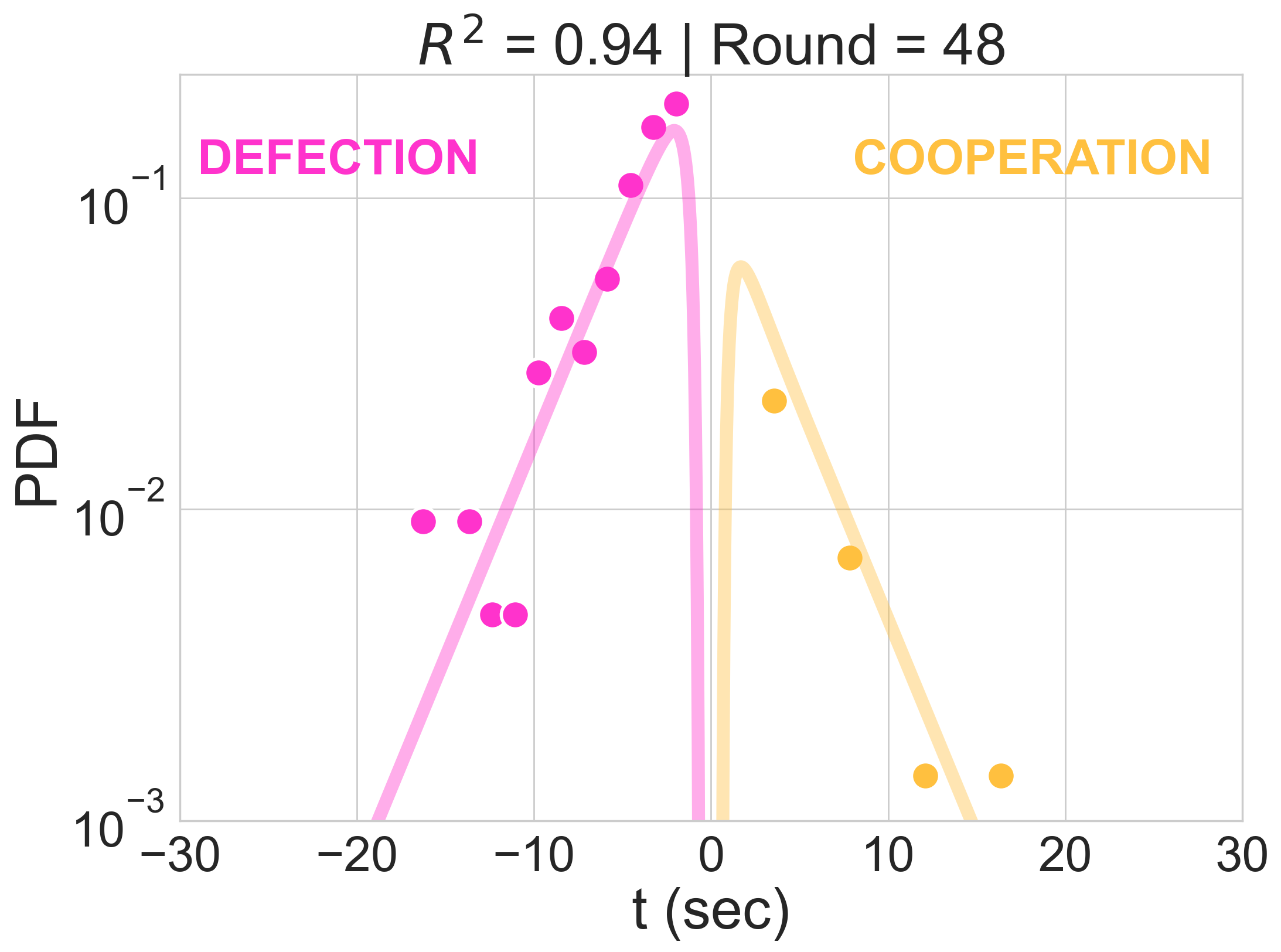}
                 \caption{}
             \end{subfigure}
             \begin{subfigure}[b]{0.45\textwidth}
                 \centering
                 \includegraphics[width=\textwidth]{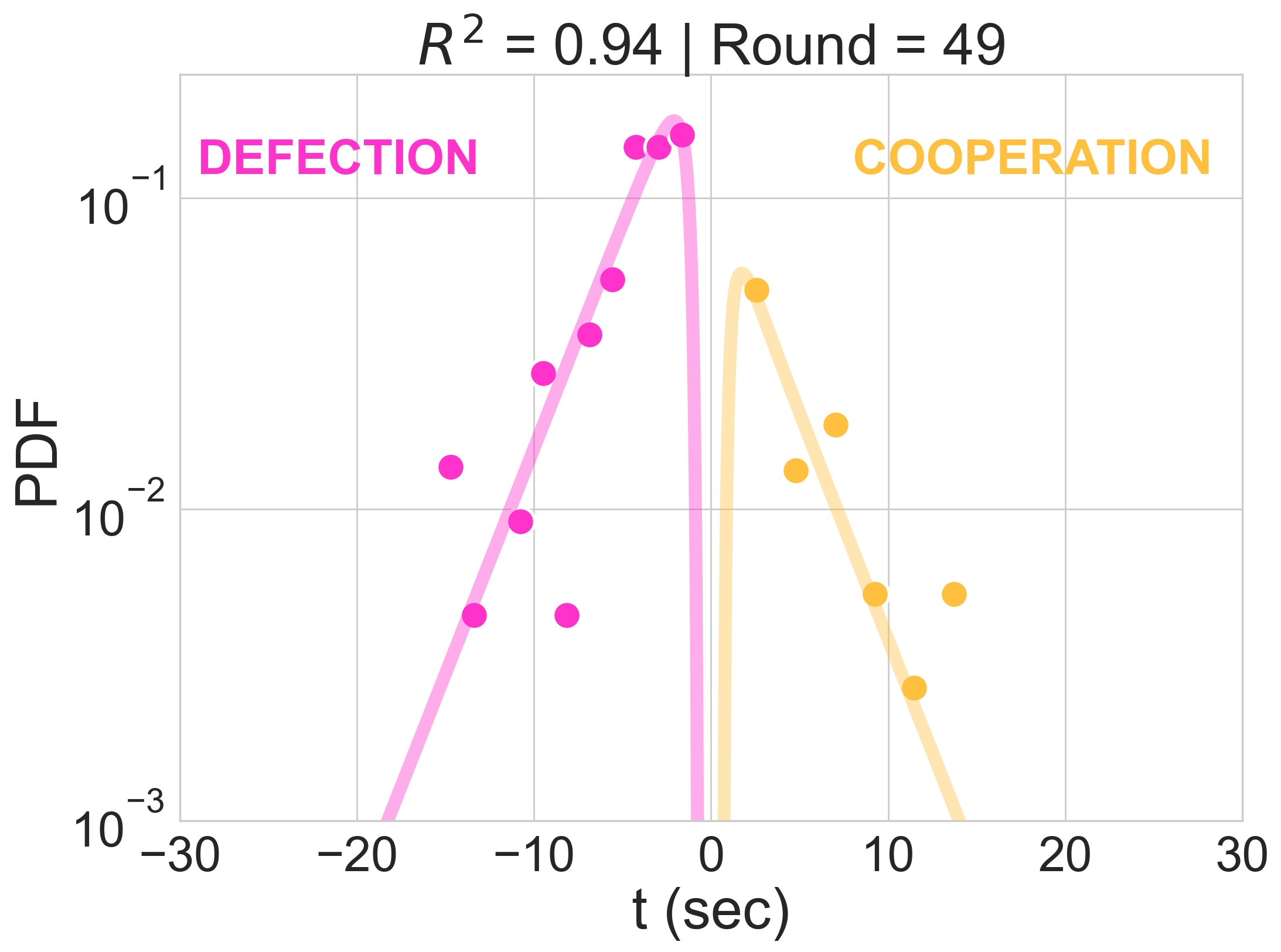}
                 \caption{}
             \end{subfigure}
             \begin{subfigure}[b]{0.45\textwidth}
                 \centering
                 \includegraphics[width=\textwidth]{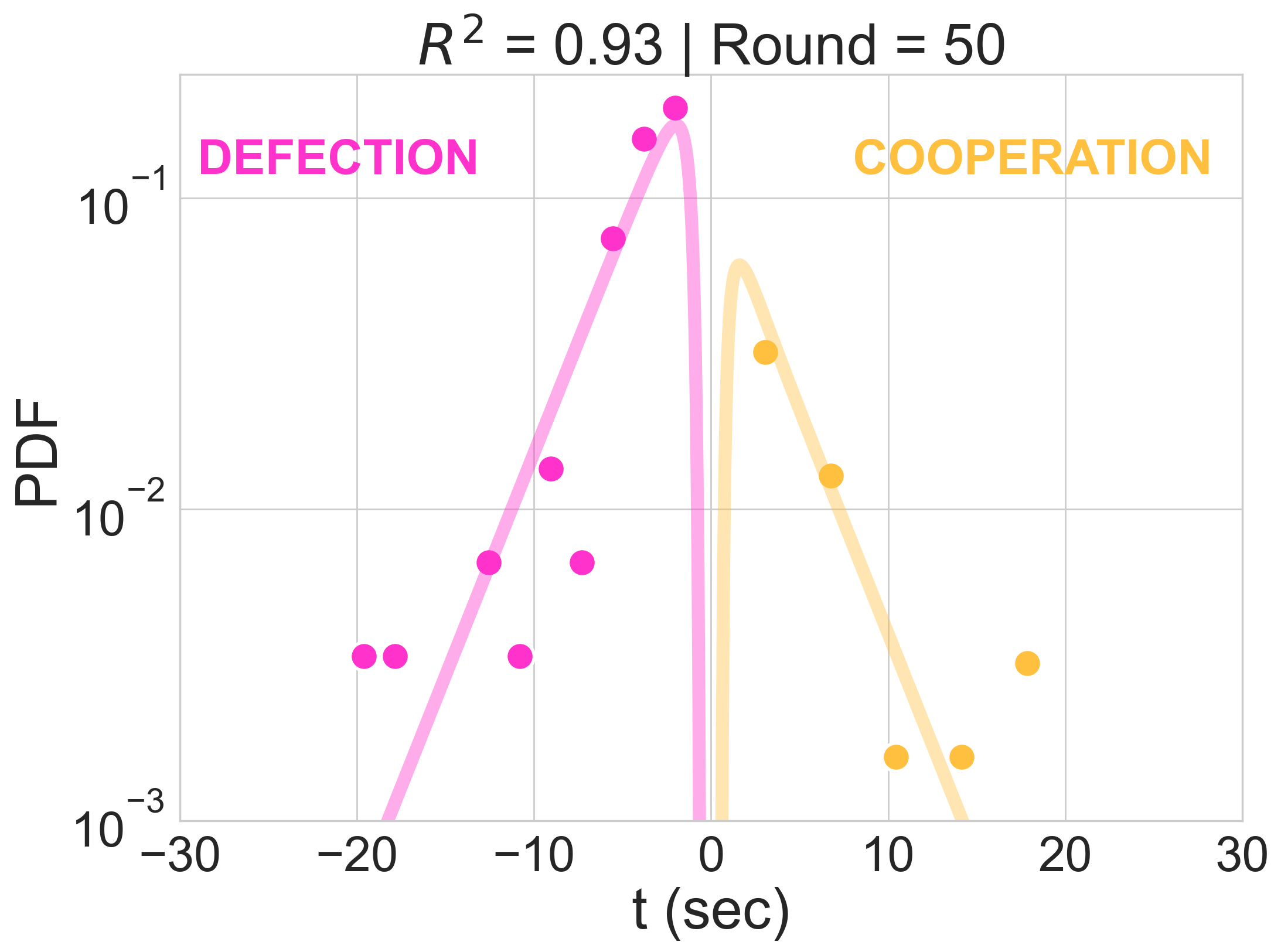}
                 \caption{}
             \end{subfigure}
             \begin{subfigure}[b]{0.45\textwidth}
                 \centering
                 \includegraphics[width=\textwidth]{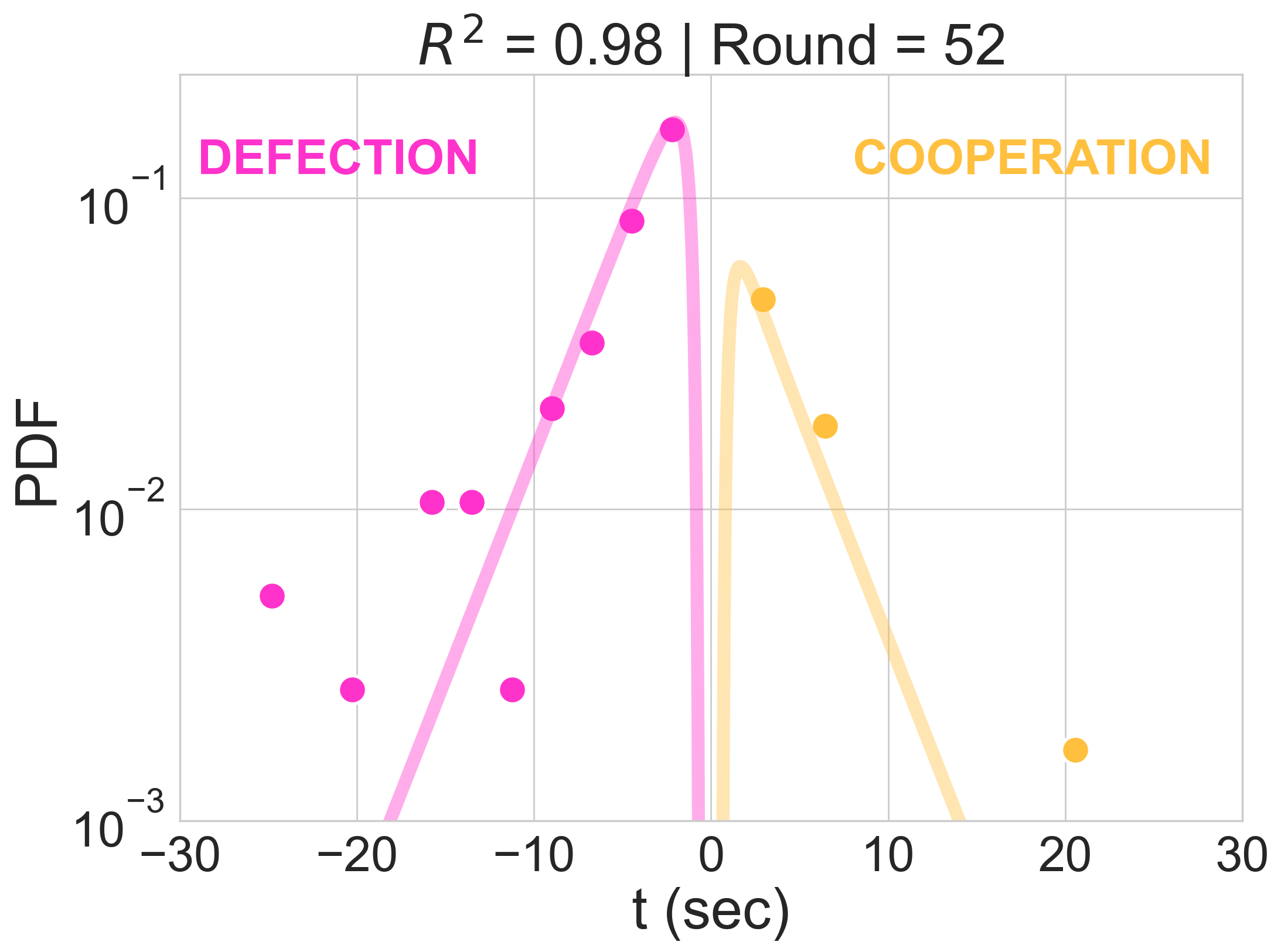}
                 \caption{}
             \end{subfigure}
             \caption{{\bf Accuracy over the testing set.} Response times PDFs for rounds $46-50$,$52$ (results on round $51$ are included in Fig.\ref{PDF-test}, in section Results of the main paper). Results are obtained using i) data (dots), and ii) our predictive model (lines). The left side (pink) corresponds to defection responses, while the right side (orange) corresponds to cooperation responses.}
             \label{SI_PDF8}
        \end{figure}
        \begin{figure}[H]
             \centering
             \begin{subfigure}[b]{0.45\textwidth}
                 \centering
                 \includegraphics[width=\textwidth]{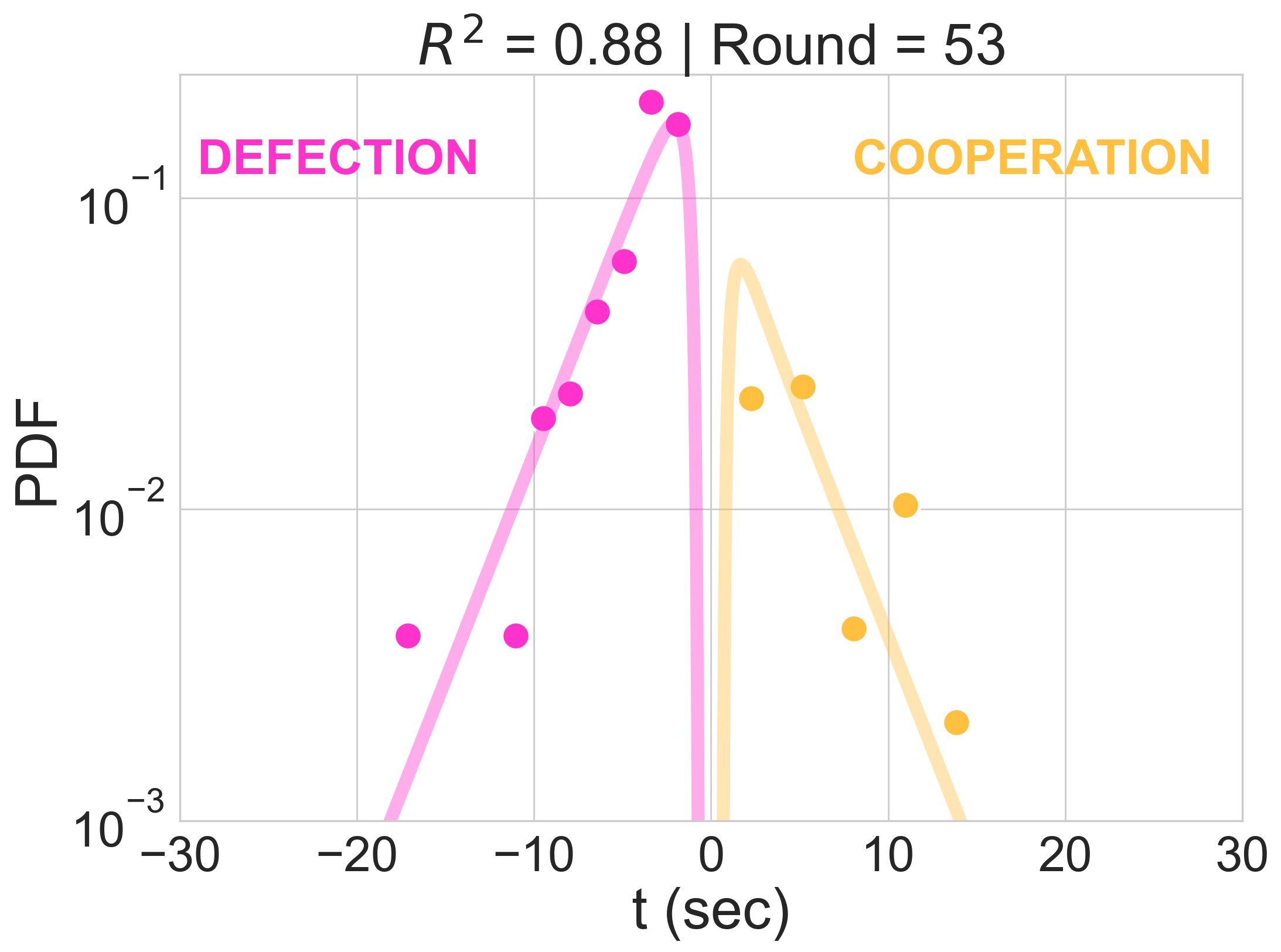}
                 \caption{}
             \end{subfigure}
             \begin{subfigure}[b]{0.45\textwidth}
                 \centering
                 \includegraphics[width=\textwidth]{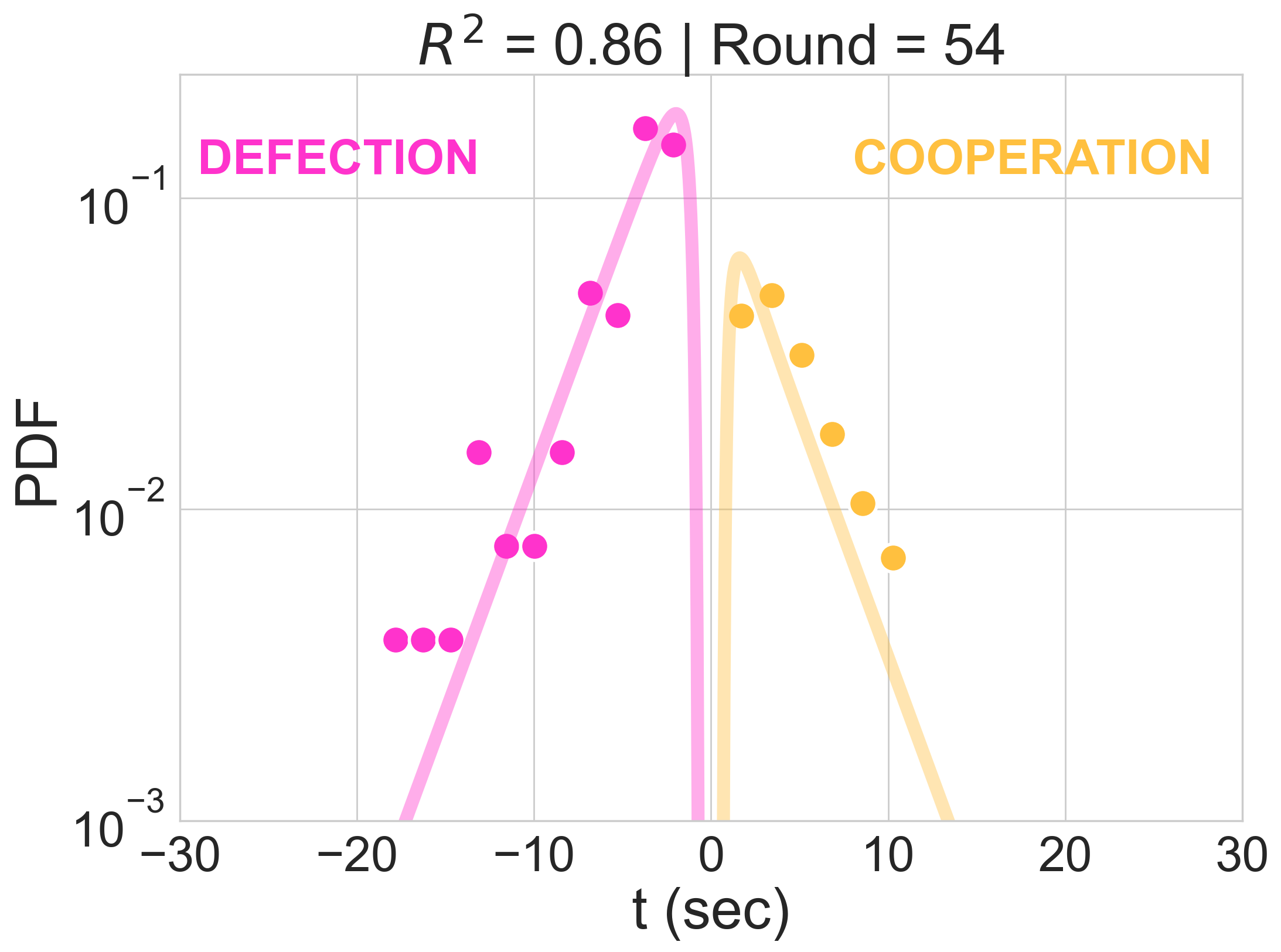}
                 \caption{}
             \end{subfigure}
             \begin{subfigure}[b]{0.45\textwidth}
                 \centering
                 \includegraphics[width=\textwidth]{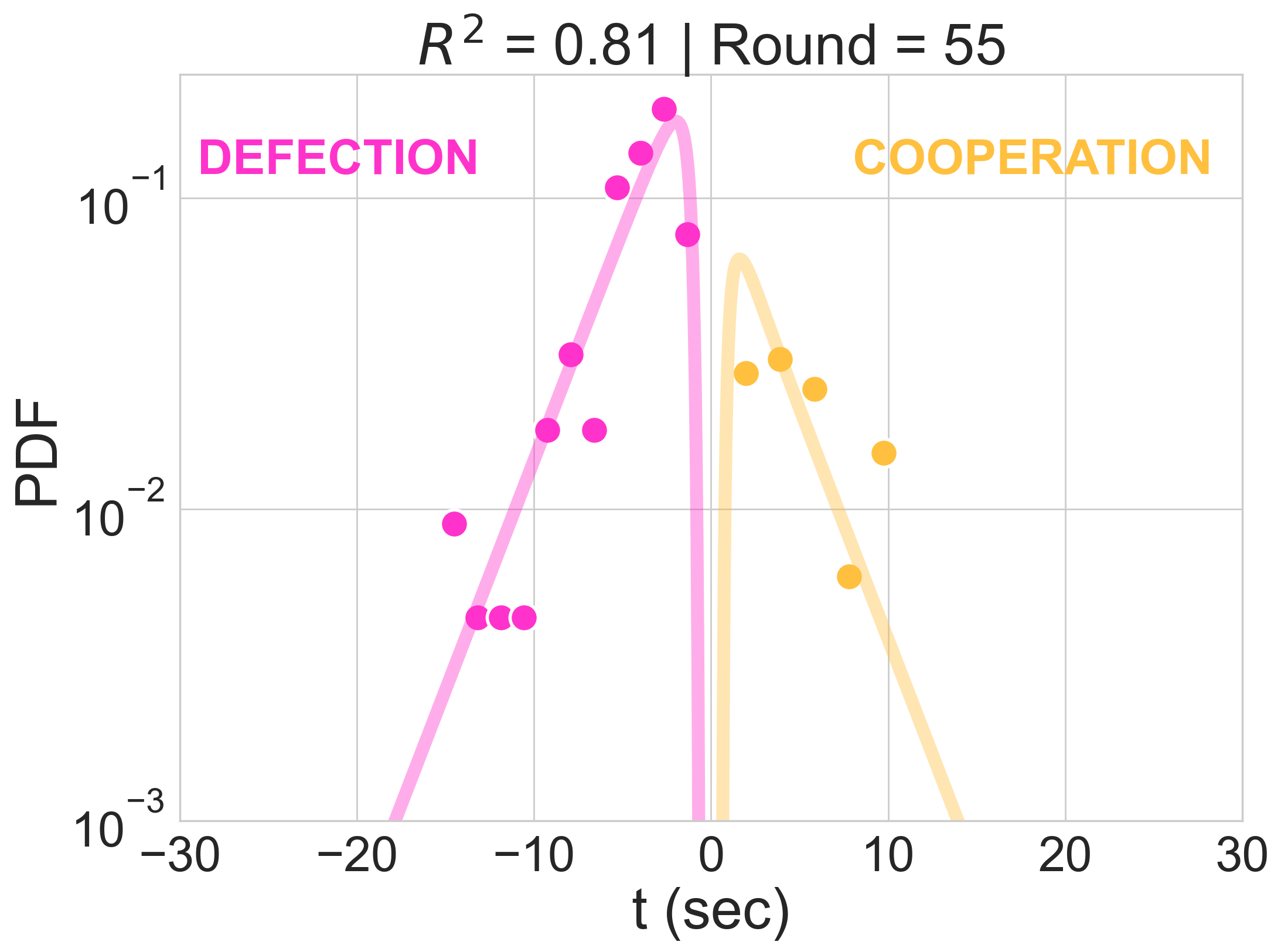}
                 \caption{}
             \end{subfigure}
             \begin{subfigure}[b]{0.45\textwidth}
                 \centering
                 \includegraphics[width=\textwidth]{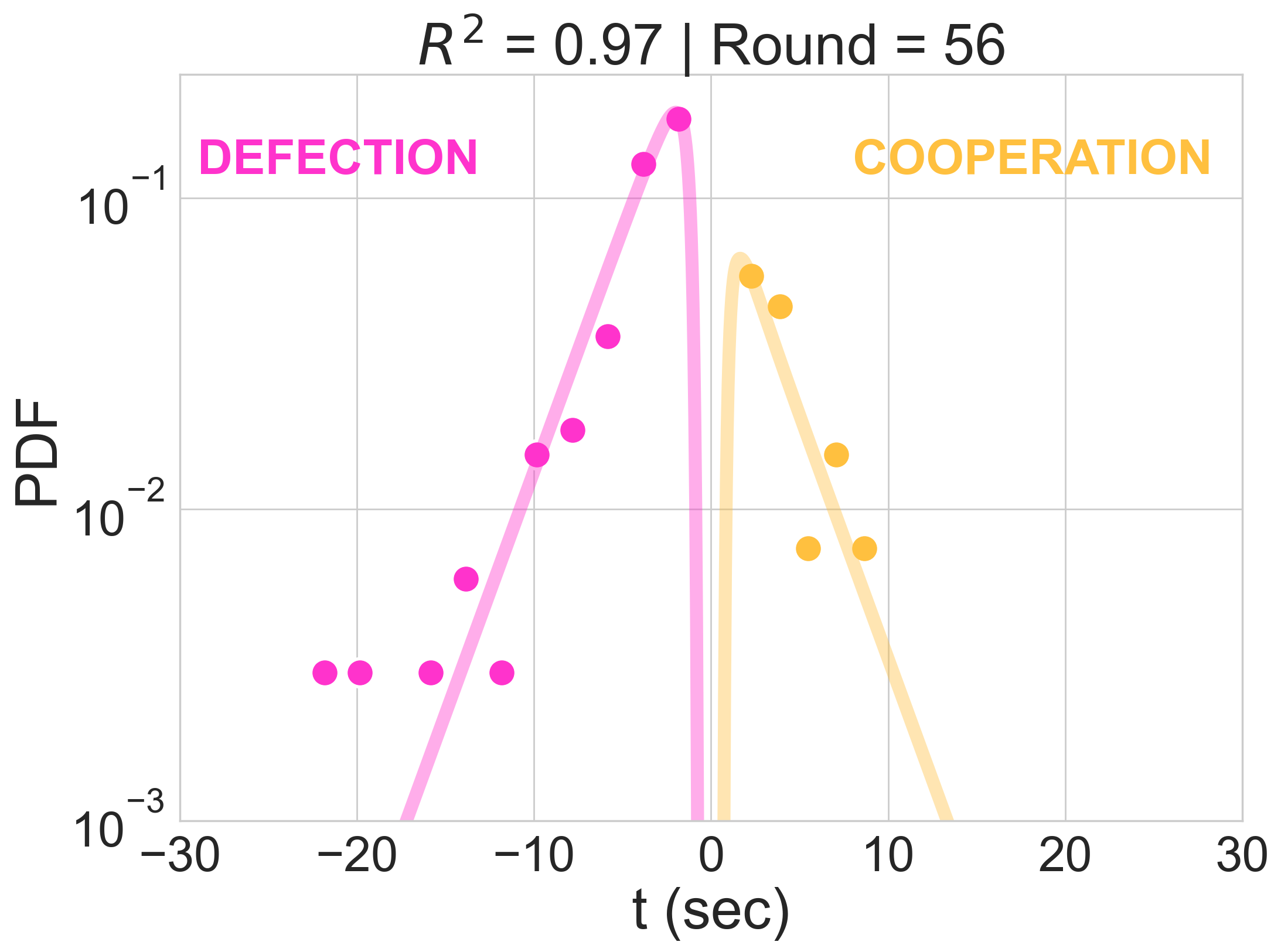}
                 \caption{}
             \end{subfigure}
             \begin{subfigure}[b]{0.45\textwidth}
                 \centering
                 \includegraphics[width=\textwidth]{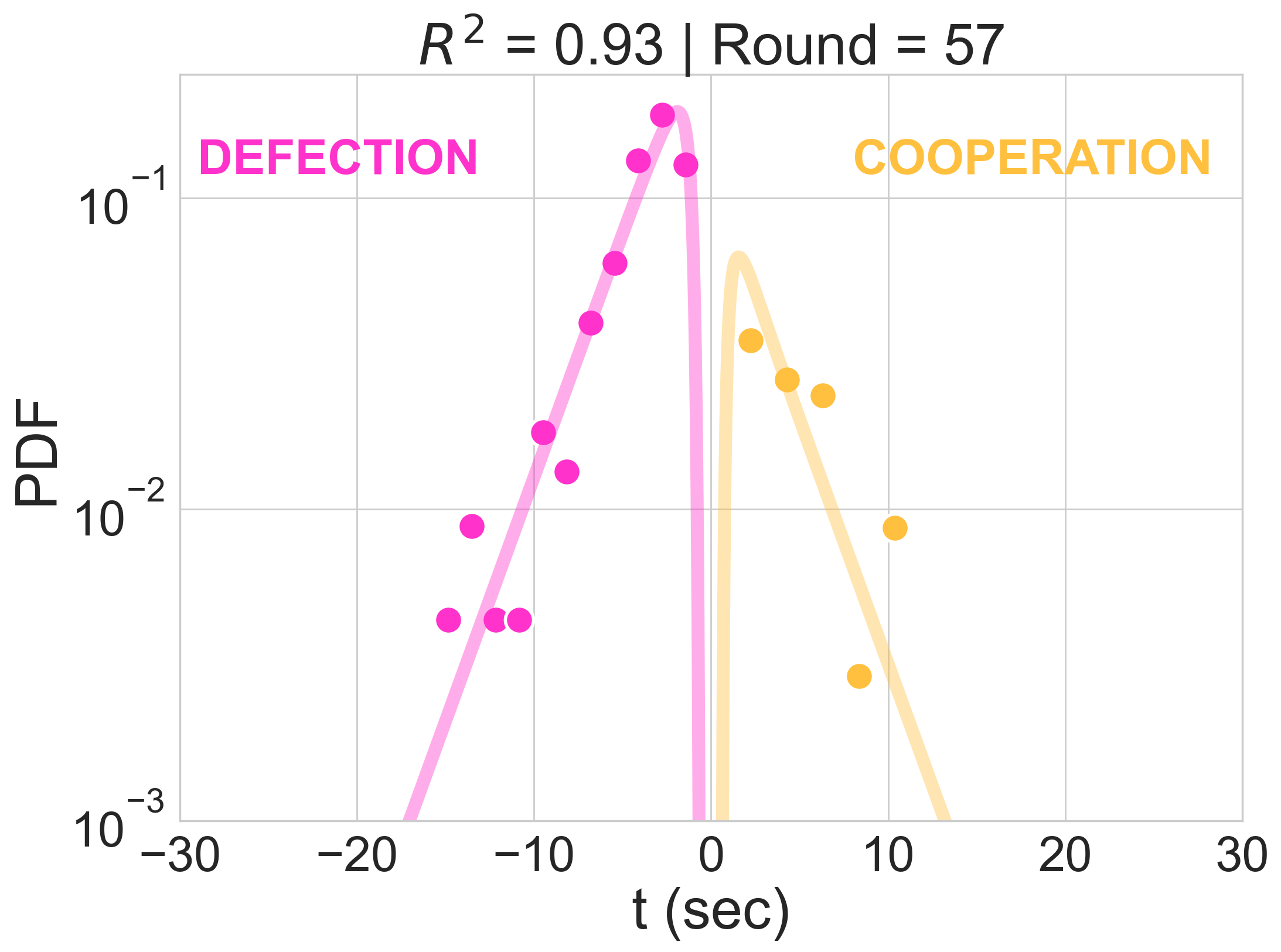}
                 \caption{}
             \end{subfigure}
             \begin{subfigure}[b]{0.45\textwidth}
                 \centering
                 \includegraphics[width=\textwidth]{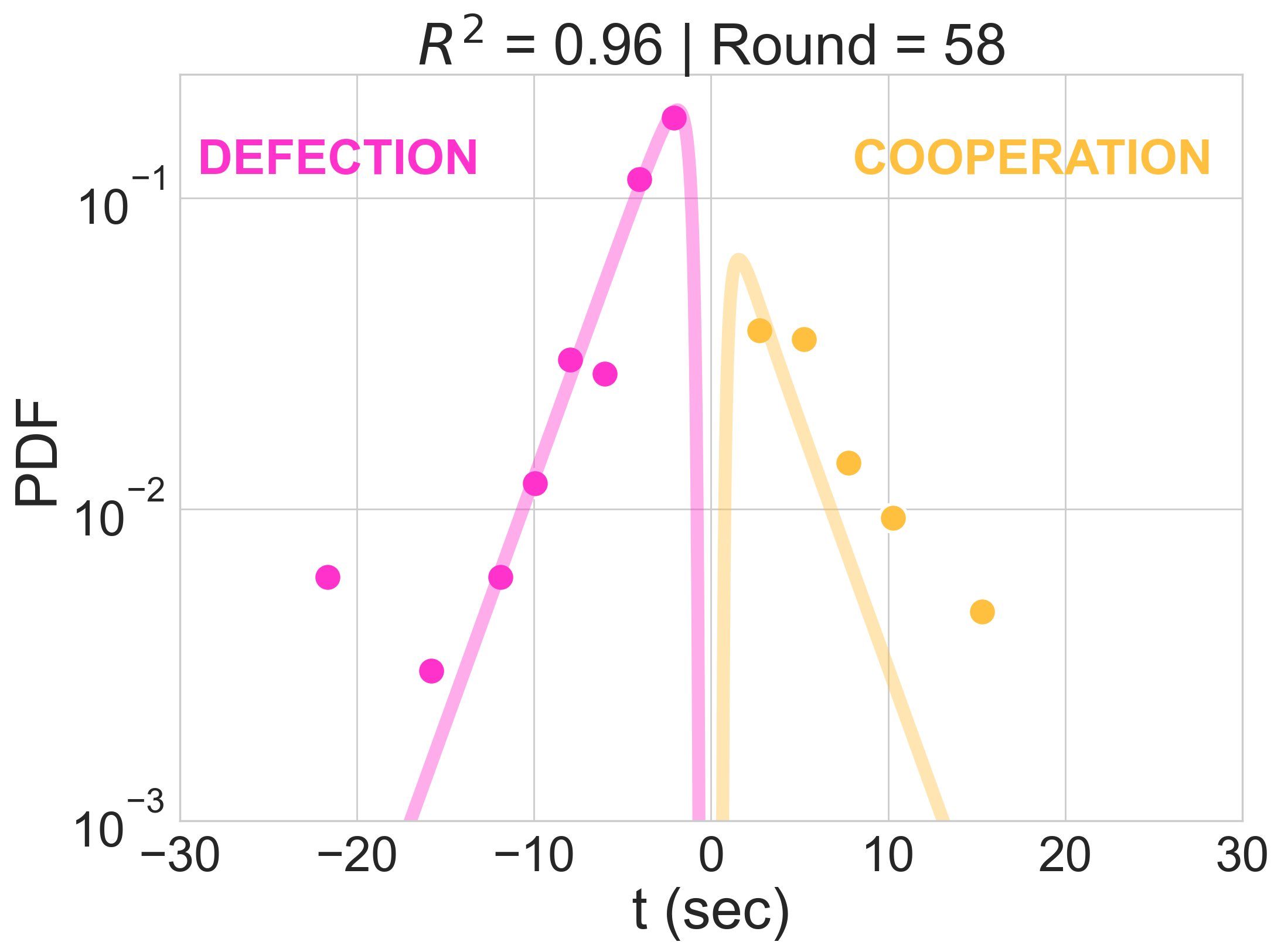}
                 \caption{}
             \end{subfigure}
             \caption{{\bf Accuracy over the testing set.} Response times PDFs for rounds $53-58$. Results are obtained using i) data (dots), and ii) our predictive model (lines). The left side (pink) corresponds to defection responses, while the right side (orange) corresponds to cooperation responses.}
             \label{SI_PDF9}
        \end{figure}
        
        \newpage
        Figures \ref{SI_PDF1}-\ref{SI_PDF9} show the response times PDFs obtained from the theoretical formula in Eq. \ref{RT-PDF}, combined with the evolving free parameters of the underlying DDM, predicted by our model at each round of the test set (represented by lines), compared with the empirical response times PDFs, obtained from the dataset (represented by dots). Notice that the figures associated with round $3$, $37$, and $51$ where included as an example in the main paper.
        
        \section*{Performance of the model on the training set}
         
\begin{figure}[H]
            \centering
            \begin{subfigure}[b]{0.45\textwidth}
                \centering
                \includegraphics[width=\textwidth]{plots/PDF_scheme.png}
                \caption{}
                \label{schemePDF}
            \end{subfigure}
            \begin{subfigure}[b]{0.45\textwidth}
                \centering
                \includegraphics[width=\textwidth]{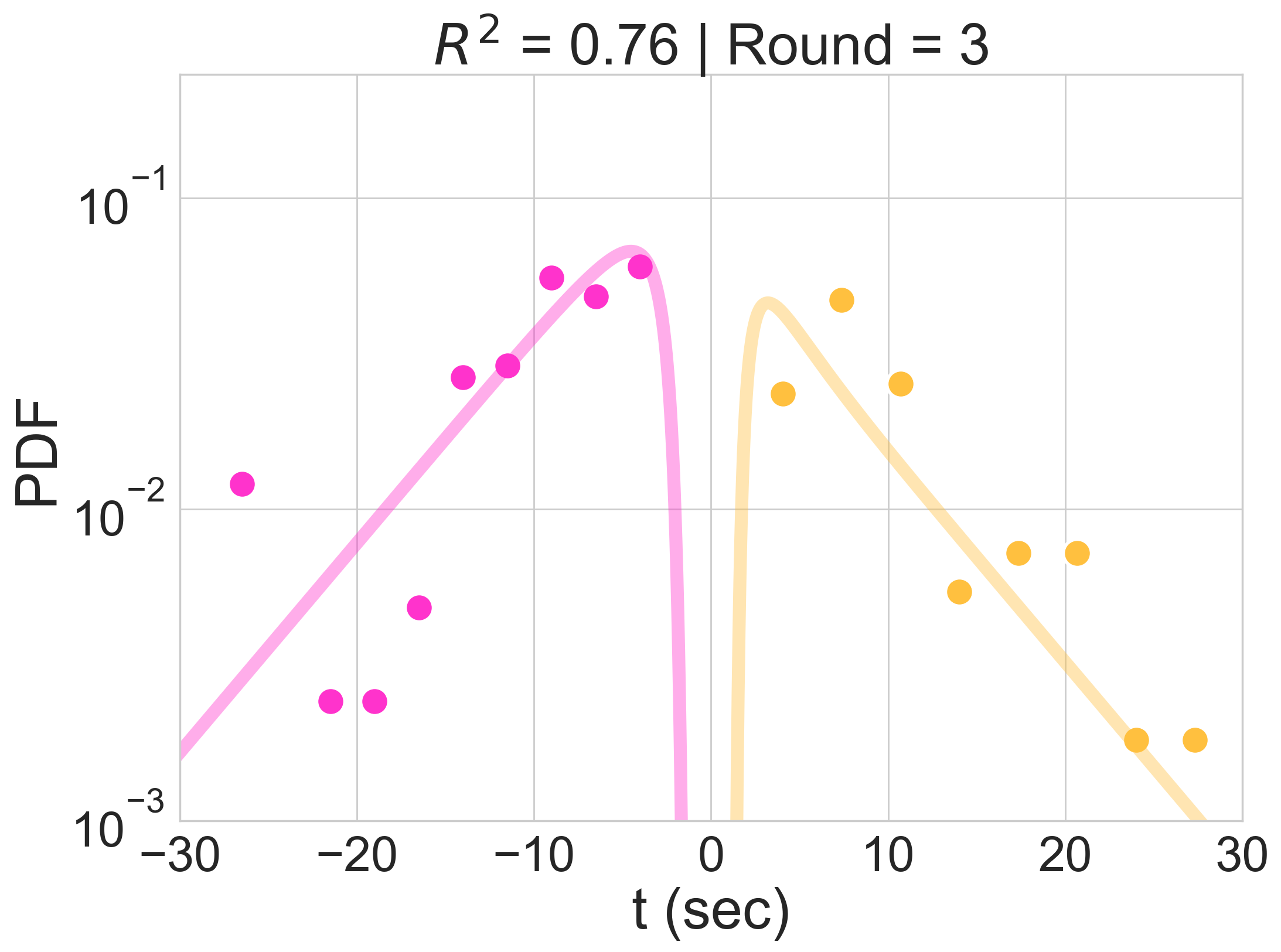}
                \caption{}
                \label{PDF_3training}
            \end{subfigure}
            \begin{subfigure}[b]{0.45\textwidth}
                \centering
                \includegraphics[width=\textwidth]{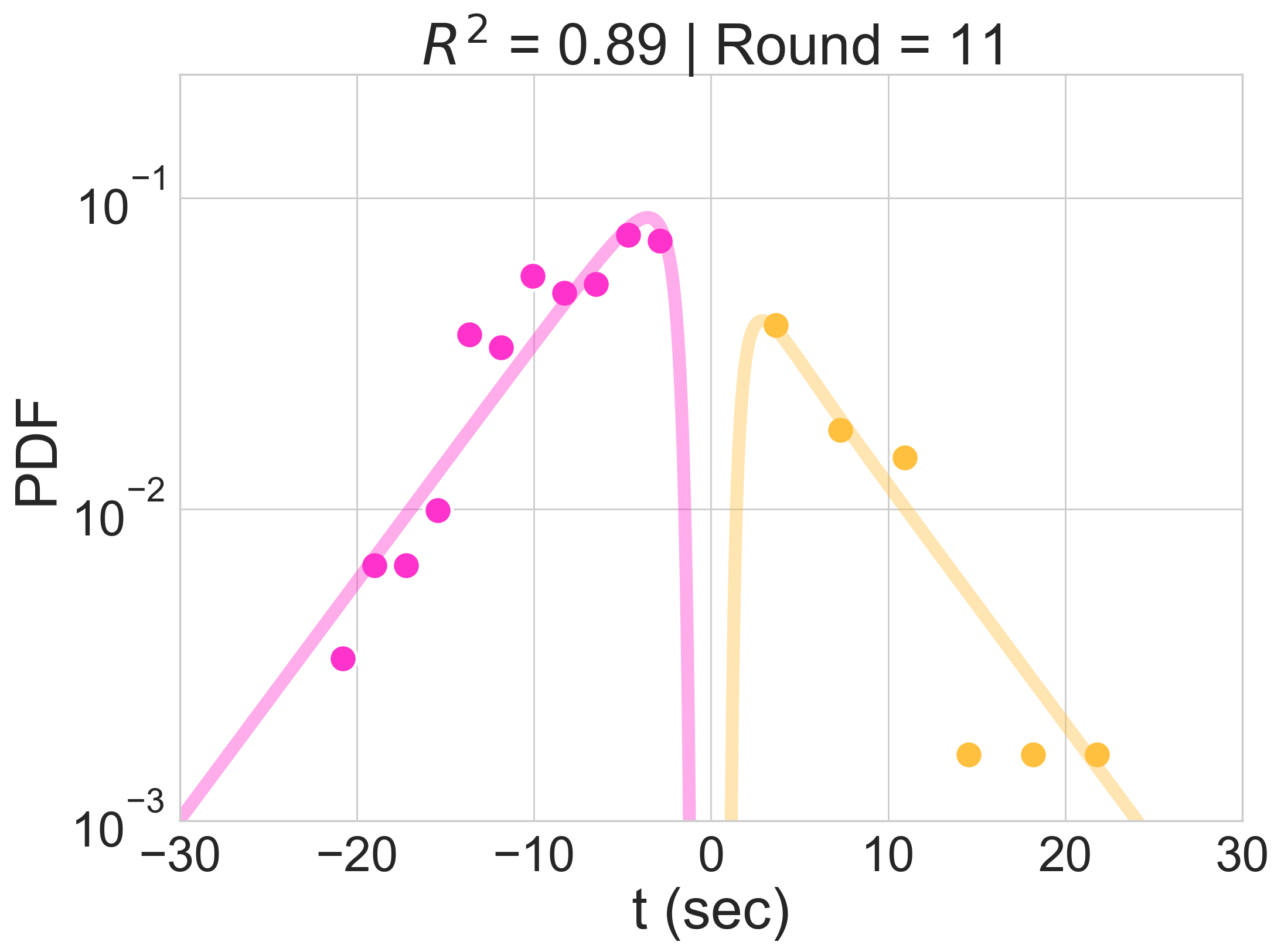}
                \caption{}
                \label{PDF_11training}
            \end{subfigure}
            \begin{subfigure}[b]{0.45\textwidth}
                \centering
                \includegraphics[width=\textwidth]{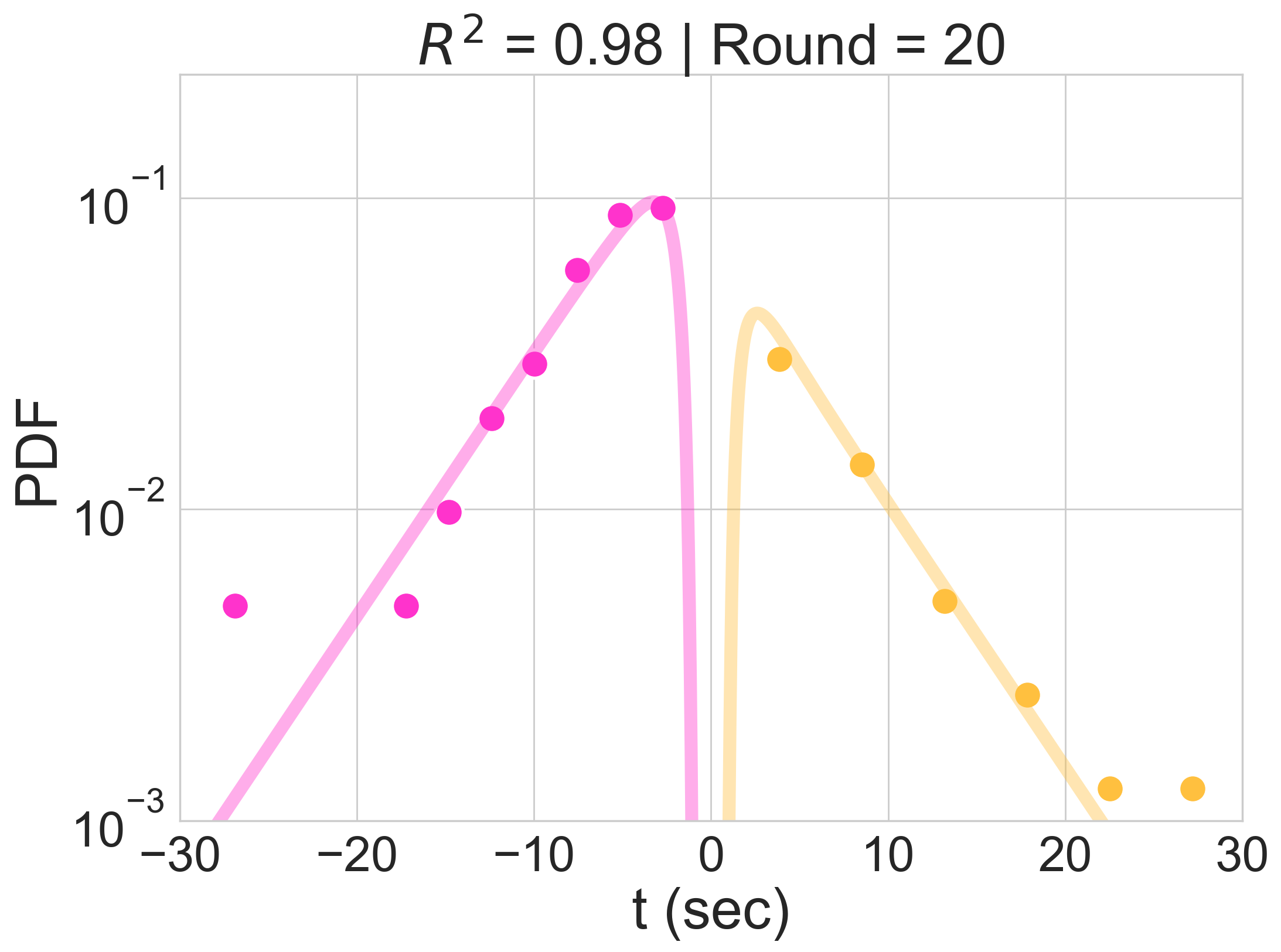}
                \caption{}
                \label{PDF_20training}
            \end{subfigure}
            \caption{\textbf{Accuracy over the training set. Panel \ref{schemePDF}} illustrates how to read the plots: the response times PDF in case of defection and cooperation are shown on the left side (pink), and right side (orange) respectively. Both curves are given by the upper equation for the PDFs, and their integral (the area below the curve) represent the expected rate (equation below). \textbf{Panels \ref{PDF_3training}-\ref{PDF_11training}-\ref{PDF_20training}} show the PDFs at rounds $3$ (up-right), $11$ (down-left), and $20$ (down-right). Results are obtained by: i) Eq.~\ref{RT-PDF} using i) data (dots), and ii) our predictive model (lines).} 
            \label{PDF-train}
        \end{figure}

\begin{figure}[H]
        \includegraphics[width=\textwidth]{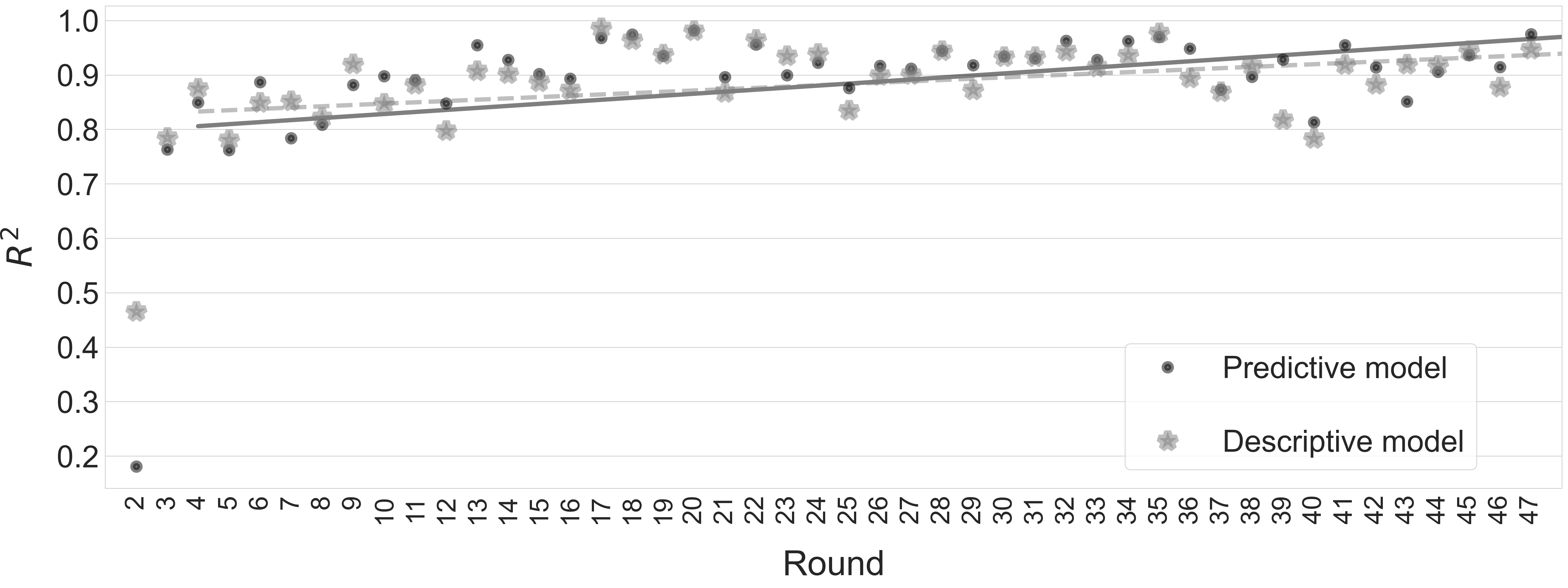}
        \caption{$R^2$ values attained on the training dataset by PDFs employing parameters directly fitted at each round via Bayesian Regression (gray), and by PDFs employing parameters predicted by our model (black)} 
        \label{rsquared_training}
        \end{figure}

\begin{figure}[H]
            \centering
            \includegraphics[width=1\textwidth]{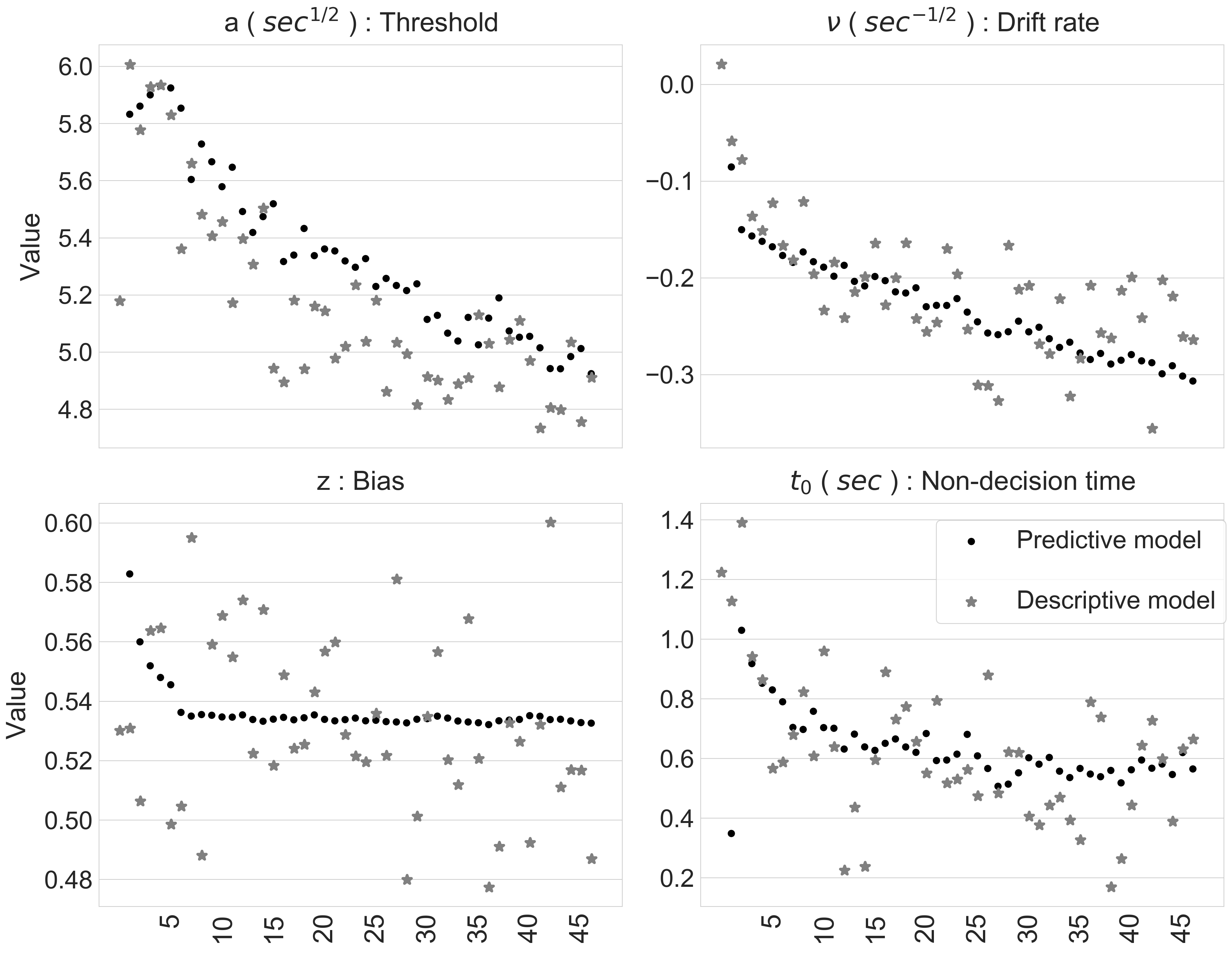}
            \caption{{\bf DDM parameters evolution over the training set}. The figure includes parameters directly fitted at each round via Bayesian Regression (grey), and parameters predicted by our model (black).}
            \label{train-param-comparison}
        \end{figure}

        In this section we include in Figures \ref{PDF-train}-\ref{rsquared_training} the results obtained by our model in replicating the training data (namely, Experiment 1 in section Data of Materials and Methods, in the main paper), in terms of response time PDFs prediction. In Figure \ref{rsquared_training}, the R-square values obtained by our predictive model are compared with the ones achievd by parameters \textit{a-posteriori} fitted after each round via Bayesian Regression, as done in literature by \cite{gallotti2019quantitative}. Moreover, we show in Fig.\ref{train-param-comparison} the evolution of both the DDM parameters predicted by our model, and the ones obtained \textit{a-posteriori} (as previously done for the results on the test set presented in section Results of the main paper).

\end{document}